\documentclass[prb,twocolumn,amsmath,amsfonts,10pt]{revtex4-1}
\usepackage{graphicx}
\usepackage{amsmath}
\usepackage[utf8]{inputenc}
\usepackage{color}
\usepackage{bbold} 

 \definecolor{darkgreen}{rgb}{0,0.5,0} 
 \usepackage{color}
 \usepackage[normalem]{ulem}
 \usepackage{cancel}
\begin{document}

\title{{Investigation of metal-insulator like transition through the \\ab initio density matrix renormalization group approach}}
\author{E. Fertitta$^{1}$, B. Paulus$^{1}$, G. Barcza$^{2}$, and \"O. Legeza$^{2}$}
\affiliation{$^1$ Institut f\"ur Chemie und Biochemie - Takustr. 3, 14195 Berlin,\\ Freie Universit\"at Berlin, Germany\\
$^2$Strongly correlated systems "Lend\"ulet" research group,\\ Wigner Research Centre for Physics, P.O.Box 49 Hungary}

\date\today

\begin{abstract}
We have studied the Metal-Insulator like Transition (MIT) in lithium and beryllium ring-shaped clusters through ab initio Density Matrix Renormalization Group (DMRG) method. Performing accurate calculations for different interatomic distances and using Quantum Information Theory (QIT) we investigated the changes occurring in the wavefunction between a metallic-like state and an insulating state built from free atoms. We also discuss entanglement and relevant excitations among the molecular orbitals in the Li and Be rings and show that the transition bond length can be detected using orbital entropy functions. Also, the effect of different orbital basis on the effectiveness of the DMRG procedure is analyzed comparing the convergence behavior.
\end{abstract}

\maketitle

\vspace{1cm}
\section{Introduction}
\begin{figure}[htb]
\centerline{
\includegraphics[width=0.9\columnwidth]{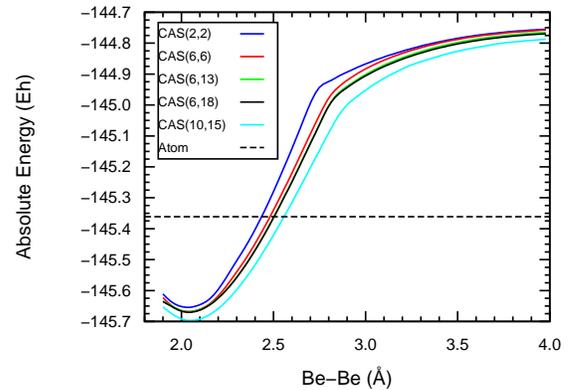}
}
\caption{(Color online) Potential energy curves calculated for the ground state of Be$_{10}$ at the Complete Active Space Self Consistent Field level of theory, using different active spaces and active electrons. The dashed line corresponds to the dissociation limit calculated within the same approach. A minimal basis set has been used in all cases.}
\label{fig:be6_cas}
\end{figure}

We present the investigation of a Metal-Insulator Transition (MIT)\cite{MIT1,MIT2,MIT3,MIT4,MIT5} in pseudo-onedimensional systems, \emph{i.e.} lithium and beryllium rings, through the ab initio Density Matrix Renormalization Group (DMRG) \cite{white92,white93}. Despite the apparent simplicity of these model systems, a meaningful description of the transition from the metallic regime, close to the equilibrium bond length, till dissociation limit, reveals a high complexity due to correlation effects which cannot be fully described using traditional multiconfigurational approaches. As an example, we report in Fig.~\ref{fig:be6_cas} different potential energy curves (PES) of Be$_{10}$ ring, calculated using a minimal atomic basis set within the Complete Active Space Self Consistent Field (CAS-SCF) method\cite{ref_CASSCF}. As one can see, despite the use of larger and larger active spaces (AS) gives a finer and finer description of the PES, more and more of the correlation energy is retrieved, the plateau always lies at much higher energy values than the dissociation limit calculated within the same method using an AS consisting of the 2$s$ and 2$p$ functions of the free atoms. The challenge of calculating the PES of Be$_n$ rings, within a multiconfigurational approach, arises indeed from the high correlation effects, due to the quasi degeneracy of valence 2$s$ and virtual 2$p$ orbitals.  This means that CAS(2$n$, 4$n$) calculations would be required and this becomes of course difficult if systems with more than four Be atoms are considered. DMRG helps us to overcome this dimensionality problem that we encounter with standard CAS methods, whose computational effort increases exponentially.\\
Lithium rings also offer a particular interesting model system and the calculation of their whole PES allows to study the transition from a metallic state with a half filled 2$s$ band to an insulting state made by isolated atoms. In such a transition, the character of the wavefunction varies dramatically which is reflected by the change of the electron correlation from predominantly dynamic correlation in the metallic case to static correlation in the dissociation limit. This can be analyzed using Quantum Information Theory (QIT), in particular studying the change in orbital entanglement. Indeed, besides reaching energies comparable to the Full-CI, DMRG gives us the chance to calculate important quantities as the one-site entropy\cite{legeza2003} and the two-sites mutual information \cite{legeza2006-qpt,rissler06,barcza2010a,boguslawski2012b,boguslawski2013a}. In this work we show how these can be employed to analyze the MIT and identify the position of the transition.\\
Finally, we will investigate the effect of the orbital basis on the DMRG results. Although, several DMRG works can be found in literature where various orbital basis were employed to study quantum chemical systems \cite{legeza-lif,rissler06,barcza2010a,boguslawski2012b,boguslawski2013a,gosh2008,luo2010,mitrushchenkov2011,ma2013,wooters2013,kurashige2013}, no rigorous analysis in terms of resulting entanglement patterns have been carried out, yet. As we will show, the use of different basis as well as the starting Hartree-Fock configuration can have a huge impact on the effectiveness of the method. Indeed, despite in all cases the result converges toward the Full-CI limit, we can observe how the choice of a local orbital basis instead of a canonical orbital basis might help to reach that level with less computational effort.\\
Moreover, despite we present quantum chemical results, it is important to underline that the chosen systems can be easily compared with physical models. In particular the use of localized function will allow us to interpret the systems in term of a Hubbard model.\\

\section{Numerical Procedures}
\subsection{Basis states}

\begin{figure}[htb]
\centerline{
\includegraphics[width=0.9\columnwidth]{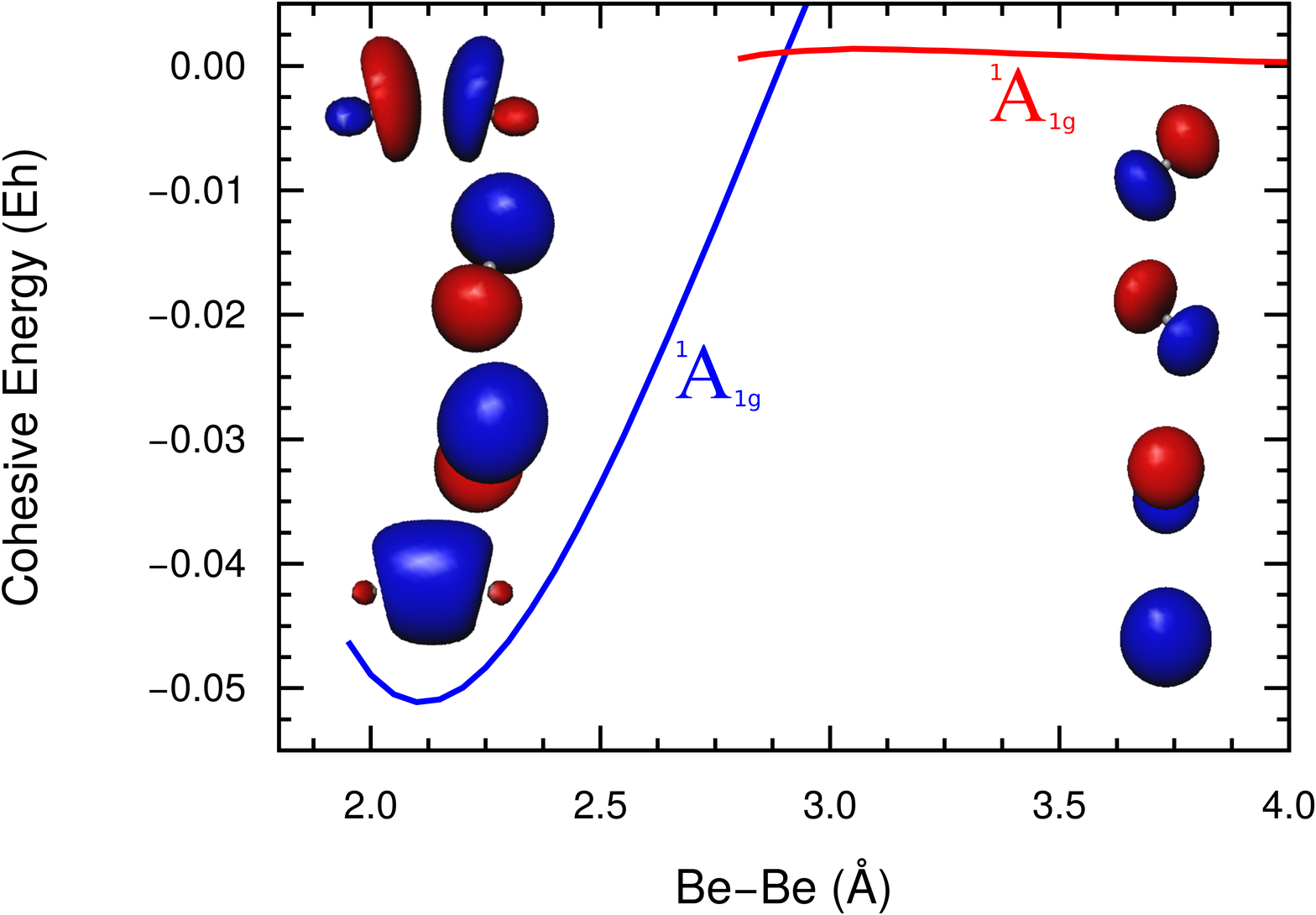}
}
\caption{(Color online) Hartree-Fock ground state potential energy curves calculated for Be$_{10}$ determined with the two leading configurations. Additionally the valence and virtual localized orbitals are shown for the two configurations, obtained by an unitary tranformation of the corresponding canonical orbital basis.}
\label{fig:configurations}
\end{figure}

As previously mentioned, we focused our interest on the use of different orbital basis to expand the Hilbert space and we investigated their effect on the effectiveness of the DMRG calculations. Indeed, even though, in principle, the same result will be obtained using different orbital basis constructed from the same atomic basis set, the quantum entanglement, which is crucial in the DMRG routine, strongly depends on this choice. The orbital basis that we considered were the canonical orbitals, used to describe the Hartree-Fock (HF) wavefunction, and the localized orbitals (LOs) obtained from a possible unitary transformation of this basis.\\
As sketched in Fig.~\ref{fig:configurations}, in the case of beryllium rings two main HF configurations are important, depending on the distance regime, in order to describe the PES till dissociation. Around the minimum, the Hartree-Fock wavefunction presents a high $p$ character while at larger distances, the HF configuration consists of doubly occupied linear combinations of (almost) pure 2$s$ orbitals. We will refer to these as configuration 1 and 2, respectively. In the two cases, the orbital localization will then yield different LOs. In the first case $\sigma$-like orbitals with a $sp$ character will emerge, while from configuration 2, one will obtain atom-like 2$s$ and 2$p$ basis (see Fig.~\ref{fig:configurations}).\\
In order to describe lithium rings, we also used a local orbital basis. In this case, the occupied LOs were obtained using not only the $n/2$ doubly occupied valence orbitals, but also the first $n/2$ virtual orbitals with a mainly $s$ character. The starting occupation for each LO was then set to one.\\
Hartree-Fock calculations, Foster-Boys localization procedure\cite{Boys} and generation of integral files for DMRG calculations were all performed using the MOLPRO quantum chemistry package\cite{MOLPRO}. The atomic basis sets employed consisted of 1$s$, 2$s$ and 2$p$ functions only, and 1$s$ orbitals were always kept frozen. In particular a STO-3G was used for lithium\cite{STO-3G} while for beryllium we employed a basis set relying on $cc$-pVDZ\cite{cc_pVDZ} where only two contracted $s$ functions and one contracted radial $p$ function were used.

\subsection{Density matrix renormalization group method}

\begin{figure}[htb]
\centerline{
\includegraphics[width=0.9\columnwidth]{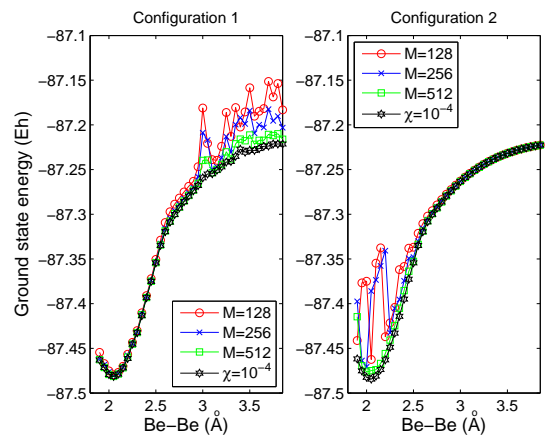}
}
\caption{Ground state energy of Be$_6$ as a function of Be-Be distance for various fixed number of block states, $M$, and for $\chi=10^{-4}$ using configuration 1 (left) and configuration 2 (right) and localized orbital basis.}
\label{fig:be6_e_dmrg}
\end{figure}

In order to study the rings built from Li and Be atoms we have employed the quantum chemical version \cite{white99} of the density matrix renormalization group (DMRG) method \cite{white92,white93}. In the past decade this method has been proved to be a rival to the conventional multiconfiguration wave function approaches and nowadays it allows us to study much larger CAS configurations than conventional methods. In our numerical procedure we also utilize various concepts inherited from quantum information theory (QIT) \cite{legeza2003,legeza2004,rissler06,barcza2010a} which allows us to use DMRG as a black box method \cite{barcza2010a,boguslawski2012b,boguslawski2013a}. In the QC-DMRG applications, the electron-electron correlation is taken into account by an iterative procedure that minimizes the Rayleigh quotient corresponding to the Hamiltonian of the system. For more detailed derivations we refer to the original papers and review articles~\cite{white99,legeza-rev,chan-rev,reiher-rev,yanai-rev}.\\
The amount of contribution to the total correlation energy of an orbital can be detected by the single-orbital entropy, $s(1)_i=-{\rm Tr} \rho_i \ln \rho_i$ where $\rho_i$ is the reduced density matrix at orbital $i$. The two-orbital entropy is constructed similarly using the reduced density matrix, $\rho_{ij}$ of a subsystem built from orbitals $i$ and $j$ and the mutual information $I_{ij}=s(2)_{ij}-s(1)_i-s(1)_j$ describes how orbitals are entangled with each other as they are embedded in the whole system. For more detailed derivations we refer to the original papers \cite{legeza2003,rissler06,barcza2010a,boguslawski2013a}. Therefore, these quantities provide chemical information about the system, especially about bond formation and nature of static and dynamic correlation \cite{boguslawski2012b,boguslawski2013a,boguslawski2013b,yanai2013}.\\
%
In order to use QC-DMRG as a black box method one has to carry out a few optimization steps. First, the arrangement of orbitals along a one-dimensional topology has to be optimized (ordering) in order to reduce the set of Schmidt ranks when the system is systematically partitioned into a left and right parts during the DMRG sweeping procedure\cite{legeza2003}. This allows us to carry out calculations with much smaller number of block states using the dynamical block state selection (DBSS) approach\cite{legeza2002,legeza2004}. This is achieved by minimizing the entanglement distance, expressed as a cost function, ${\hat I}_{\rm dist}=\sum_{i,j} I_{ij} |i-j|^\eta,$ where the entanglement between pair of orbitals is weighted by the distance in the chain between the orbitals. Using $\eta=2$ has the advantage that this optimization task can be carried out using concepts of spectral graph theory~\cite{atkin98}. It follows that the so called Fiedler vector $x=(x_1, \dots x_N)$ is the solution that minimizes $F(x)=x^\dagger L x=\sum_{i,j} I_{ij} (x_i-x_j)^2$ subject to the following constraints that $\sum_i x_i=0$ and $\sum_i x_i^2=1$, where the graph Laplacian is $L_{ij}=D_{ij}-I_{ij}$ with $D_{ii}=\sum_j I_{ij}$. The second eigenvector of the Laplacian is the Fiedler vector \cite{fiedler73,fiedler75} which defines a (1-dimensional) embedding of the graph on a line that tries to respect the highest entries of $I_{ij}$ and the edge length of the graph. Ordering the entries of the Fiedler vector by non-decreasing or non-increasing way provides us a possible ordering. Based on our numerical experiences from the past ten years, we have found, although no rigorous proof is given yet, that the best ordering obtained with small number of block states also provide almost the best ordering for calculation performed with large number of block states.\\
An other optimization task is performed in order to speed up the warm-up sweep of the DMRG procedure. Therefore, in order to achieve fast and stable convergence we also utilize the configuration interaction based on a dynamically extended active space (CI-DEAS) procedure\cite{legeza2003,legeza2004-leiden,barcza2010a}. In this method the active space is expanded iteratively using orbitals with largest one-orbital entropy values. The sequence by which orbitals are taken into account is determined by the so called CAS-vector which is simply a rendered sequence of orbital indices with decreasing one-orbital value.\\
Therefore, our black-box QC-DMRG is composed of two phases: the {\em preprocessing phase} in which the ordering and CAS-vector are optimized using fixed small number of block states and the {\em production phase} in which an accurate calculation is performed using the DBSS procedure in order to reach an a priory set error margin. In the preprocessing phase, we first use the ordering for which the integral files were generated and a random CAS vector using $M=64$ block states. We calculate the one-orbital entropy from which we obtain the CAS vector and we also determine the two-orbital mutual information and the optimal ordering by calculating the Fiedler vector. Next a DMRG calculation is carried out with the optimized ordering and CAS-vector and the whole cycle is repeated until we obtain lower total energy. In the next step this procedure is repeated, but with $M=256$ states. In the present study, we have performed the accurate calculations in the production phase using the the DBSS procedure with an a priory set value of quantum information loss $\chi=10^{-4}$ in each DMRG renormalization and truncation step and using a minimum number of block states $M_{\rm min}=512$. The preprocessing phase takes only a small fraction of the total computational time.\\
In Fig.~\ref{fig:be6_e_dmrg} we show results for Be$_6$ using configuration 1 and configuration 2. It is clear that the accuracy of DMRG fluctuates strongly as a function of Be-Be distance when the number of block states are kept fixed while the error can be controlled very efficiently using the DBSS approach.\\
Finally, we want to underline that, since the mutual information is orbital basis dependent, besides orbital ordering, the entanglement distance $I_{\rm dist}$ can be manipulated by changing the orbital basis. Therefore, performance of QC-DMRG can be optimized by using proper choice of the orbital basis, {\emph i.e.}, the same state can be obtained with much smaller number of block states\cite{legeza2005}. Our entanglement analysis can be used in this respect as well as will be shown below.

\subsection{Dependence of entanglement on orbital basis and ordering}

\begin{figure}[htb]
\centerline{
\includegraphics[width=0.5\columnwidth]{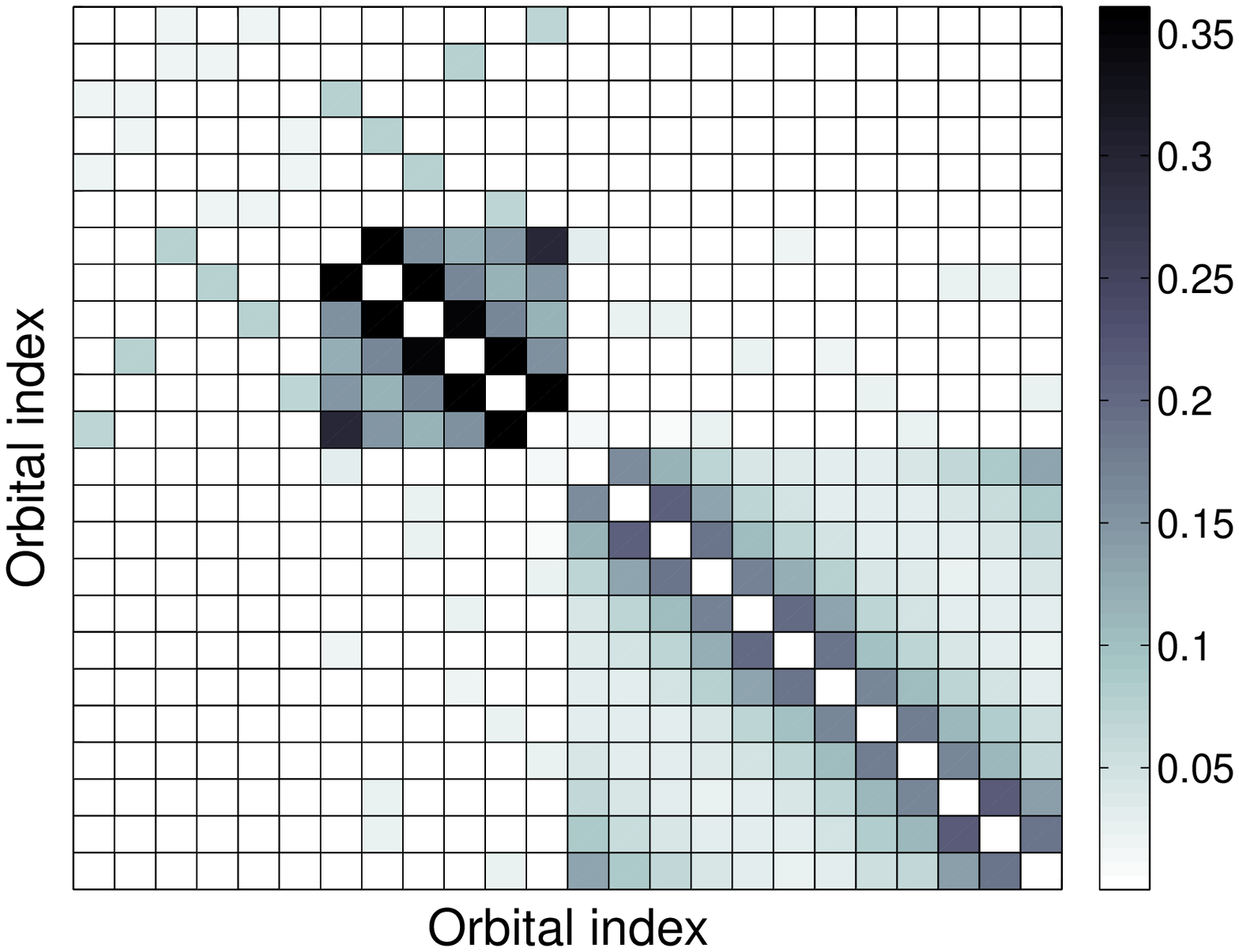}
\includegraphics[width=0.5\columnwidth]{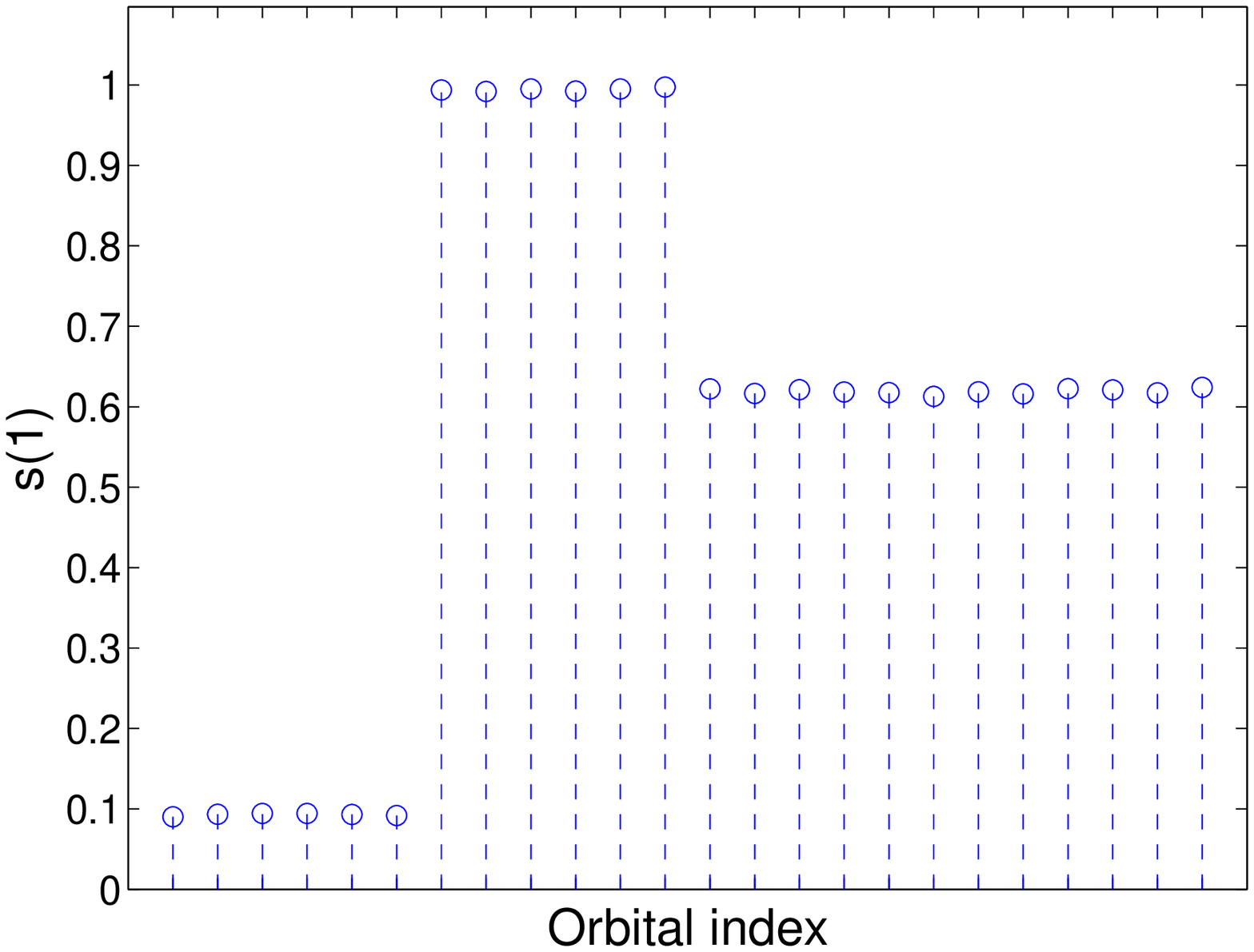}
}
\caption{(Color online) Orbital ordering optimization using the Fiedler vector for the ground state for Be$_6$ close to the equilibrium structure, $d_{\rm Be-Be}=2.15$\AA, using the DMRG method with local orbitals using configuration 2. (a) Colorscaled plot of two-orbital mutual information and (b) single-orbital entropy profile. $I_{\rm tot}=13.9, I_{\rm dist}=365.58$.}
\label{fig:I_Be6_2.15_localized}
\end{figure}

In this section we report data obtained for Be$_6$ using $M=512$ block states. Both configurations were employed and a comparison between local and canonical orbital basis is shown. As expected, regardless the chosen orbital basis, DMRG reaches the same state, but some choices can bring better results. This depends of course on the entanglement of the orbitals in the different situations.\\
Let us consider the dissociation limit first. Both in canonical and local description we have different sets of many degenerate orbitals, but while the degenerate canonical orbitals, which are delocalized over the whole system, are highly entangled, the localized functions sitting on different centers present a low correlation between each other. Moreover, how it could be expected, the LOs derived by the configuration 2 converges more effectively in this range.\\
\begin{figure}[htb]
\centerline{
\includegraphics[width=0.5\columnwidth]{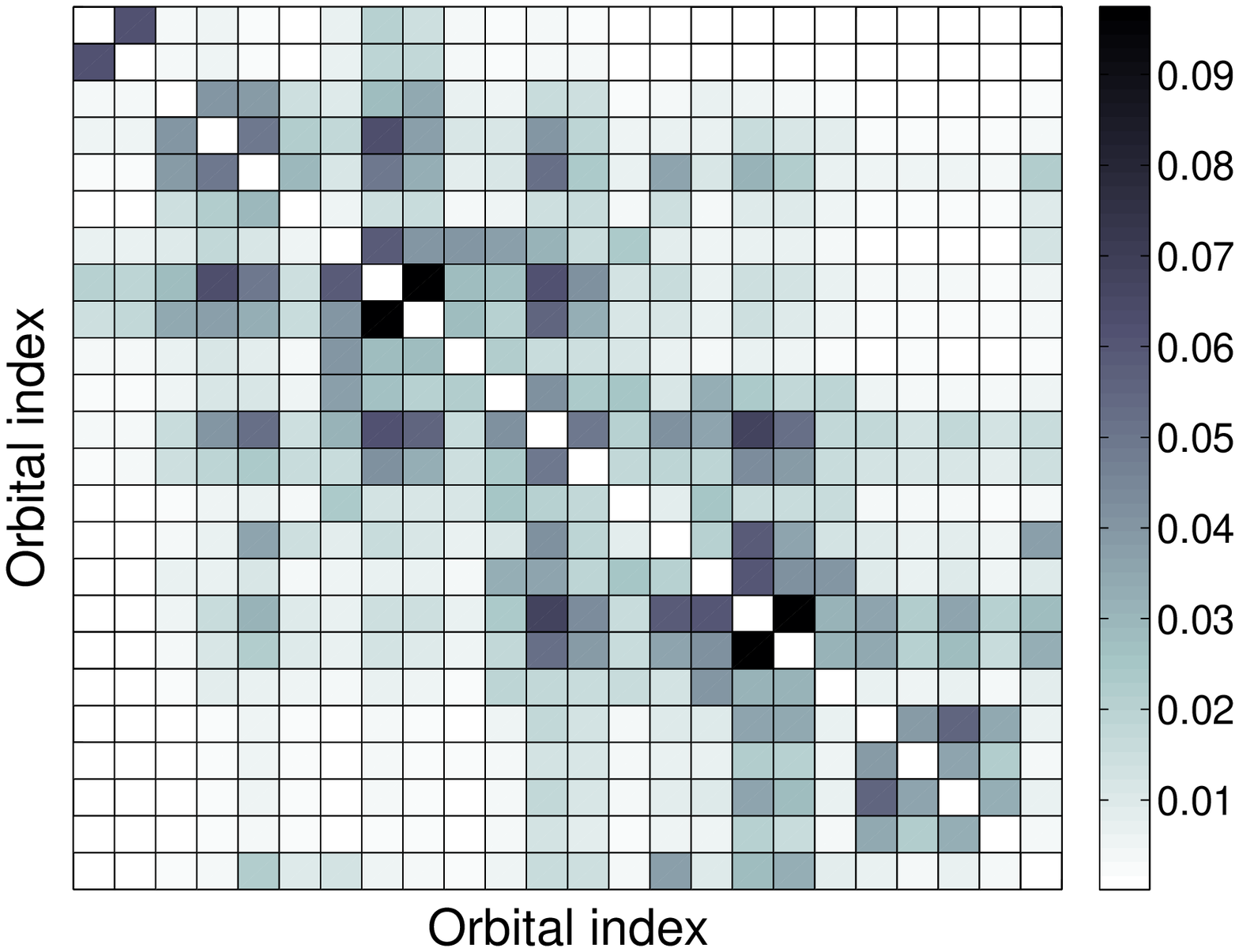}
\includegraphics[width=0.5\columnwidth]{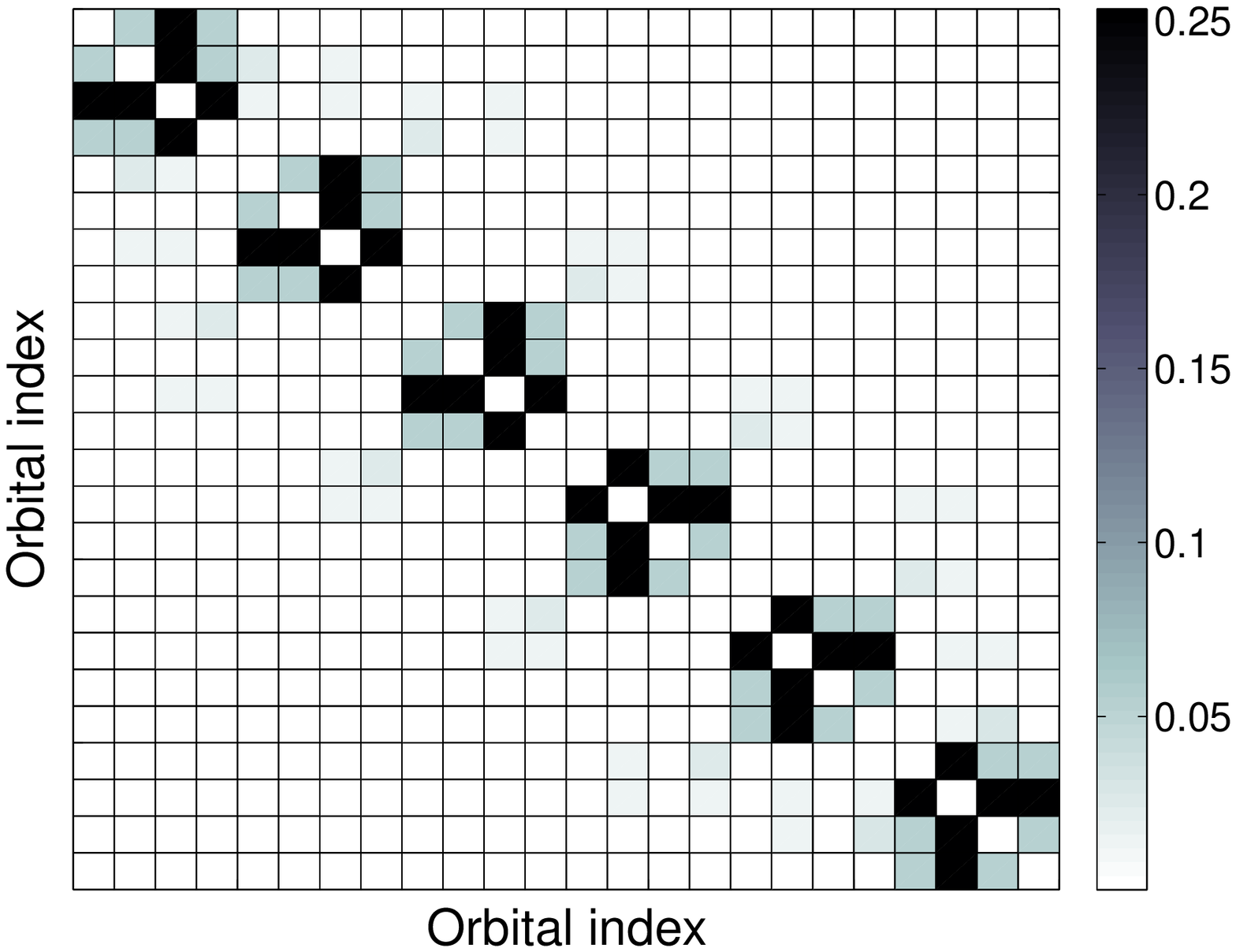}
}
\centerline{
\includegraphics[width=0.5\columnwidth]{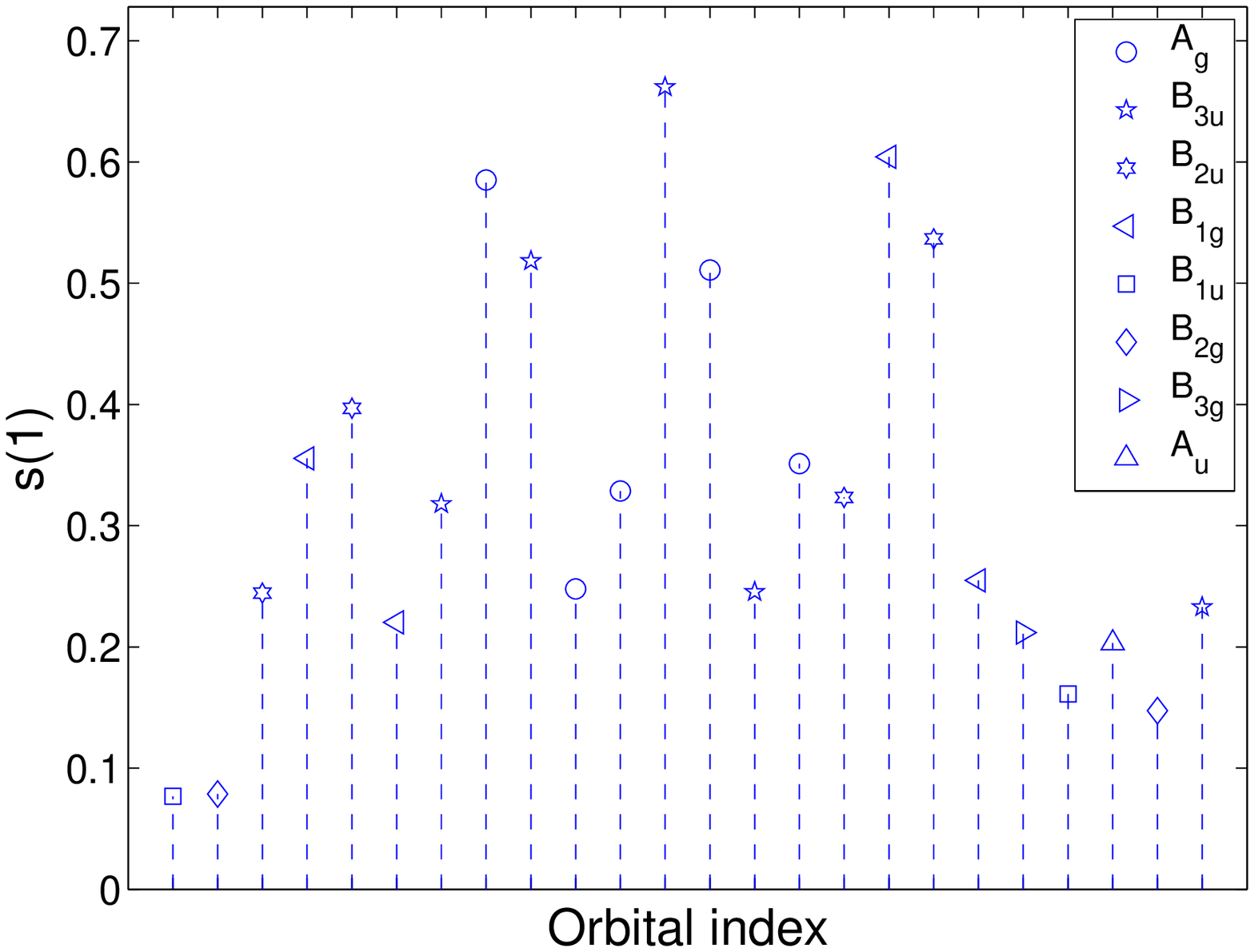}
\includegraphics[width=0.5\columnwidth]{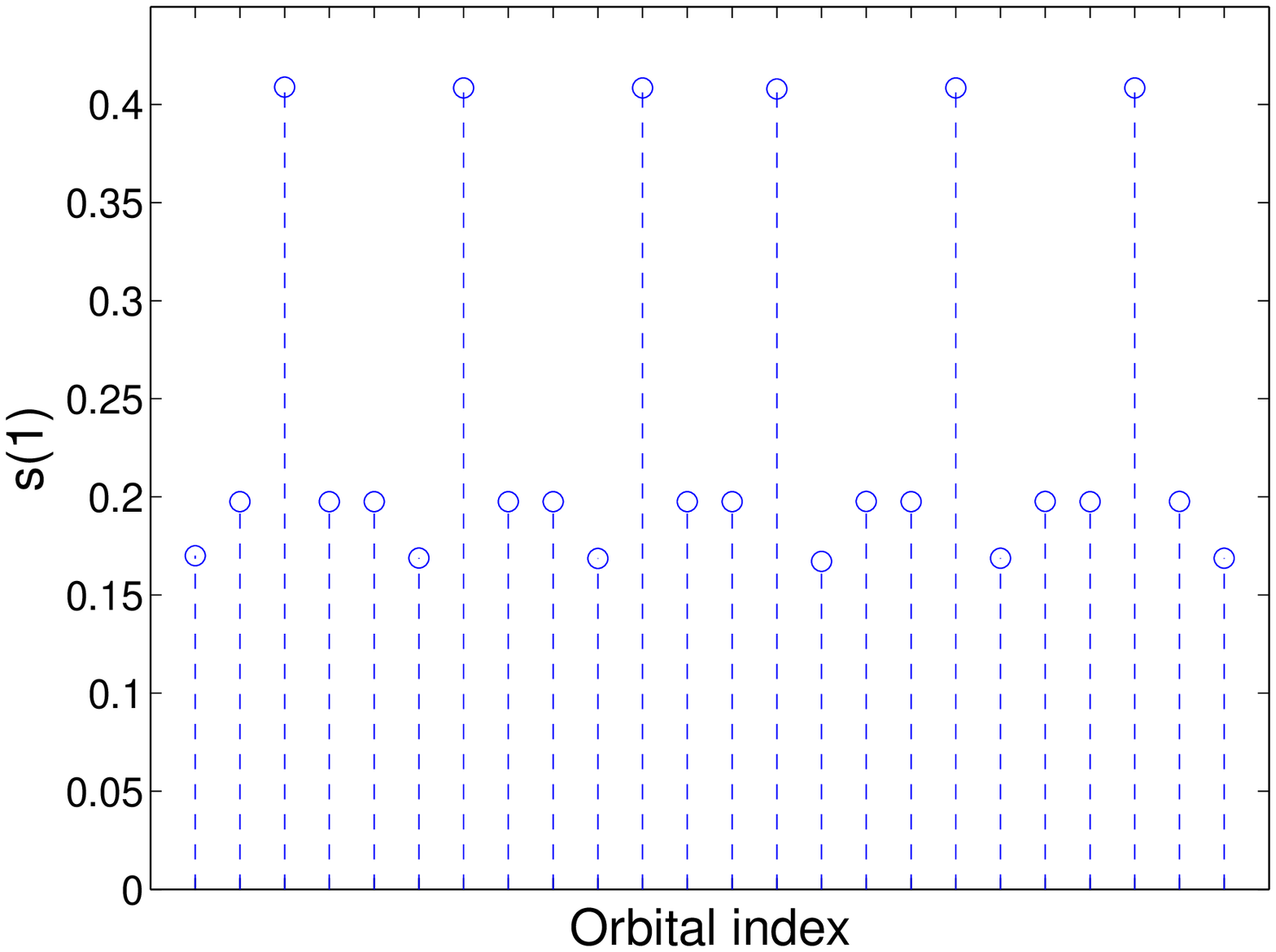}
}
\caption{(Color online) Orbital ordering optimization using the Fiedler vector for the ground state for Be$_6$ for a stretched structure, $d_{\rm Be-Be}=3.30$\AA, using the DMRG method with canonical (left) and local (right) orbitals using configuration 2. Colorscaled plot of two-orbital mutual information (upper) and single-orbital entropy profile (lower). $I_{\rm tot}=7.81 , I_{\rm dist}=332.38$ with the canonical basis and $I_{\rm tot}=5.83 , I_{\rm dist}=58.1$ with the local basis.}
\label{fig:I_Be6_3.30_canon_conf2}
\end{figure}
\begin{figure}[htb]
\centerline{
\includegraphics[width=0.5\columnwidth]{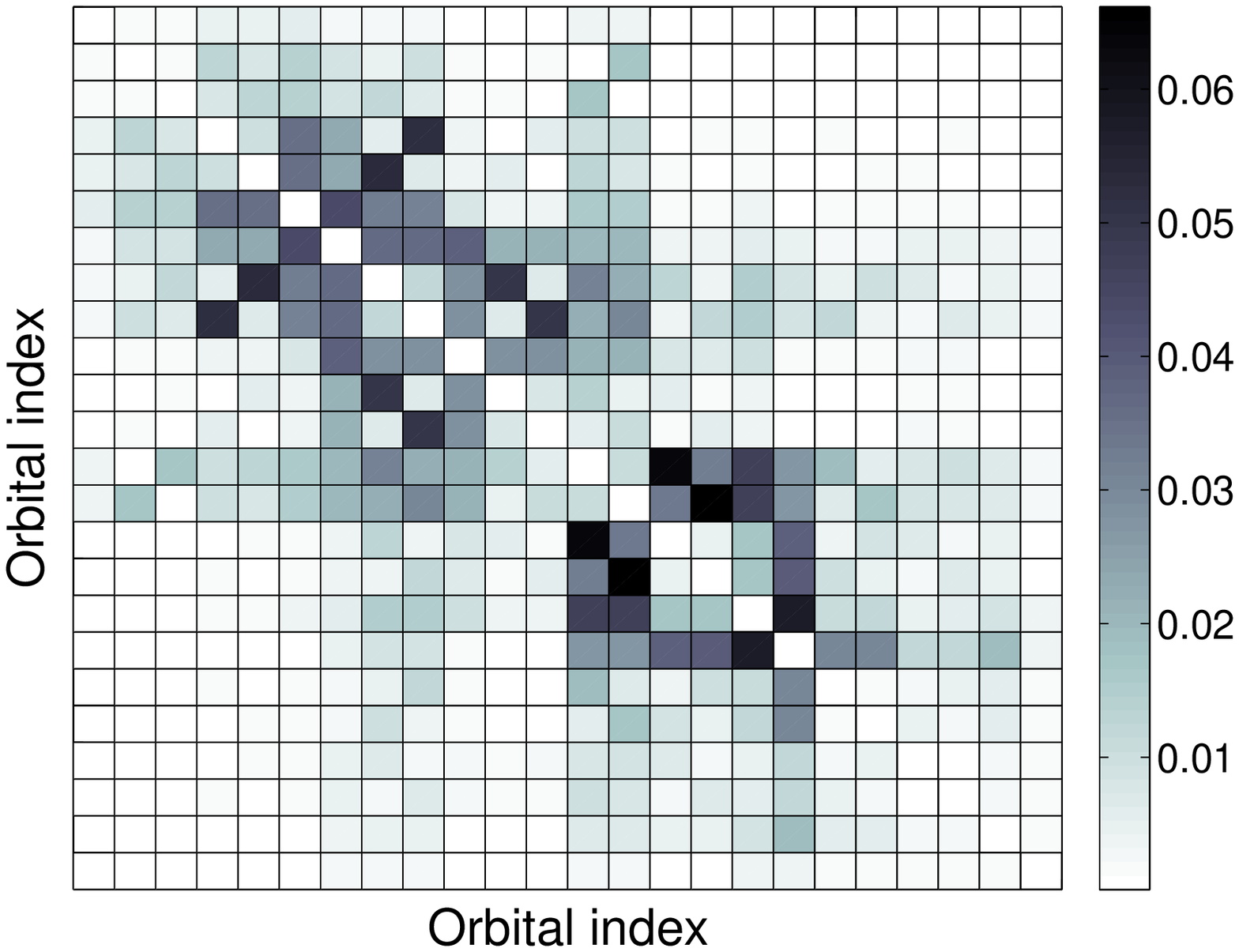}
\includegraphics[width=0.5\columnwidth]{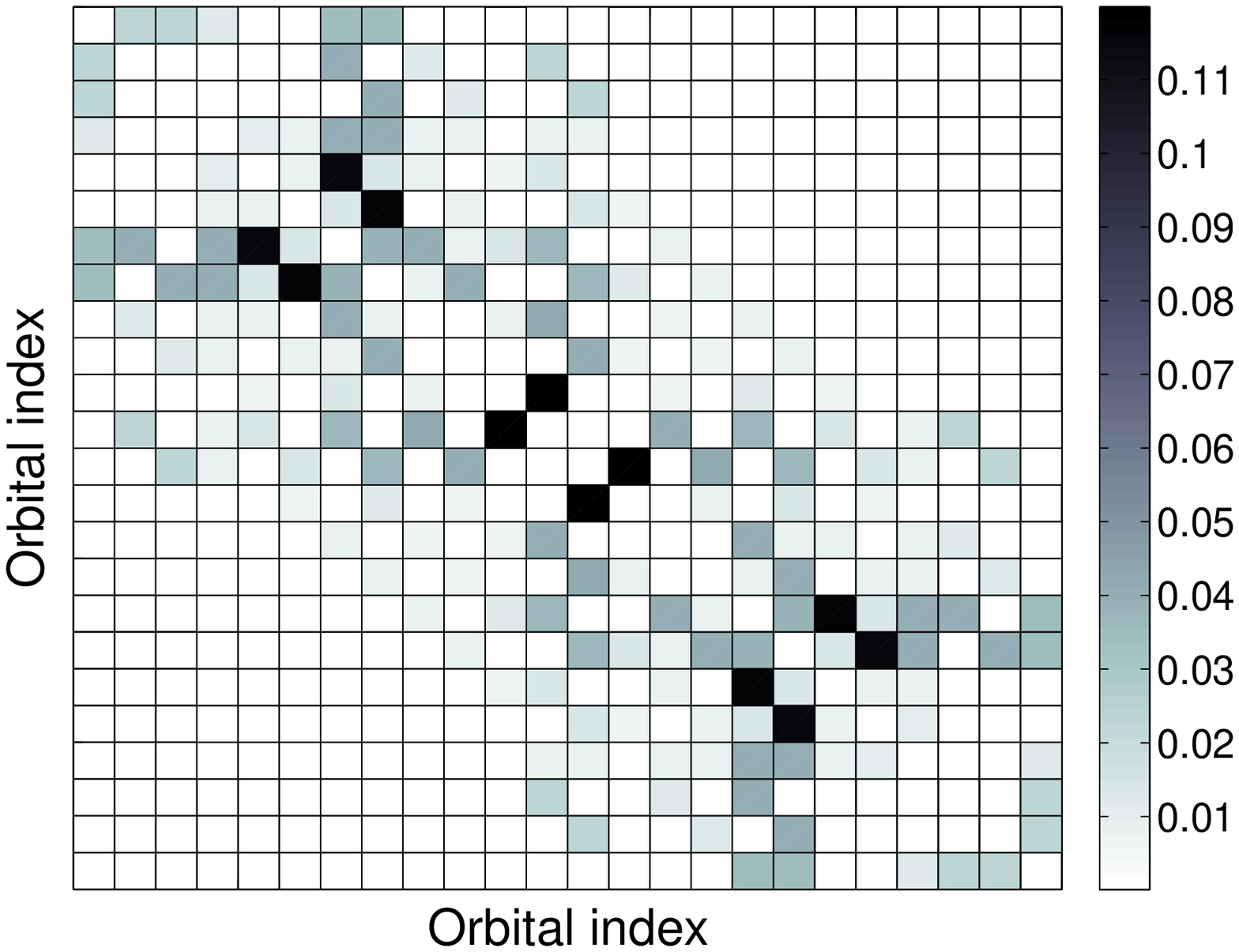}
}
\centerline{
\includegraphics[width=0.5\columnwidth]{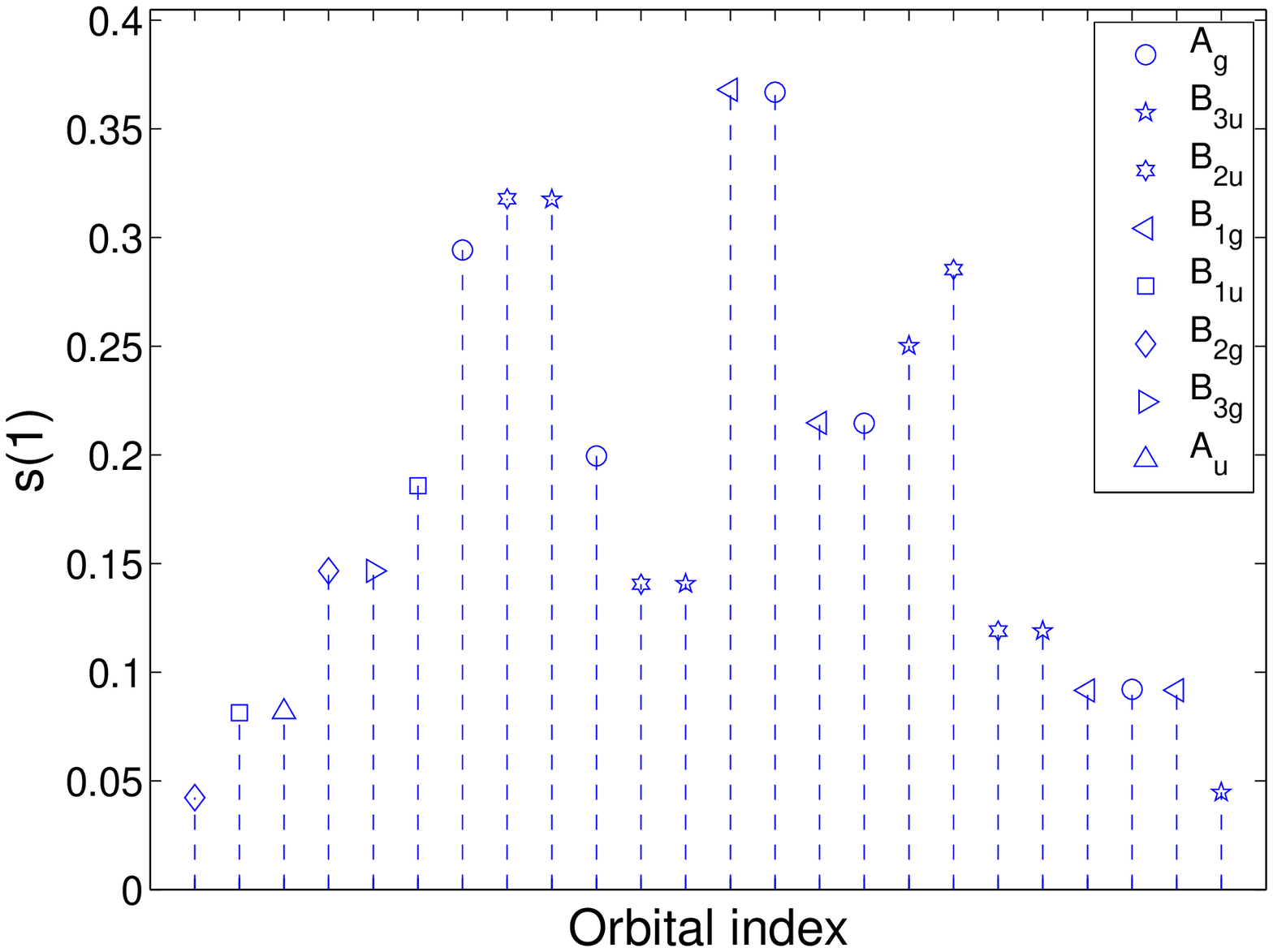}
\includegraphics[width=0.5\columnwidth]{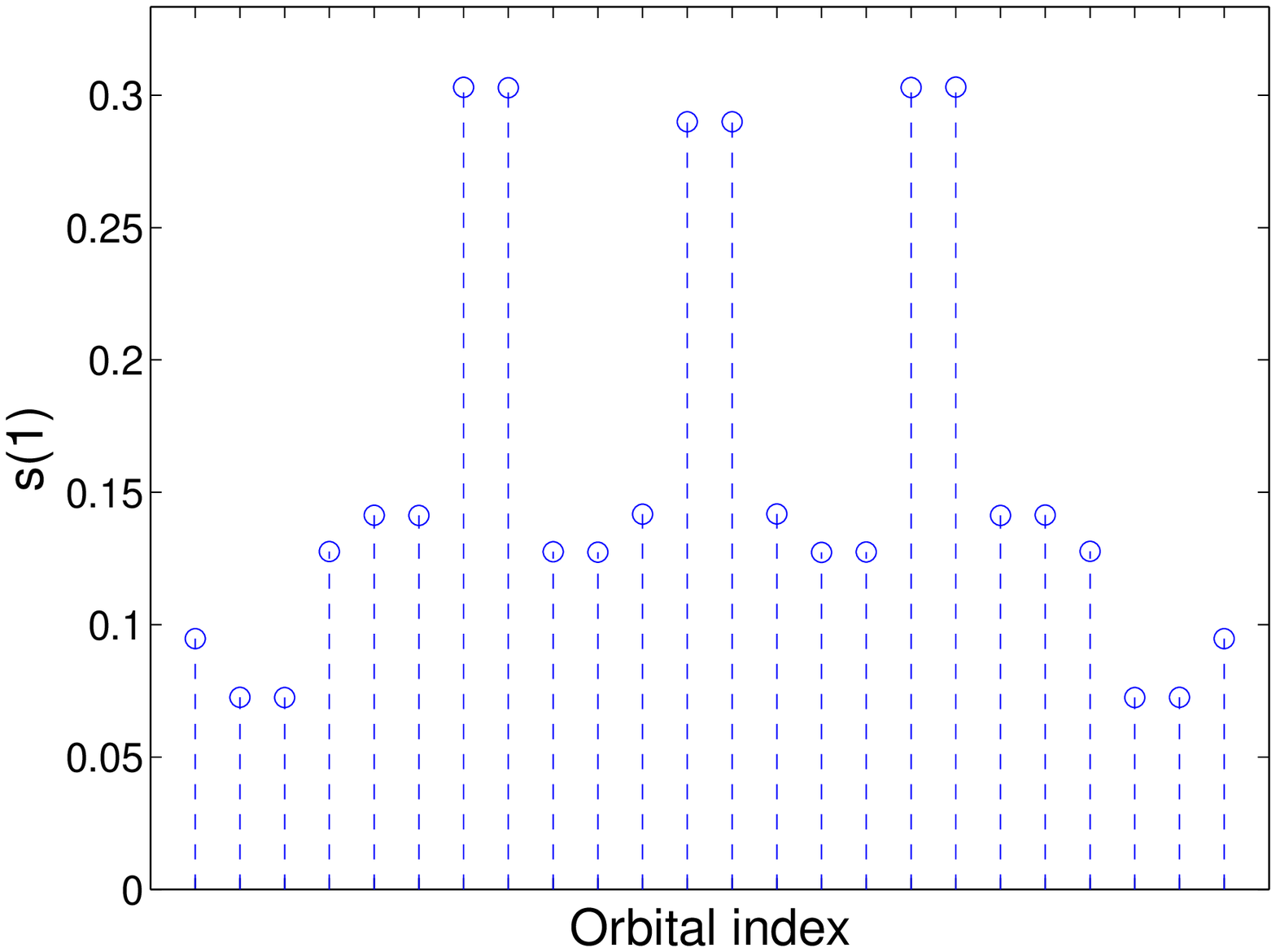}
}
\caption{(Color online) Orbital ordering optimization using the Fiedler vector for the ground state for Be$_6$ close to the equilibrium structure, $d_{\rm Be-Be}=2.15$\AA, using the DMRG method with canonical (left) and local (right) orbitals using configuration 1. Colorscaled plot of two-orbital mutual information (upper) and single-orbital entropy profile (lower). $I_{\rm tot}=4.35, I_{\rm dist}=149.70$ with the canonical basis and $I_{\rm tot}=3.88, I_{\rm dist}=100.38$ with the local basis.}
\label{fig:I_Be6_2.15_canon_conf1}
\end{figure}
As an example, the two-orbital mutual information and the one-orbital entropy using the Fiedler vector based ordering optimization are shown in Fig.~\ref{fig:I_Be6_3.30_canon_conf2} (left panels) for the ground state of Be$_6$ in the insulating regime, $d=3.30$\AA, with canonical orbitals and configuration 2. The same are plotted for the case where localized orbitals are used in the right panels. It is evident from the figure that the two-orbital mutual information is more diagonally dominant when localized orbitals were used, and $I_{\rm dist}$ dropped from 332 (canonical orbitals) to 58 (localized orbitals). This of course has a tremendous effects on the performance of DMRG since for canonical orbitals the number of required block states to reach the a priory set error margin $\chi=10^{-3}$ has reached $M=3000$ while for localized orbitals it did not grow above $M=512$. It is also remarkable to note that the ordering provided by the Fiedler vector satisfies all intrinsic demands for DMRG with open boundary condition (OBC). Although the Be$_6$ ring is a rotationally invariant system, the orbitals of Be atoms are arranged along the 1-D chain-like order in a way that the one-orbital entropy profile has a left-right symmetry (see lower-right panel in Fig. \ref{fig:I_Be6_3.30_canon_conf2}) and there is no coupling between its two ends. Such configuration is the best choice for DMRG with OBC.\\
On the other side, in the metallic state, the situation is inverted since the canonical orbitals are distributed in larger energy range and only some of them present a high orbital entropy and are important for the construction of configurations with higher weight. This does not happen if local function are used, since all of them have important weight. Nevertheless with a proper choice of ordering and block states the local orbitals offer a valid choice of orbital basis for any structure. As an example, we report results using configuration 1 for the equilibrium structure in Fig.~\ref{fig:I_Be6_2.15_canon_conf1} for the canonical (left panels) and localized orbitals (right panels), respectively. In this case, $I_{\rm dist}$ is again reduced when localized orbitals were used but with a much smaller rate. The optimal ordering provided by the Fiedler vector again gives a symmetric one-orbital entropy distribution and the couplings between the two ends of the chain is minimized. Using the DBSS approach a slightly larger number of block states were needed to reach the same accuracy threshold using canonical orbitals but the difference was much less significant than for the insulating case.\\
After showing a comparison between canonical and local orbital basis within the DMRG approach, we want to analyze the behavior of the same localized orbitals for different interatomic distances. In Fig.~\ref{fig:I_Be6_2.15_localized} we report mutual information and one-orbital entropy for Be$_6$ close to the ground state minimum and using configuration 2. It is important to remember that this starting HF configuration is not an ideal starting point for this structure, but DMRG yields the same state whatever basis is employed. Nevertheless this choice has a tremendous effect on the entanglement and then on the efficiency of the method. In this case, functions on different orbitals are more correlated than orbitals on the same orbitals while, as stated above, using the same configuration for the insulating state only orbitals sitting on the same Be atom are highly entangled. As one can see, comparing Figs.~\ref{fig:I_Be6_3.30_canon_conf2} and ~\ref{fig:I_Be6_2.15_localized}, the mutual information has a block diagonal form for the optimized ordering for both structure, but these represent a totally different situation. At interatomic distance $d=2.15${\AA } we find three blocks: in the first block 2$p_z$ orbitals of all Be atoms are grouped; the second block is formed from the 2$p_x$ and 2$p_y$ orbitals of the Be atoms, while the third block is composed from the 2$s$ orbitals of the Be atoms. In contrast to this, for $d=3.30${\AA } we find six blocks, each formed by the 2$s$, 2$p_x$, 2$p_y$ and 2$p_z$ of a given Be atom. This is very similar to what observed for lithium, since it shows how close to the minimum, the wavefunction is highly delocalized, while the more the structure is stretched the more it can be described as a state product of orbitals of the individual atoms.\\
If we now consider Configuration 1 once again, we observe in Fig.~\ref{fig:I_Be6_2.15_canon_conf1}(upper right) three groups of orbitals with the same orbital entropy which one easily identify as the valence $\sigma$-like orbitals, the 2$p_z$ functions and the remaining virtual orbitals. The first group present the highest orbital entropy since they are the doubly occupied orbitals of the starting Hartree-Fock configuration.

\section{Entanglement analysis}
The energy is not the only interesting result that can be achieved from DMRG calculations. The entanglement pictures used to optimize the DMRG procedure can also be used to extract information about the system. In this section we show entanglement analysis using localized orbitals for the Li and Be rings. This information can be exploit to learn something about the evolution of the wavefunction along the PES.\\
In addition, as has been shown in Ref.~[\onlinecite{barcza2014}], one can also analyze the sources of entanglement encoded in $I_{ij}$ by studying the behavior of the matrix elements $\rho_{ij}$. These are expressed as the expectation values of {\em  generalized correlation functions} $\langle \Psi |{\cal T}_{i}^{(m)}\,{\cal T}_{j}^{(n)}|\Psi\rangle$ where the transition operator ${\cal T}_{i}^{(m)}$ with $m=1,\dots16$ describes a possible transition between the four initial, $|\alpha\rangle$, and four final states, $|\alpha^\prime\rangle$ of orbital $i$. Therefore, ${\cal T}^{(m)}$ transforms the state $|\alpha\rangle$ with $\alpha = (m-1)({\rm mod } \;\, 4)$ into state $|\alpha^\prime\rangle$ with $\alpha^\prime=\left \lfloor (m-1)/4 \right \rfloor +1$, ${\cal T}^{(m)}|\alpha\rangle=|\alpha^\prime\rangle$, where $\left \lfloor x \right \rfloor$ denotes the floor function, the integral part of $x$. A given generalized correlation function measures the expectation value of the resonance amplitude between the initial and final states within a particular environment and can be expressed as
$
\langle {\cal T}_{i}^{(m)}\,{\cal T}_{j}^{(n)}\rangle =
\sum_{\beta} C^*_{l(m),l(n),\beta} C_{r(m),r(n),\beta} \, .
$
Here the wavefunction of the tripartite system is written as
$
\left|\Psi\right\rangle=\sum_{\alpha_i,\alpha_j,\beta}
C_{\alpha_i,\alpha_j,\beta}\left|\alpha_i,\alpha_j,\beta\right\rangle \, ,
$
where $\alpha_i$ and $\alpha_j$ label the basis of the orbitals $i$ and $j$, and $\beta$ labels the basis of the environment, which is composed of the remaining orbitals. In general, $\langle {\cal T}_{i}^{(m)}\,{\cal T}_{j}^{(n)}\rangle$ contains both connected and disconnected contributions between subsystems $i$ and $j$. In order to circumvent this behavior, we generally study the connected part of the generalized correlation functions,
$
\langle {\cal T}_{i}^{(m)}\,{\cal T}_{j}^{(n)}\rangle_{\rm C}= 
\langle {\cal T}_{i}^{(m)}\,{\cal T}_{j}^{(n)}\rangle- 
\langle {\cal T}_{i}^{(m)}\rangle \langle{\cal T}_{j}^{(n)}\rangle \, , 
$
where the disconnected part, given by the product of the expectation values of the local transition operators, is subtracted out. Note that the mutual information is formulated in such a way that the disconnected parts of the generalized correlation functions do not contribute. As we will see below, these can be used to identify the relevant physical processes that lead to the generation of the entanglement.

\subsection{Li rings with only 2s functions}

\begin{figure}[htb]
\centerline{
\includegraphics[width=0.5\columnwidth]{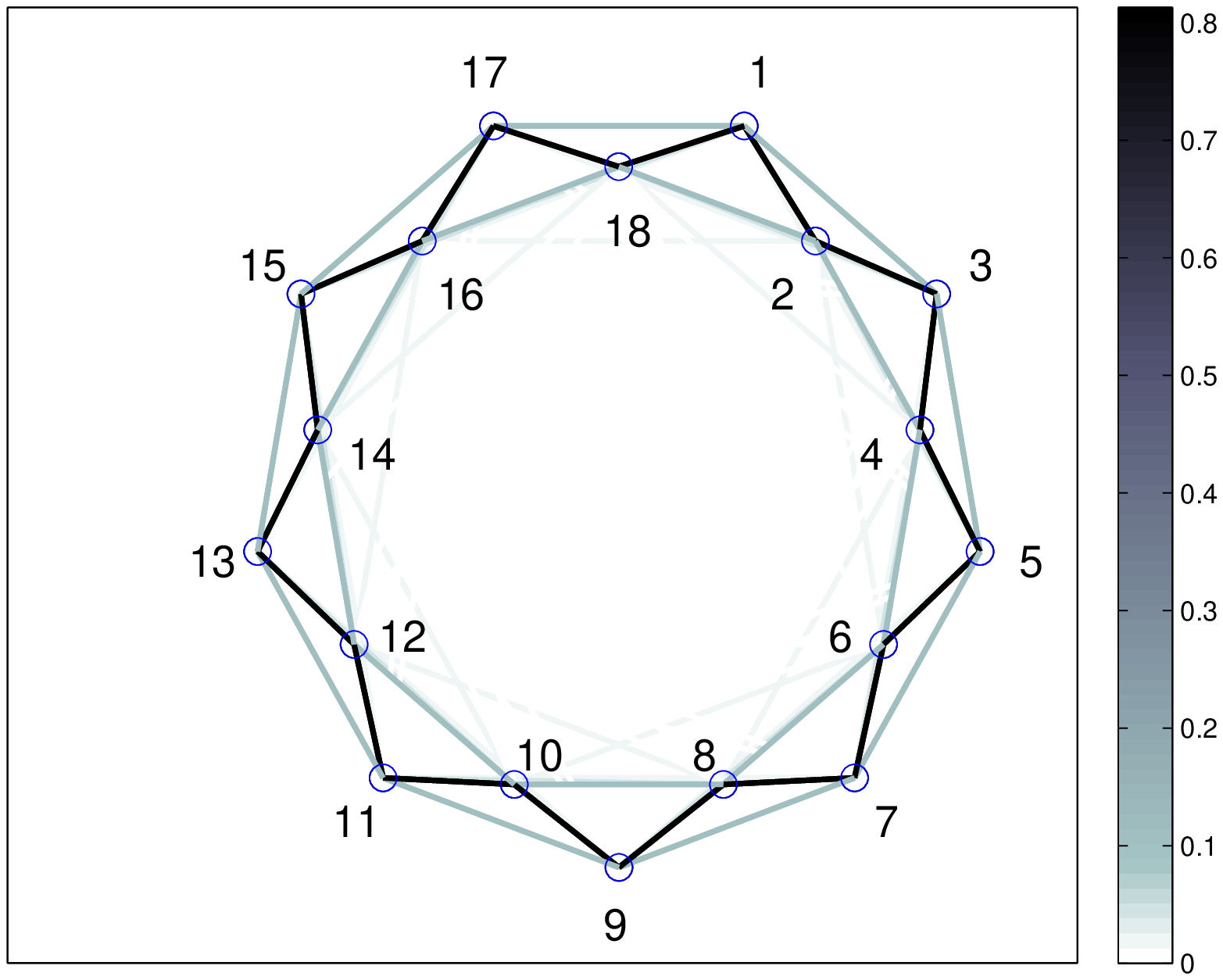}
\includegraphics[width=0.5\columnwidth]{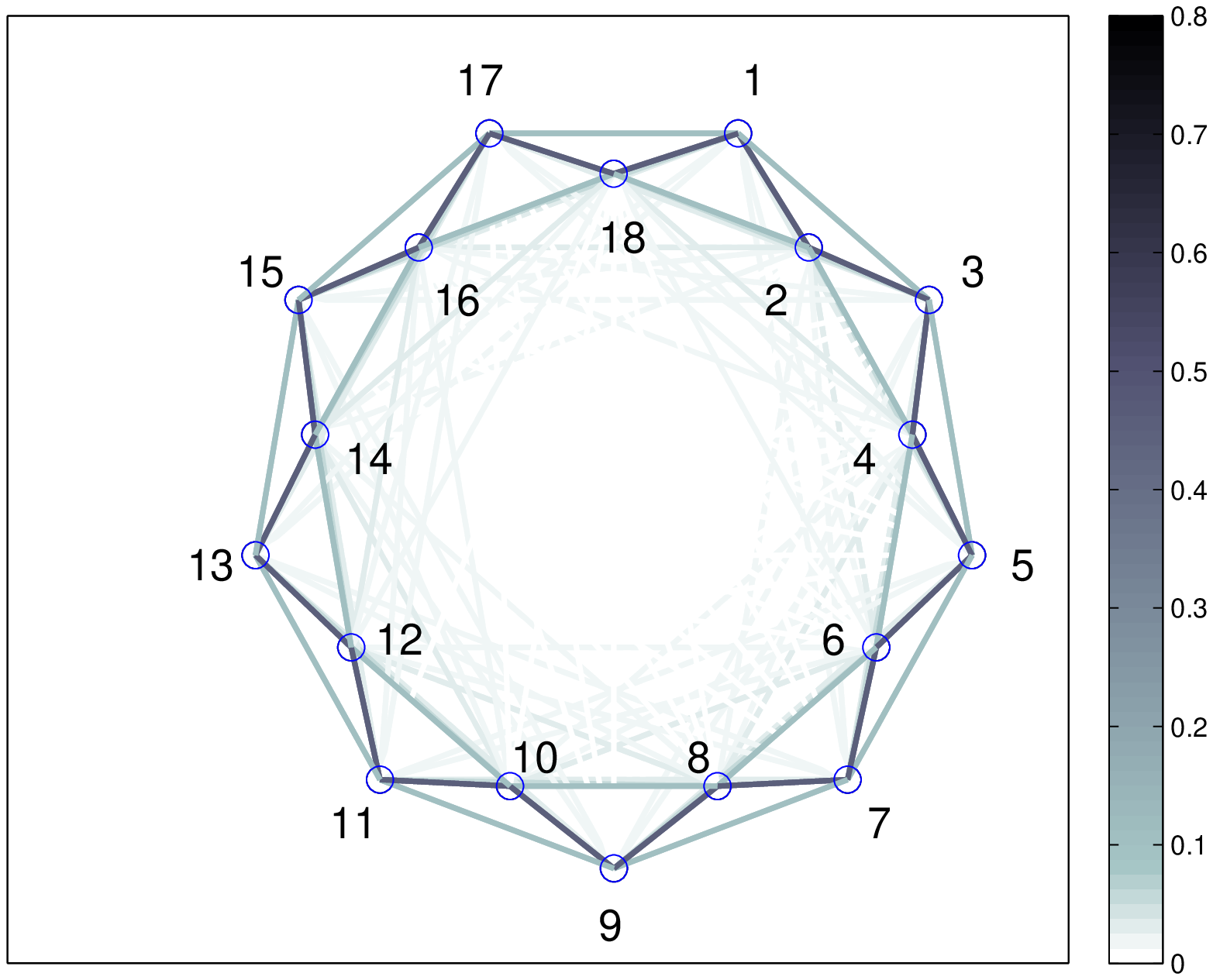}
}
\caption{(Color online) Schematic plot of the two-orbital mutual information for a ring cluster built from 18 Li atoms but using only 2s functions of the Li atoms for bond length $d_{\rm Li-Li}=3.05$\AA (a) and for $d_{\rm Li-Li}=6.00$\AA (b).}
\label{fig:I_Li_only_S}
\end{figure}

\begin{figure}[htb]
\centerline{
\includegraphics[width=0.9\columnwidth]{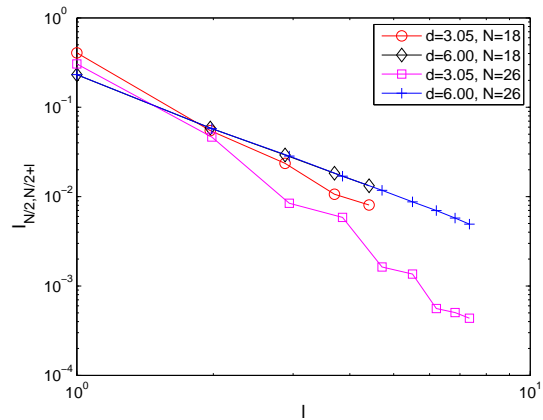}
}
\caption{(Color online) Decay of the $n^{\rm th}$ neighbor entanglement bond strength measured as a function of the interatomic distance for the Li-ring represented with only 2$s$ orbitals for a metallic state ($d_{\rm Li-Li}=3.05$\AA) and for insulating situation at $d_{\rm Li-Li}=6.00$\AA.
}
\label{fig:I_bond_decay_3.05}
\end{figure}

Despite $p$ functions are fundamental for a fair description of the chemistry of lithium, we can learn important information about the physics of correlated electrons in metallic systems using only 2$s$ atomic orbitals for Li rings. Using a localized orbital basis to describe such a system through any ab initio method is analogous to a Hubbard model with one function per site. In this sense stretching the bond length in the lithium rings is equivalent to change $t$ and so the $U/t$ of a half-filled Hubbard model.\\
As expected and independently on the size of the system, all localized orbitals have the same site entropy respecting the rotational invariance of the system. The way these value changes with the structure resembles what one can observe in a Hubbard model. In the metallic case the delocalization of the $n$ electrons in $n$ equivalent orbitals leads to the conclusion that the four possibilities (empty, up-spin, down-spin and doubly occupied) have the same weight for each orbital (1/4), from which $s(1)=\ln4=1.386$. The calculated eigenvalues of the one-orbital reduced density matrix are $\omega_0=0.13, \omega_{\downarrow}=0.37, \omega_{\uparrow}=0.37$ and $\omega_{\downarrow\uparrow}=0.13$
corresponding to the empty, down-spin, up-spin and doubly filled basis states. It follows that the orbital entropy is $s(1)=1.266$ which indicates that all the four basis states gain finite weight being a clear sign of a metallic behavior analogous to the Hubbard model at small Hubbard $U$. Note that in the half-filled Hubbard model a finite gap opens in the charge sector for arbitrary small $U>0$ value.\cite{lieb1968}\\
On the other hand, at the dissociation limit the wavefunction can be described as a state product of 2$s$ orbitals of Li atoms in $^2S$ states which means that the only possibilities are up-spin and down-spin electrons per each site. This is clearly reflected by the eigenvalues of the one-orbital reduced density matrix obtained for an insulating situation ($d_{\rm Li-Li}=6.00$\AA) which are almost zero for the empty and doubly occupied orbital (0.002(1)) and 0.500(4) for up- and down-spin. The orbital entropy $s(1)=0.720(2)$ being close to $\ln 2=0.693$ indicates again that only two basis states gain finite weight so the empty and doubly filled basis states are excluded from the wavefunction.\\
Let us now consider the two-orbital mutual information. In Fig.~\ref{fig:I_Li_only_S} we report its schematic representation for Li$_{18}$ in a metallic and insulating regime. In the first case, at interatomic distance 3.05\AA, all orbitals are highly entangled among each other, because of the high delocalization of the wavefunction in this regime. Of course, the value of $I_{ij}$ reduces if more distant neighbors are considered. We show this decay for both regimes and for Li$_{18}$ and Li$_{26}$ in Fig.~\ref{fig:I_bond_decay_3.05} in a log-log plot. As one can see, in the insulating state the linear trend is not affected by the size of the system, while at shorter interatomic distance, the slope of the linear decay increases if going from Li$_{18}$ and Li$_{26}$.\\
It has to be underlined that at in the insulating regime, other orbitals are also highly entangled, but the nature of this correlation is totally different than in the previous case. In order to see this, one has to observe the different elements of the two-orbital density matrix shown in Fig.~\ref{fig:li_rho_3.05} in the Appendix. While in a metallic situation different hopping and spin-flipping terms have a large contribution (left panels), only the spin flipping plays a role in a stretched structure (right panels). This is because even at dissociation the separated atoms present a strong quantum entanglement in the singlet state we are considering. Confirming this, is the fact that at the same structure, the almost degenerate state with highest spin multiplicity (ferromagnet state) present no entanglement among the 2$s$ orbitals, since the wavefunction can be described as a pure state product.

\subsection{Li ring with 2$s$ and 2$p$ functions}

\begin{figure}[htb]
\centerline{
\includegraphics[width=0.5\columnwidth]{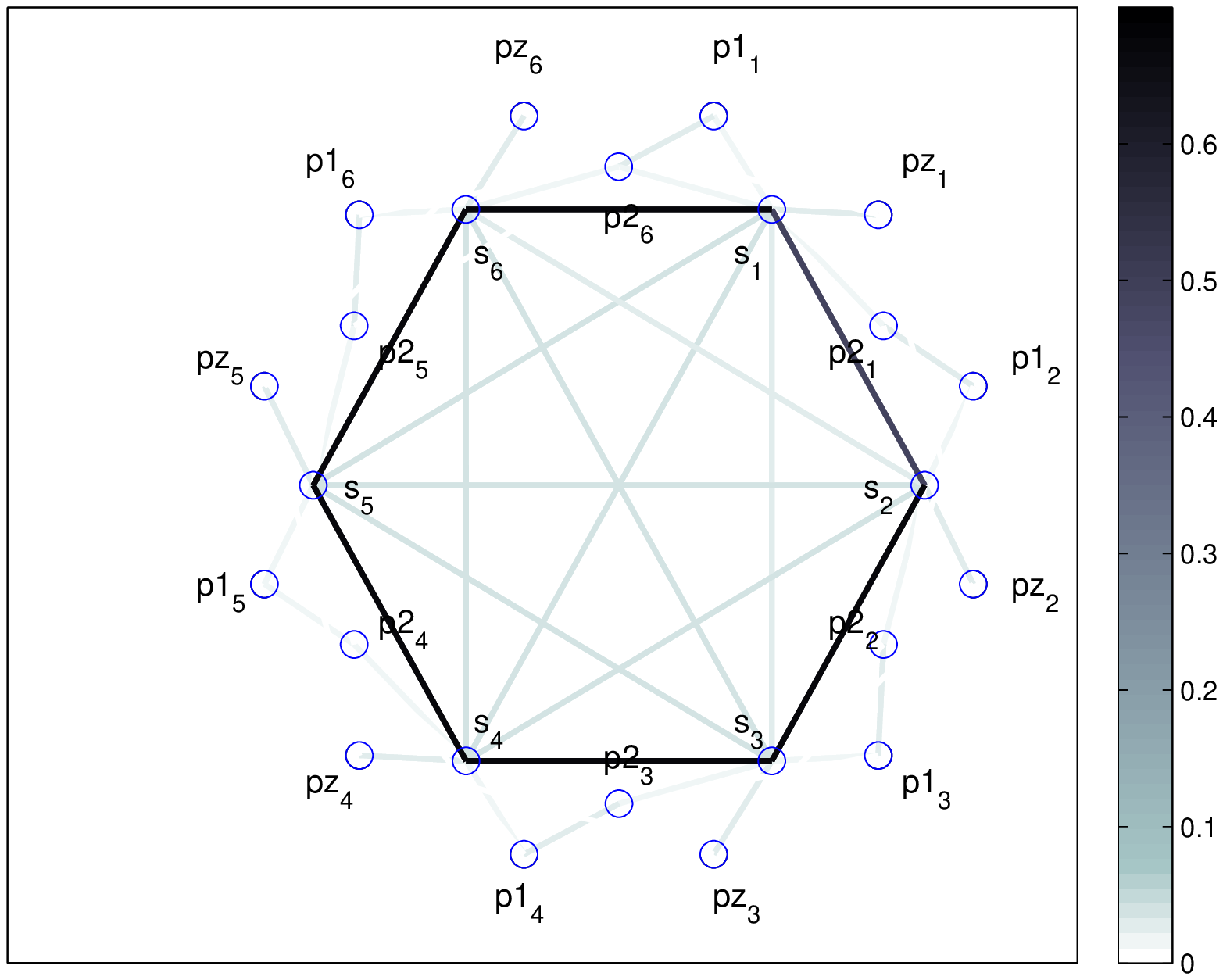}
\includegraphics[width=0.5\columnwidth]{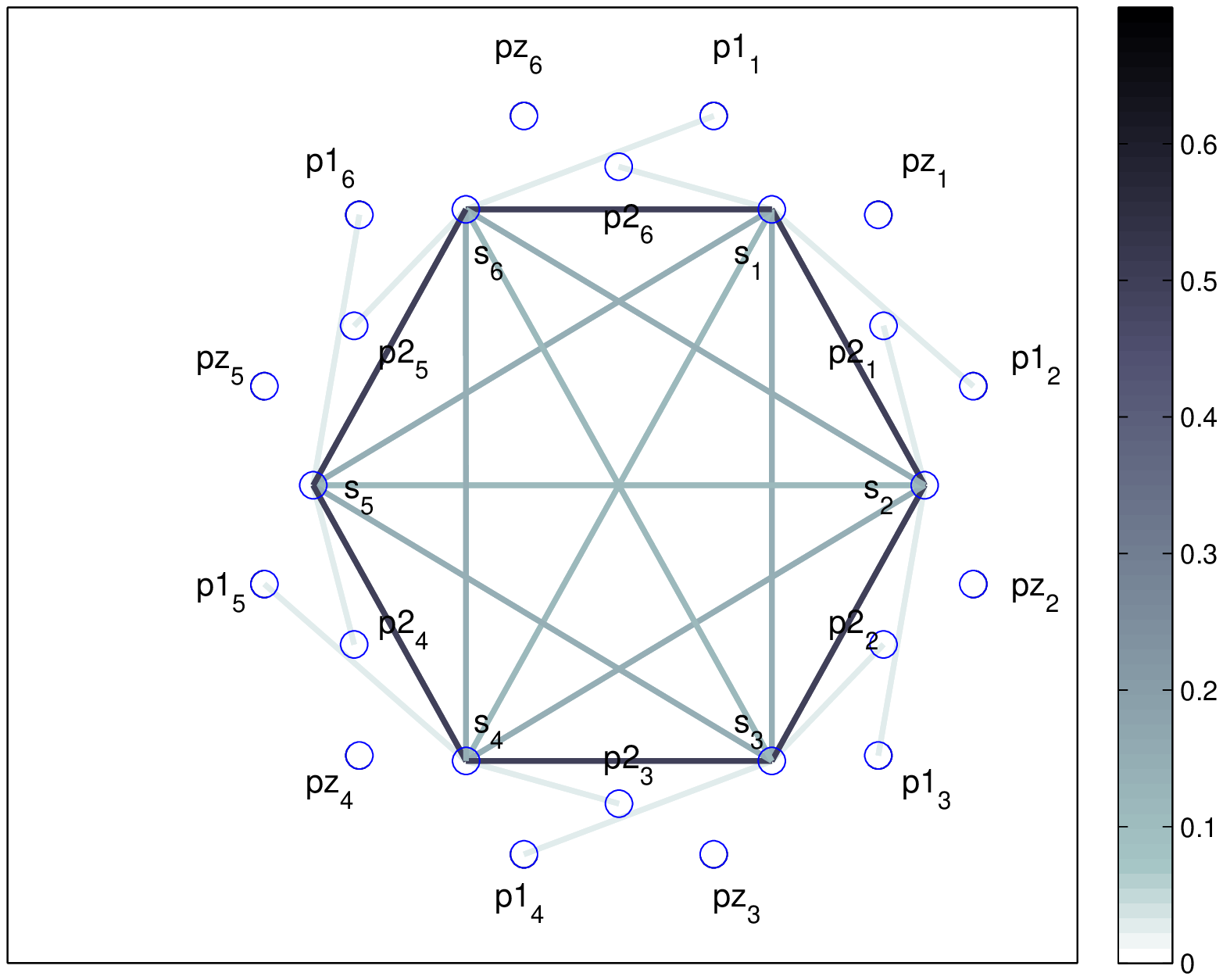}
}
\centerline{
\includegraphics[width=0.5\columnwidth]{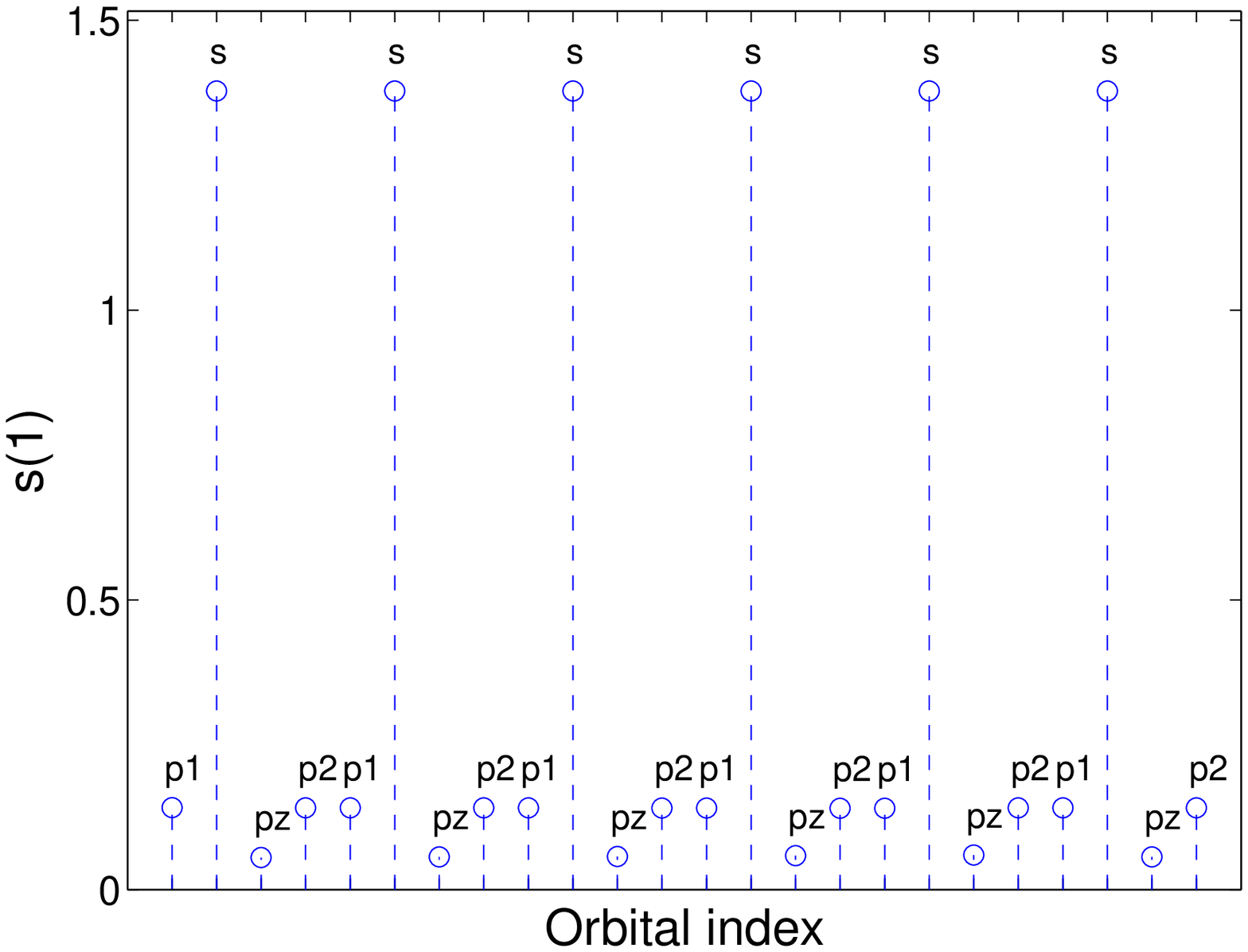}
\includegraphics[width=0.5\columnwidth]{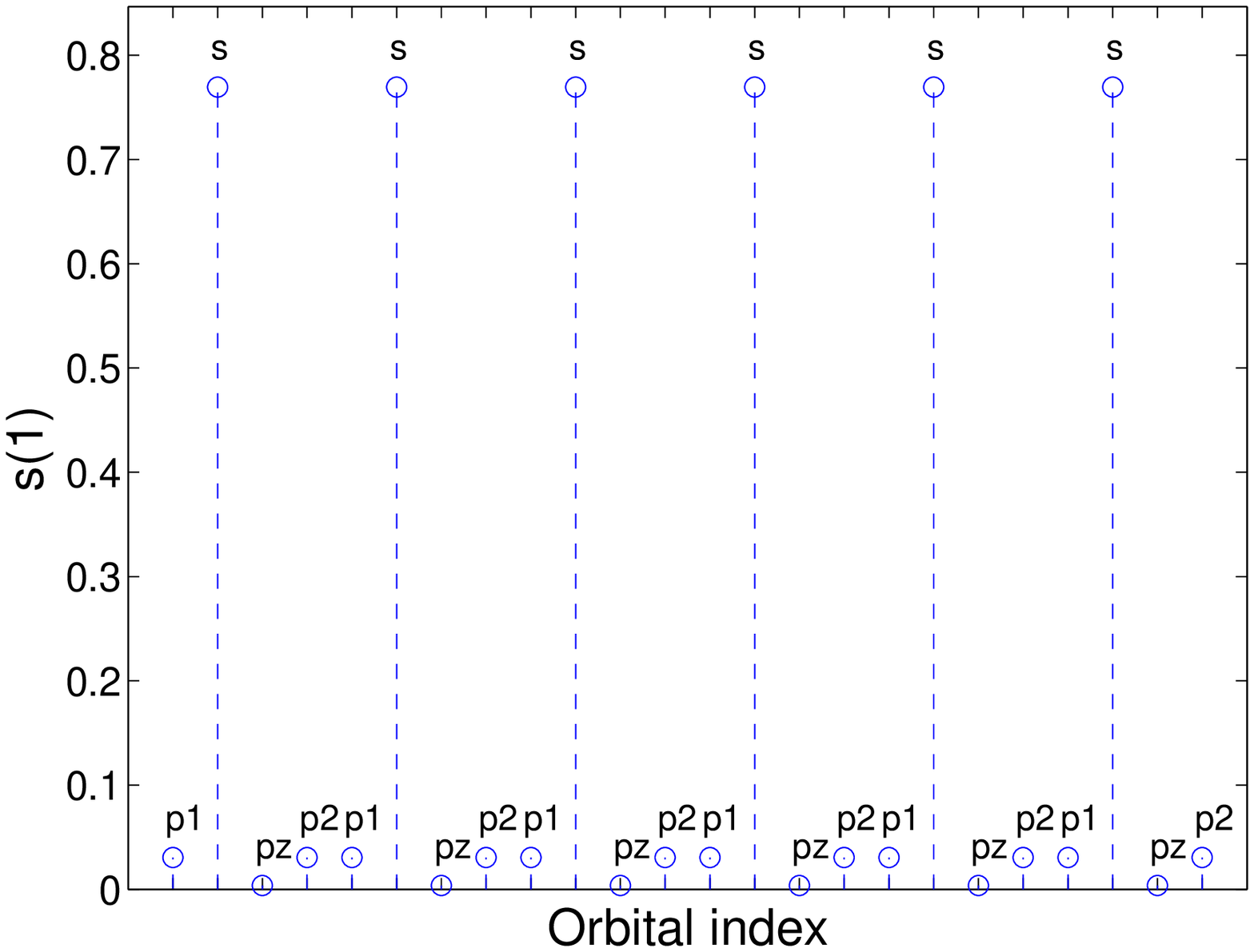}
}
\caption{(Color online) Schematic plot of the two-orbital mutual information (upper panels) and one-orbital entropy (lower panels) for Li$_6$ ring calculated using 2$s$ and 2$p$ atomic functions for bond lengths $d_{\rm Li-Li}=3.05$\AA (left) and for $d_{\rm Li-Li}=6.00$\AA (right).
}
\label{fig:Li6_sto3g}
\end{figure}

The next step in the study of lithium system is the analysis of the effect of $p$ atomic functions. Because of the obvious increase of the active space we focused our attention on the smaller ring Li$_6$, but the reduced size does not have drastic effects on the nature of the system.\\
In Fig.~\ref{fig:Li6_sto3g} the two-orbital mutual information and the one-orbital entropy is reported for both the metallic and the insulating state. The presence of the three 2$p$ atomic functions per site increases, of course, the complexity of the entanglement picture, but the main information is analogous to what observed for larger Li rings with 2$s$ orbitals only. Firstly, once again, independently on the structure, the rotational symmetry is respected in the degeneracy of the one-site entropy values. Moreover, the highest entanglement still occurs among the 2$s$ orbitals especially at long interatomic distances because of the spin-flipping (see Fig.~\ref{fig:li_rho_sto-3g}) and the $s(1)$ are just slightly effected by the presence of 2$p$ orbitals.\\
The entanglement between 2$s$ and 2$p$ or between 2$p$ orbitals is order of magnitude smaller then between 2$s$ atomic functions and decreases sensibly going from a metallic to an insulating state. This reflects the fact that intraband transitions within the half filled 2$s$ band are more important than interband transitions. Moreover we observe that the 2$p_z$ orbitals are only entangled with 2$s$ orbitals, while the 2$p$ atomic functions lying in the plane of the ring would seem to be correlated between each other only if centered on the same Li atom. Moreover, comparing the correlation functions reported in Fig.~\ref{fig:li_rho_3.05}~and~\ref{fig:li_rho_sto-3g} of the appendix, one can see that because of the presence of the 2$p$ orbitals, the hopping terms gain importance also at large interatomic distance. \\
As expected, toward dissociation the correlation between $s$ and $p$  orbitals on the same site drops even more. Indeed, one has to remember that in free Li atom (if core electrons are kept frozen as we did) the correlation energy is zero and the HF wavefunction with one electron in the 2$s$ orbital represent the correct solution within the limit of the atomic basis set. So it is clear that the $s-p$ entanglement has to get to zero with increasing distance.

\subsection{Be ring}

\begin{figure}[htb]
\centerline{
\includegraphics[width=0.5\columnwidth]{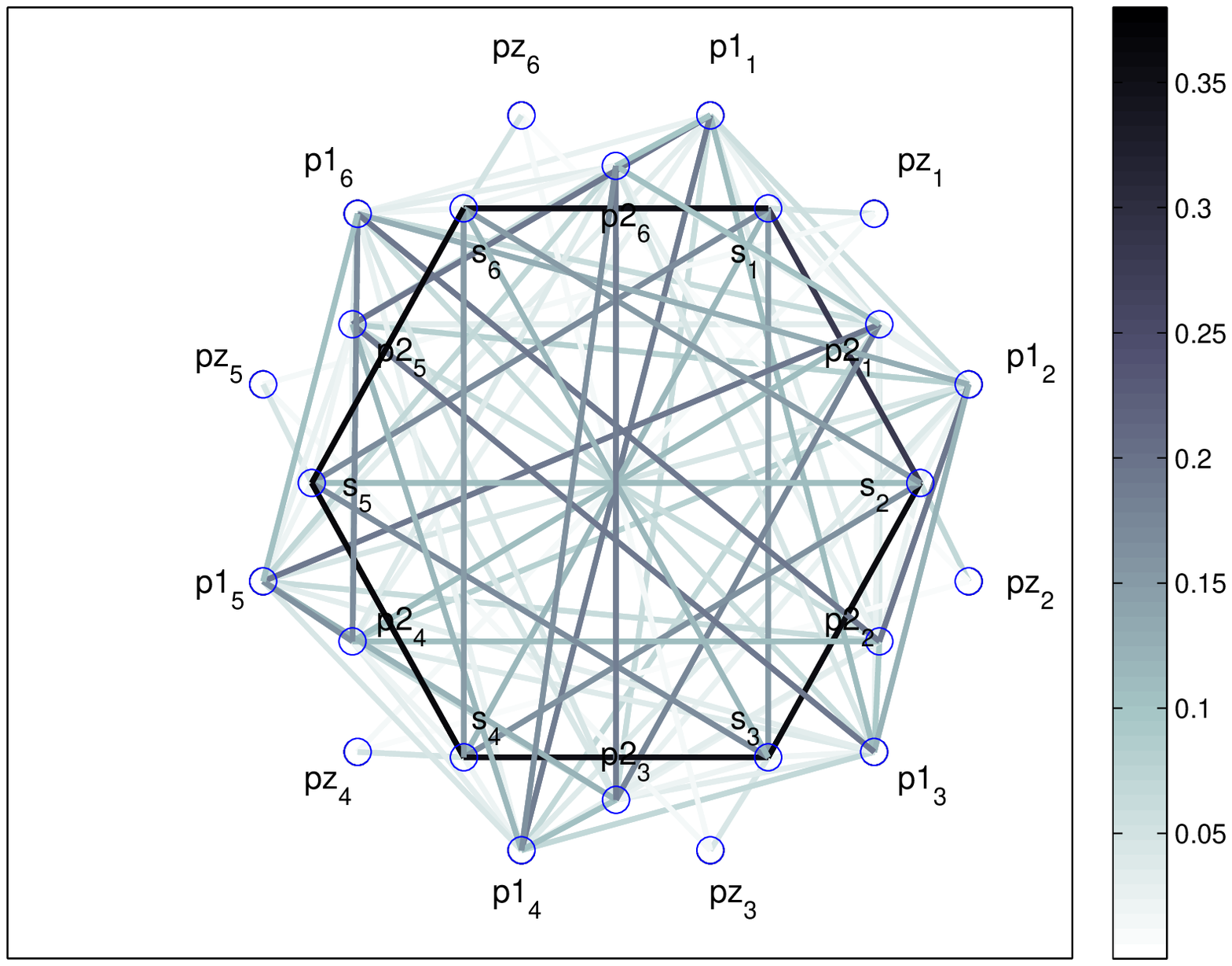}
\includegraphics[width=0.5\columnwidth]{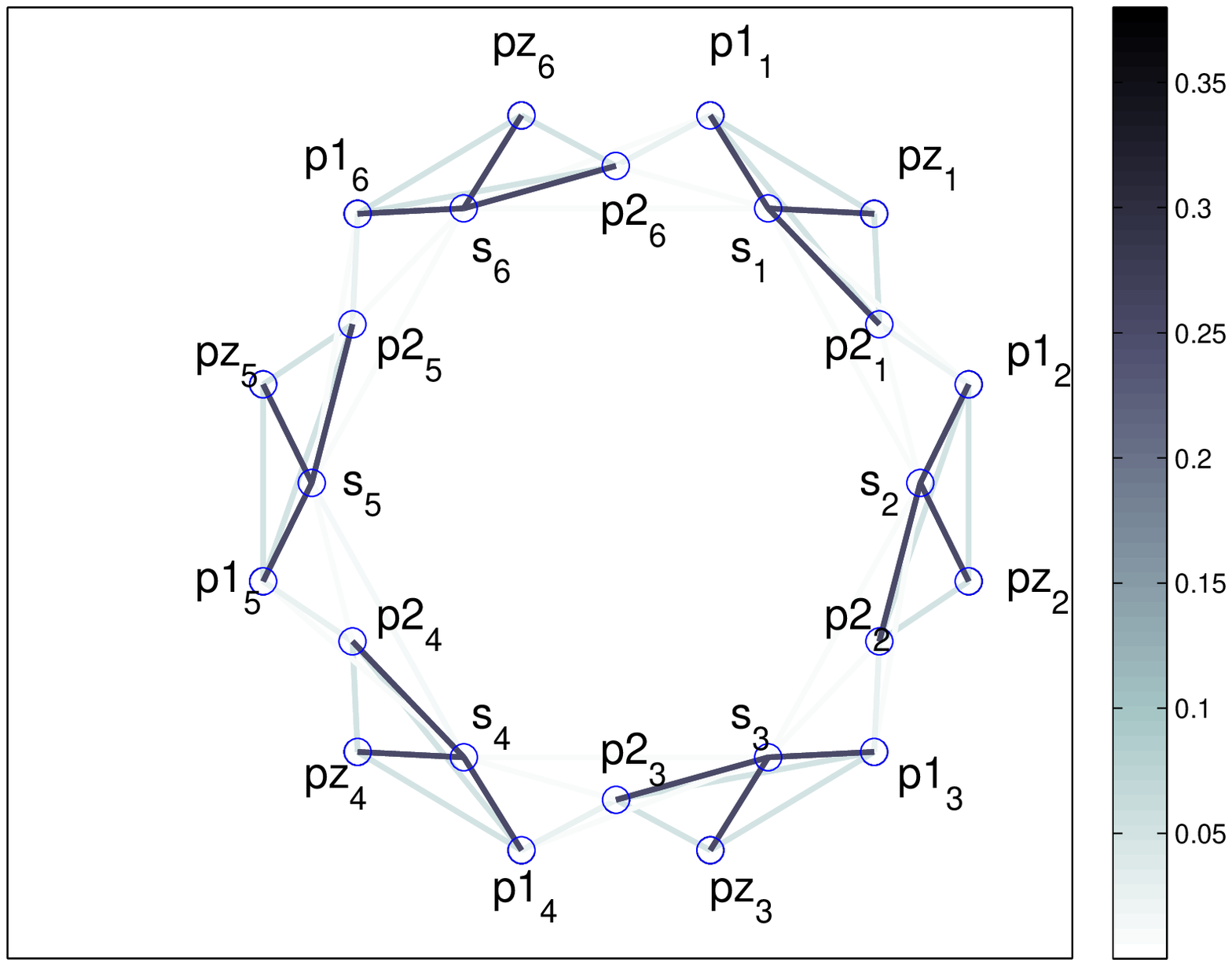}
}
\centerline{
\includegraphics[width=0.5\columnwidth]{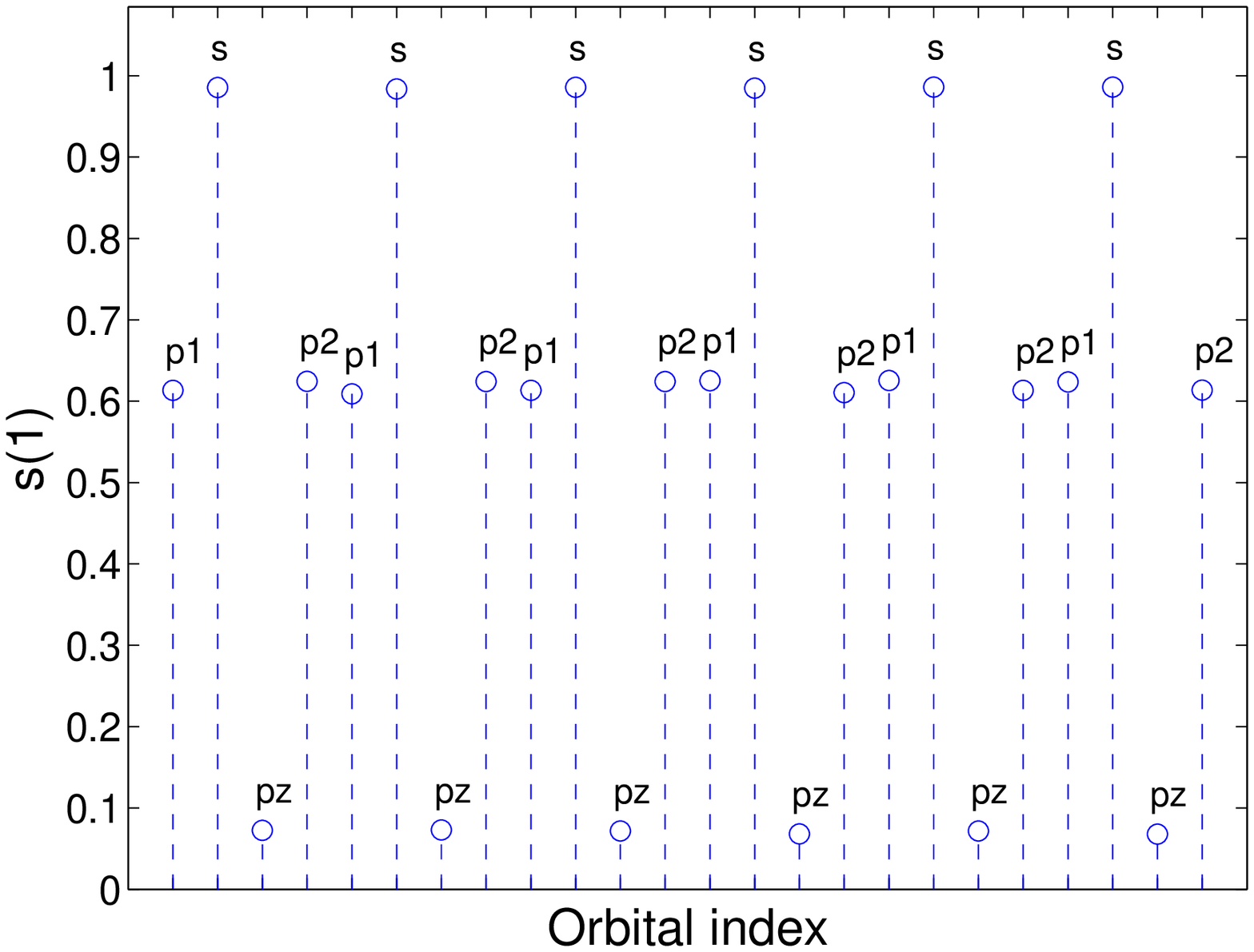}
\includegraphics[width=0.5\columnwidth]{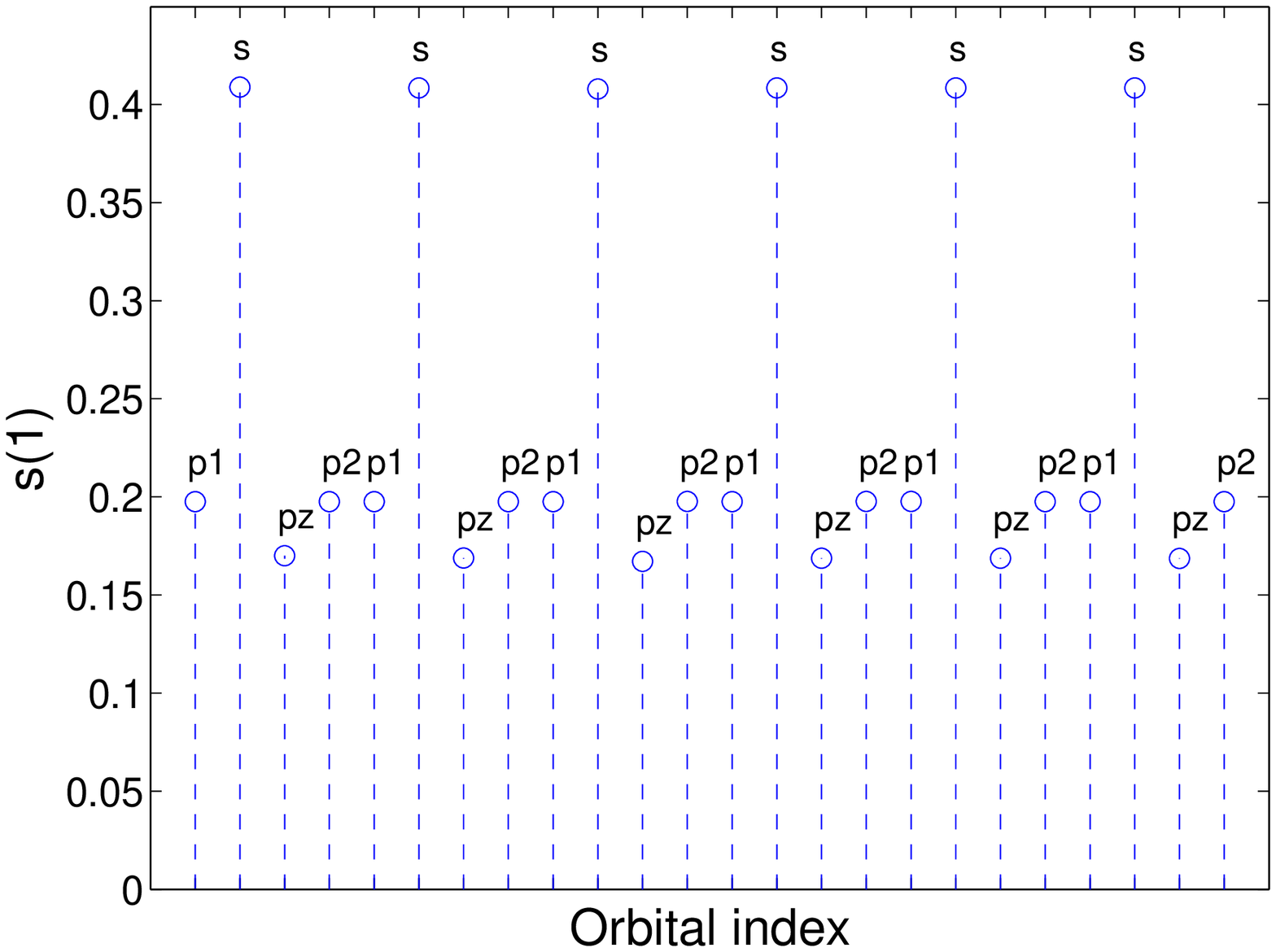}
}
\caption{(Color online) Schematic plot of the two-orbital mutual information (upper panels) and one-orbital entropy (lower panels) for Be$_6$ ring calculated using 2$s$ and 2$p$ atomic functions for bond lengths $d_{\rm Be-Be}=2.15$\AA (left) and for $d_{\rm Be-Be}=3.30$\AA (right) using configuration 2.
}\label{fig:Be6_pcolor_entro}
\end{figure}

\begin{figure}[htb]
\centerline{
\includegraphics[width=0.9\columnwidth]{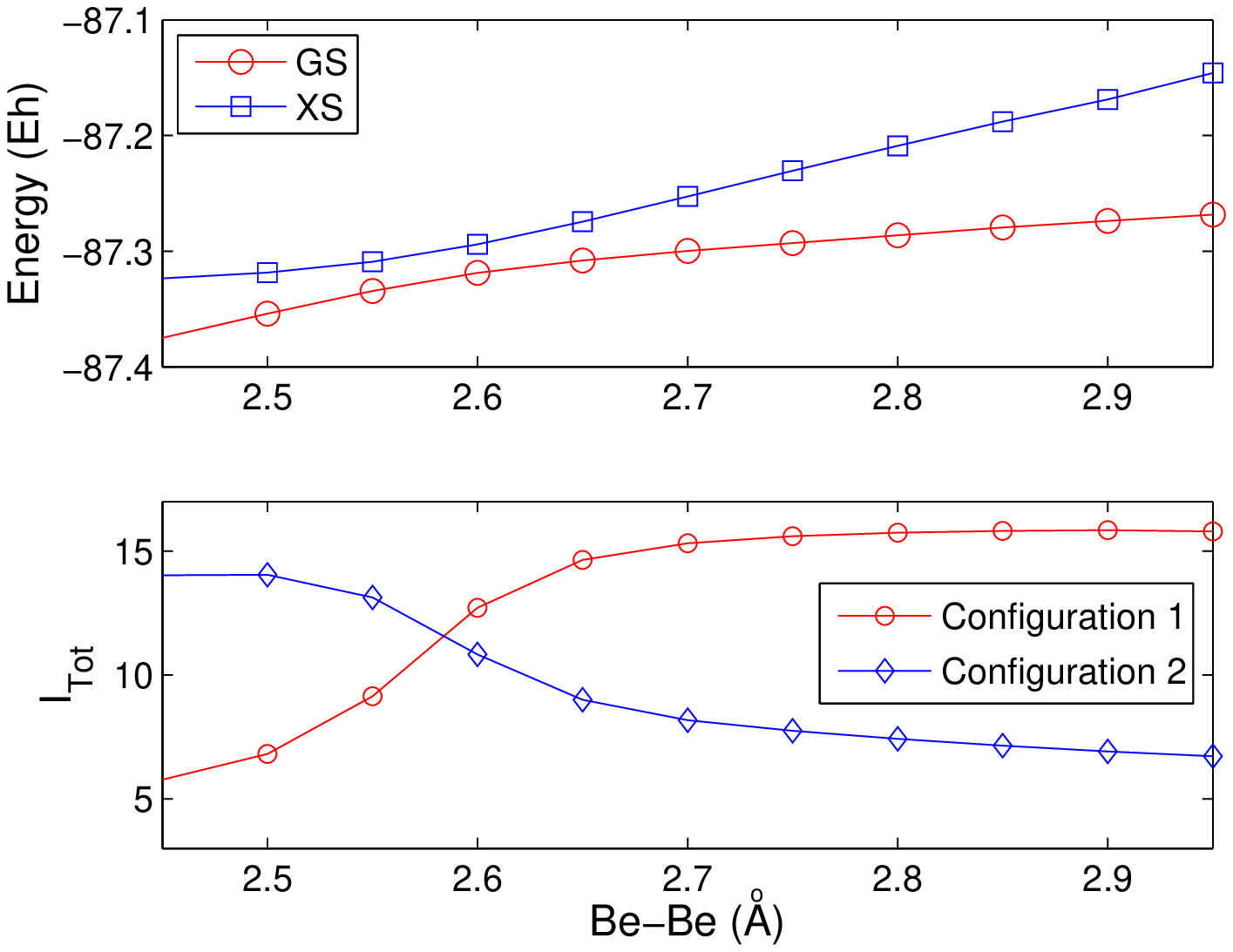}
}
\centerline{
\includegraphics[width=0.9\columnwidth]{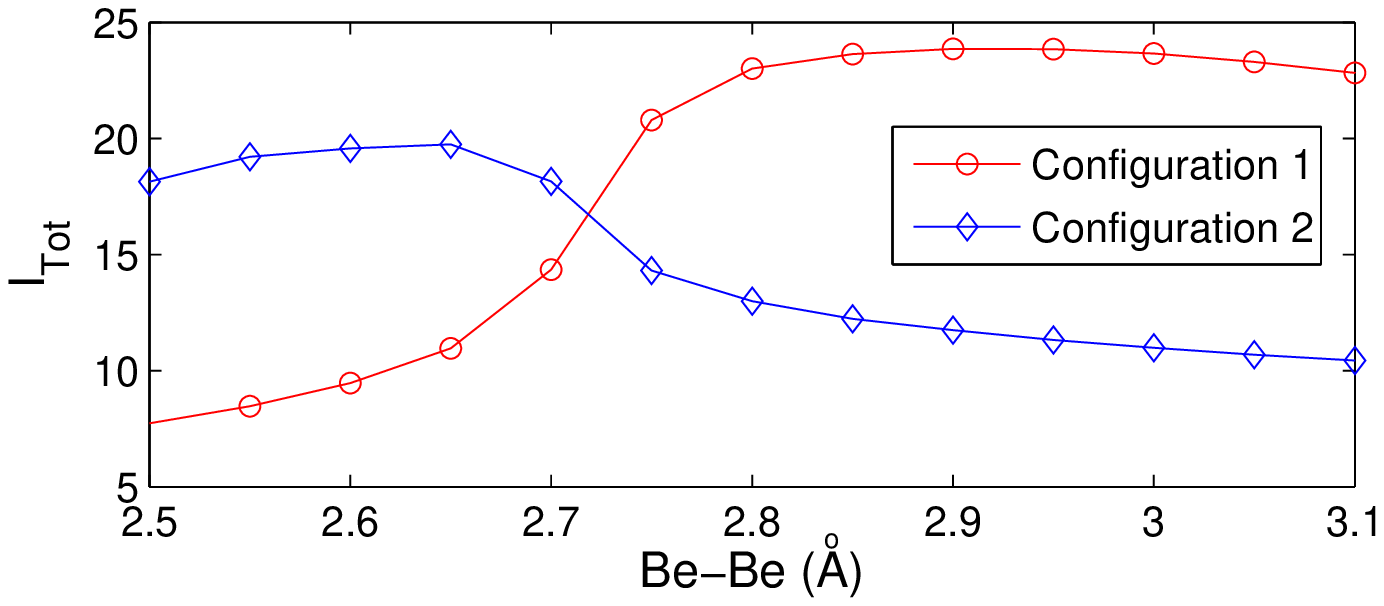}
}
\caption{Energies of the two lowest lying $^1A_g$ singlet states and total quantum information $I_{\rm tot}$ for Be$_6$ and Be$_{10}$ (in this case only $I_{\rm tot}$ is considered) as a function of $d_{\rm Be-Be}$, calculated using localized orbitals for configuration 1 and configuration 2. The sudden change in $I_{\rm tot}$ indicating the position of the avoided crossing, occurs at $d_{\rm Be-Be}\simeq 2.55${\AA } for Be$_6$ and 2.72{\AA } for Be$_{10}$.}
\label{fig:Be6_I_tot}
\end{figure}

In this last section we will analyze the results obtained for beryllium rings. For this system, the use of both 2$s$ and 2$p$ atomic basis is mandatory to have a minimal meaningful description.\\
As one can see, observing Fig.~\ref{fig:Li6_sto3g} and~\ref{fig:Be6_pcolor_entro}, the main differences between Be$_6$ and Li$_6$, besides the strong correlation between 2$s$ and 2$p$ orbitals, occurs at dissociation. As 2$s$ orbitals are doubly occupied in the HF solution, hopping and flipping between them are less important that in the case of lithium and the quantum entanglement between isolated Be atoms can be considered zero. On the other hand, the quasi-degeneracy of 2$s$ and 2$p$ orbitals causes that the static correlation constitutes about 93\% of the total correlation energy. This can be deduced by the strong entanglement shown in the pictorial representation of the mutual information.\\
Analyzing the correlation functions reported in Fig.~\ref{fig:be_rho} of the appendix, it is evident that in comparison to lithium the hoppings between neighboring 2$p$ atomic functions play a more important role for short interatomic distances. This can be interpreted as an index of a metallic character, since it highlights the delocalized character of the wavefunction. Despite from a band structure ({\emph i.e.} one electronic) picture, one could conclude that Be rings are insulating, when static correlation is rightfully described, one can observe a metallic-like behavior analyzing indicators other than the band-gap.\\
As already discussed in Ref.~[\onlinecite{murg2014-lif}], the one-orbital entropy function can also be used to locate an avoided crossing. Since a metal-insulator like transition can be identified as a state transition in finite systems, we can use strong change in $I_{\rm tot}$ to identify it. This is a more advanced approach than calculating low lying energy spectrum since it requires to calculate ground state properties only \cite{legeza2006-qpt}. Therefore, the behavior of $I_{\rm tot}$ as a function of interatomic distance can be used to detect and locate transition points where the wavefunction changes dramatically. In first panel of Fig.~\ref{fig:Be6_I_tot} energies of the two lowest lying $^1A_g$ singlet states of Be$_6$ as a function of $d_{\rm Be-Be}$ are shown around the avoided crossing. In the second panel the total quantum information calculated using configuration 1 and 2 is reported. Despite $I_{\rm tot}$ strongly depends on the orbital basis employed, it always show strong fluctuations around a state transition, like the one proposed. In our case, we observe how if one configuration or the other is used, a jump or a drop of its value occurs. In both cases this dramatic change occurs at $d_{\rm Be-Be}\simeq 2.55${\AA } which indicates the position of the avoided crossing. The same observation is true also for the larger cluster Be$_{10}$. As can be seen in the lower panel of Fig.~\ref{fig:Be6_I_tot}, in this case, the avoided crossing can be localized at $d_{\rm Be-Be}\simeq 2.70${\AA } by studying $I_{tot}$.

\section{Conclusion}

An ab initio Density Matrix Renormalization Group study has been performed on pseudo-onedimensional systems in order to investigate Metal-Insulator like Transition. Through DMRG we overcame the active space size problem encountered for Beryllium rings and we were able to perform very accurate calculations reaching the Full-CI level which is generally prohibitive for these systems even with a minimal basis set.\\
Also the analysis of generalized correlation functions was used to study the change occurring in the wavefunction as a function of the interatomic distance. Underlying the differences in the entanglement between different orbitals, we showed how the systems evolve from a metallic state to an insulating one. Furthermore we discussed that, despite the total quantum information strongly depends on the orbital basis, it can be employed to locate the position of the MIT in these systems. The use of this quantity as an index for characterizing this kind of transitions will be further invastigated in future works.\\
Finally, we focused on the effect of the orbital basis on the DMRG procedure. Comparing the results obtained using canonical and localized orbital basis, we observed that using the latter less computational effort was necessary to reach the same level of accuracy because of the different orbital entanglement which, as discussed, is a crucial in quantum chemical DMRG.


\acknowledgments{This research was supported in part by the Hungarian Research Fund (OTKA) under Grant No. NN110360 and K100908, the Agence Nationale de la Recherche (ANR) and the German Research Foundation (DFG) via the project ``Quan\-tum-chemical investigation of the metal-insulator transition in realistic low-dimensional systems (MITLOW)''. Financial support by the Max-Planck scociety is appreciated.}



\section*{Appendix}
In this section we report the pictorial representation of the generalized correlation functions used to construct the off-diagonal elements of the two-orbital reduced density matrix for the different systems analyzed above (see Fig.~\ref{fig:li_rho_3.05},~\ref{fig:li_rho_sto-3g} and~\ref{fig:be_rho}).\\
In Figs.~\ref{fig:li_rho_3.05} of the Appendix we show strength of transition amplitudes between initial and final states on orbital $i$ and $j$ with different line colors. For the equilibrium distance the one-electron hopping and spin exchange are highly probable processes while pair hopping can happen with a such smaller rate. The amplitudes of such processes also drop by an order of magnitude as the distance between $i$ and $j$ is increased. In contrast to this, for the stretched structure at $d_{\rm Li-Li}=6.00${\AA } only the spin-flip process gain large probability while one-electron hopping is probably between nearest neighbors only and it is an order of magnitude smaller. The anti-ferromagnetic ground state is well reflected through the osciallation in the sign of the spin-spin correlation function with increasing distance between $i$ and $j$.
\begin{figure*}
\begin{minipage}{20cm}
\hskip -2.0cm
\begin{minipage}{8cm}
\centerline{
\includegraphics[scale=0.25]{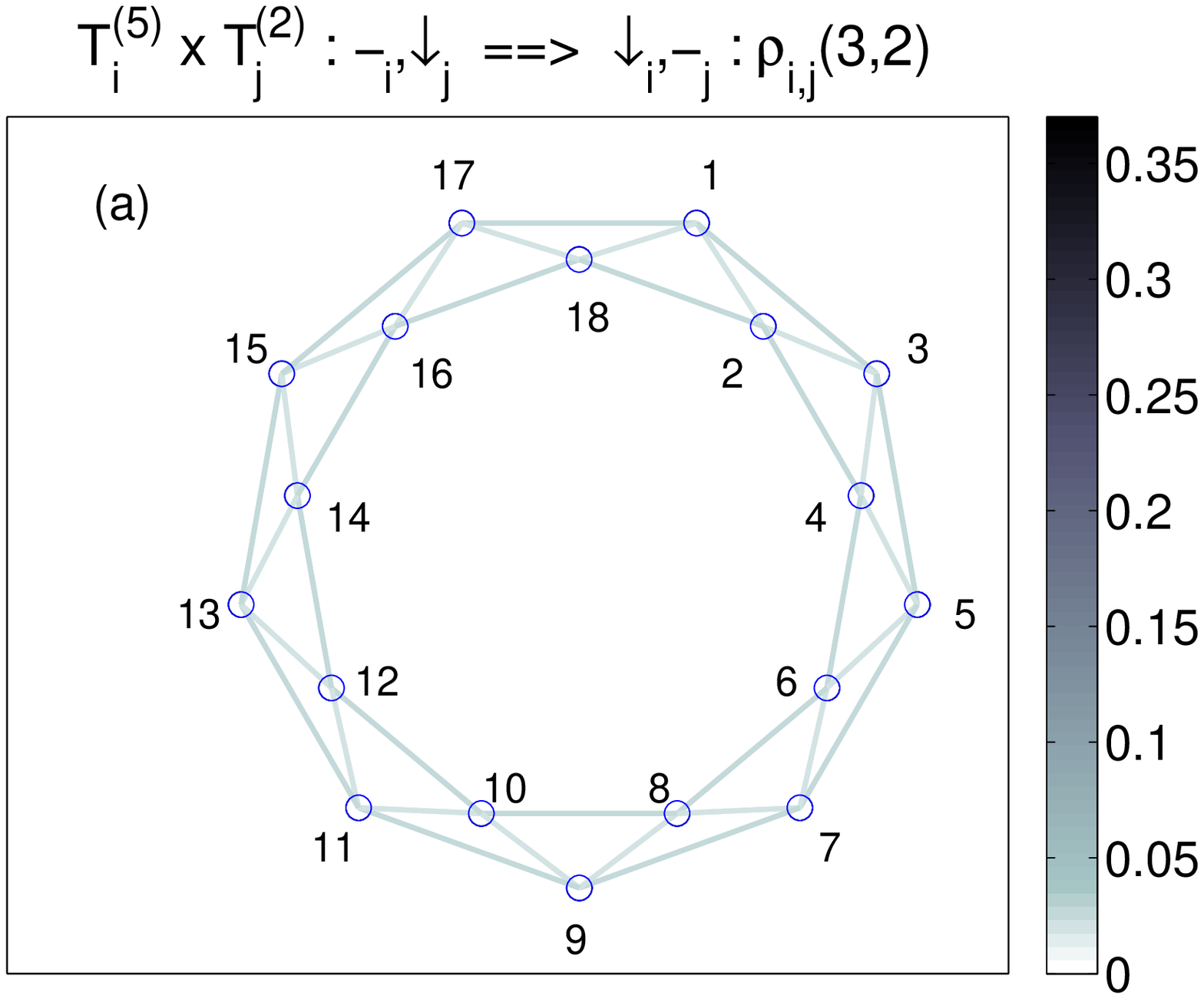}
\includegraphics[scale=0.25]{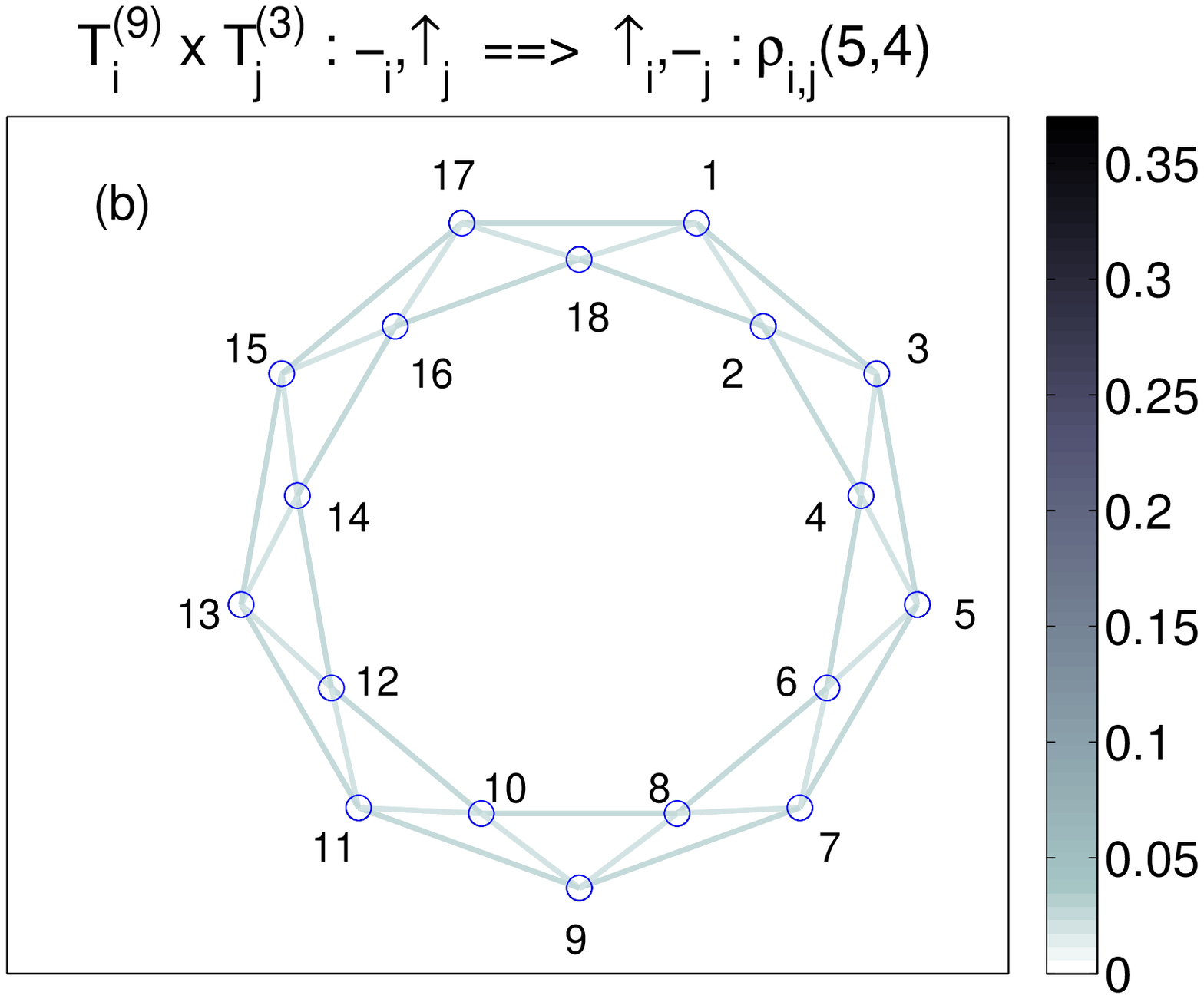}
}
\centerline{
\includegraphics[scale=0.25]{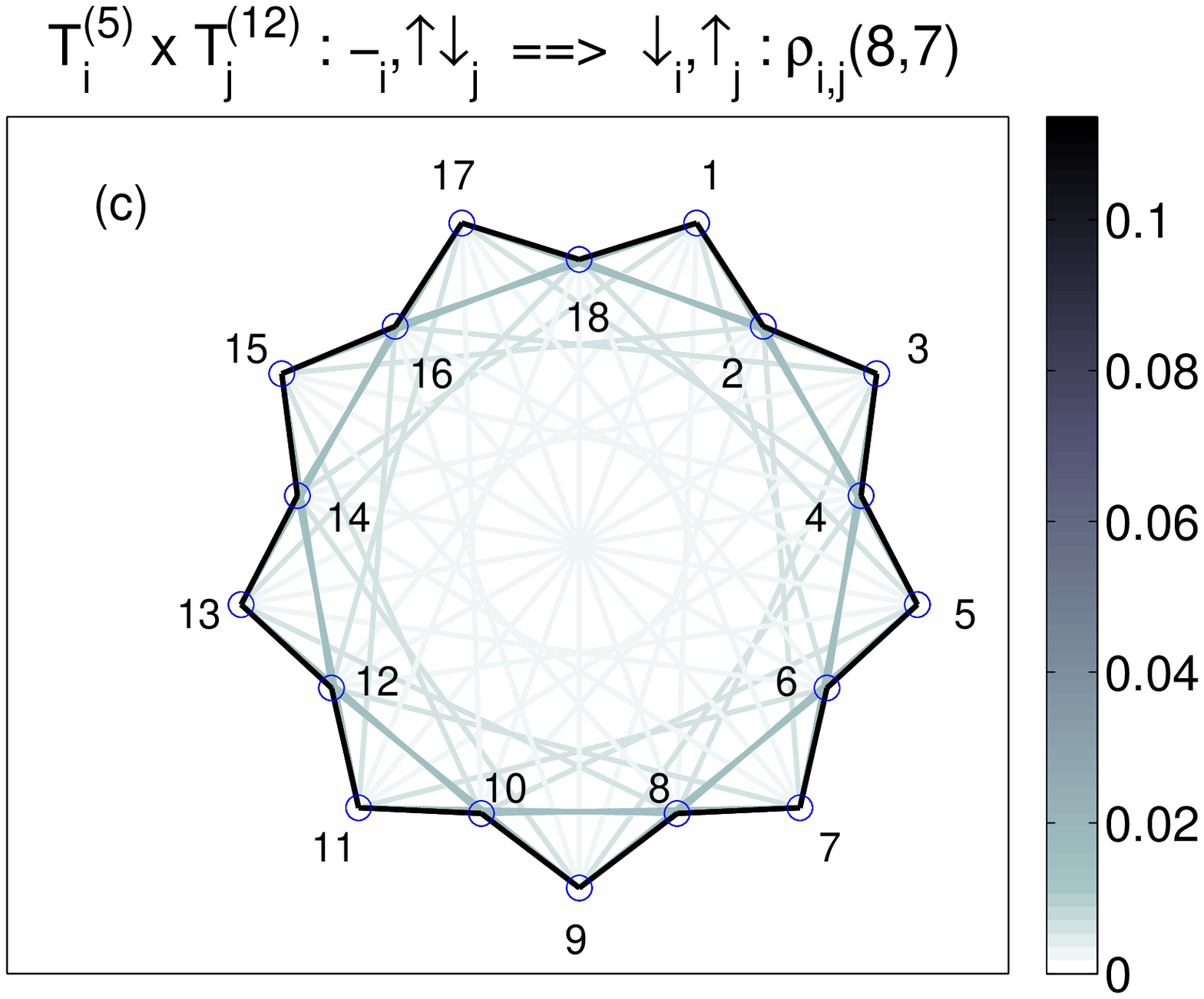}
\includegraphics[scale=0.25]{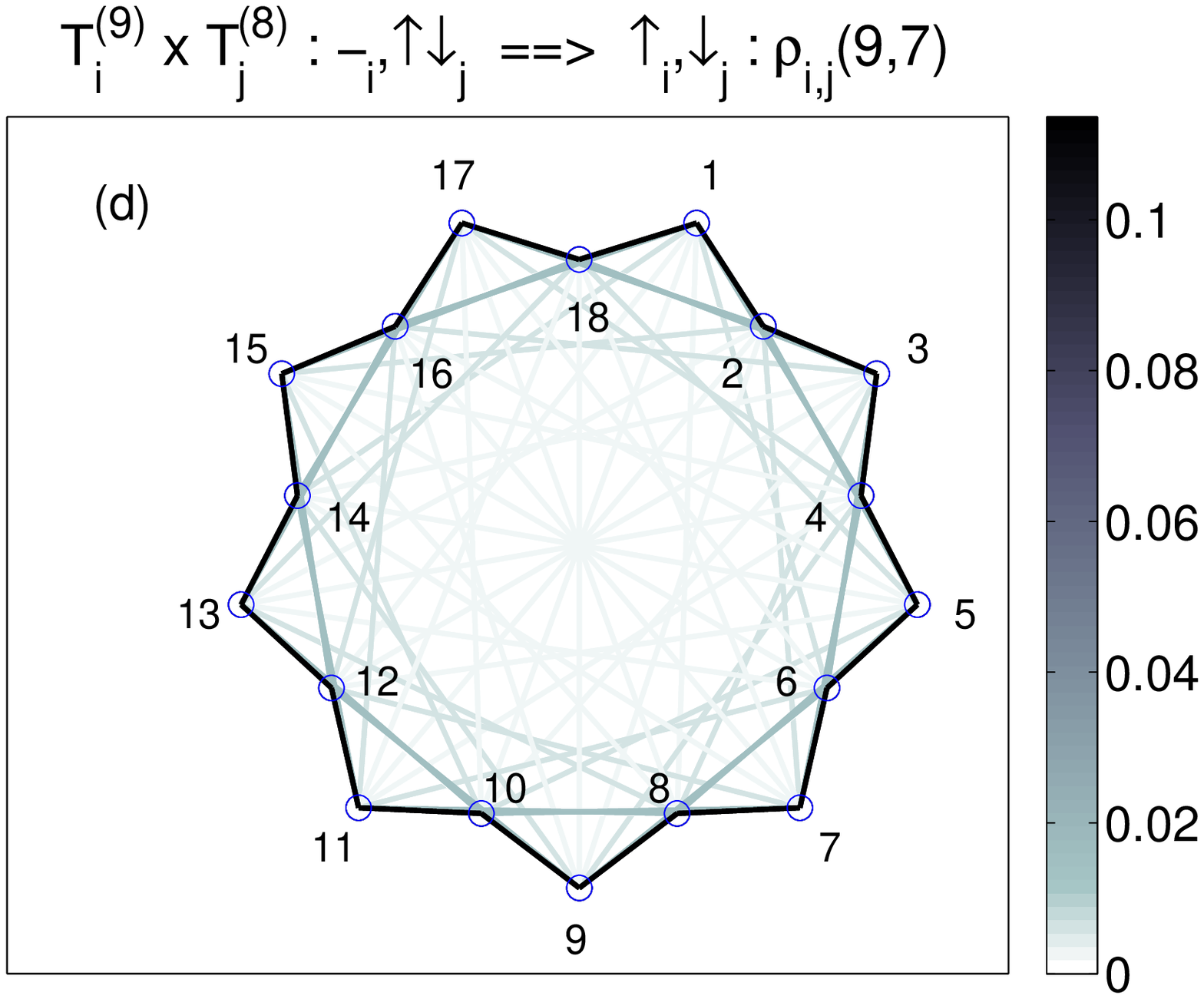}
}
\centerline{
\includegraphics[scale=0.25]{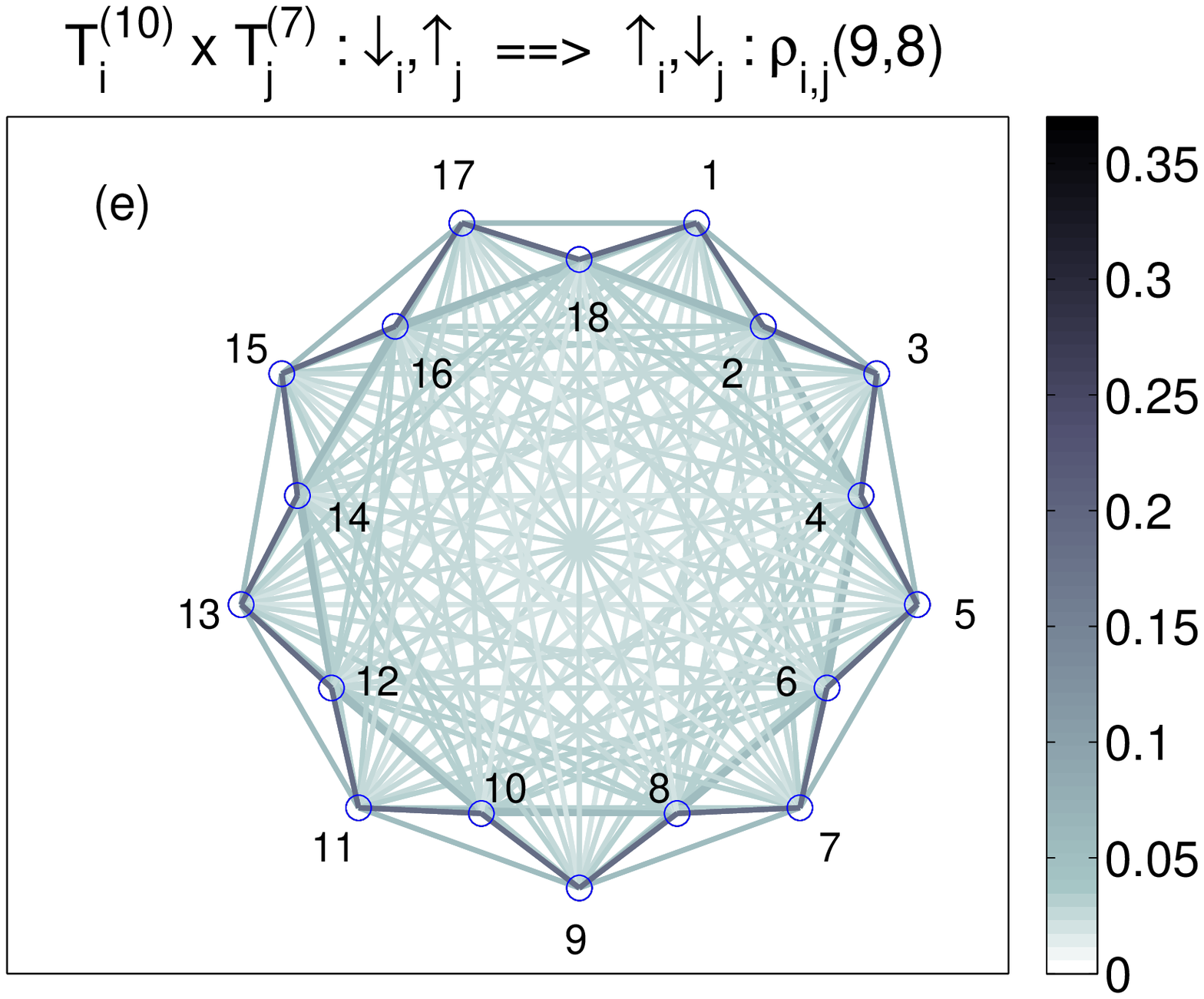}
\includegraphics[scale=0.25]{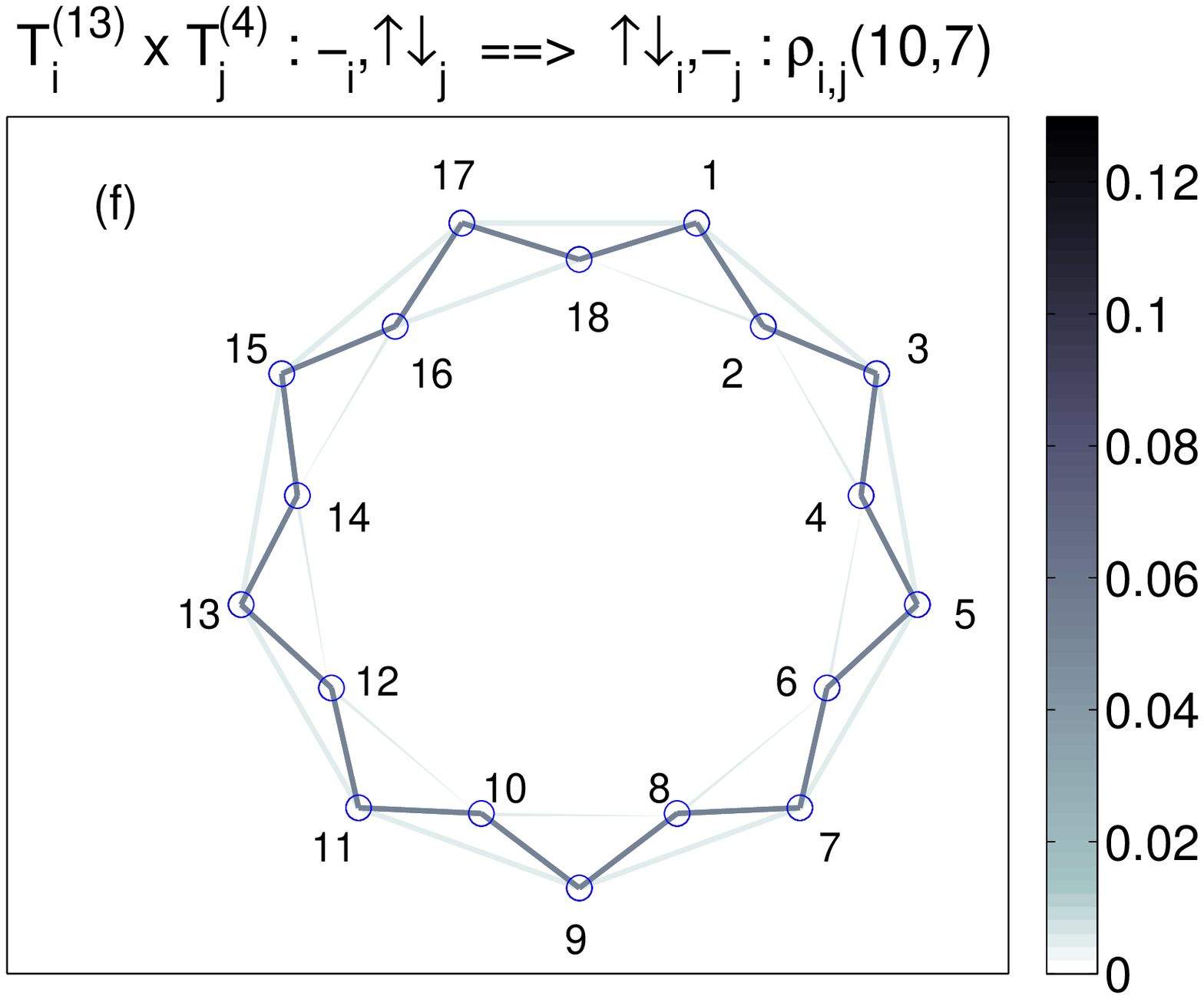}
}
\centerline{
\includegraphics[scale=0.25]{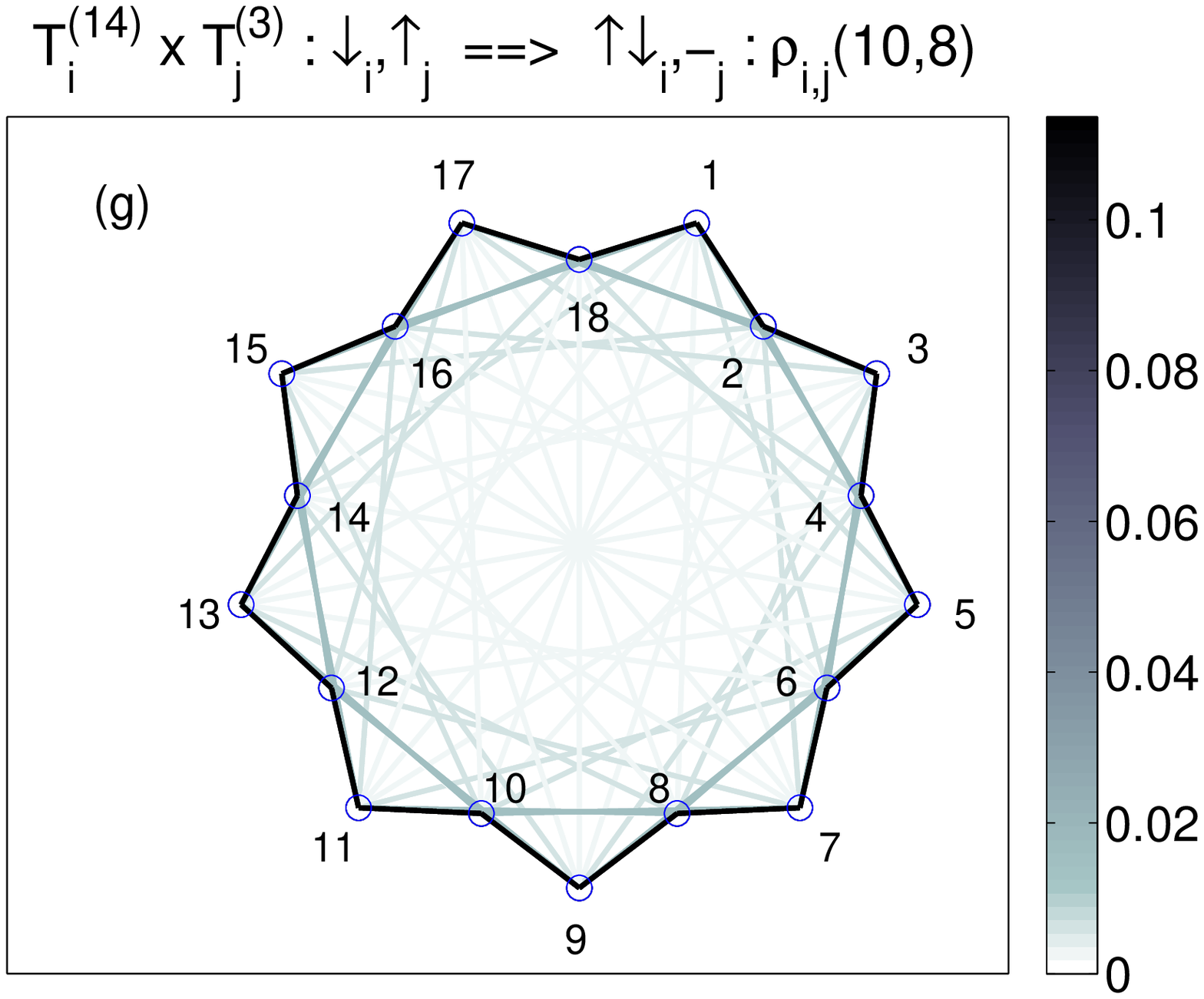}
\includegraphics[scale=0.25]{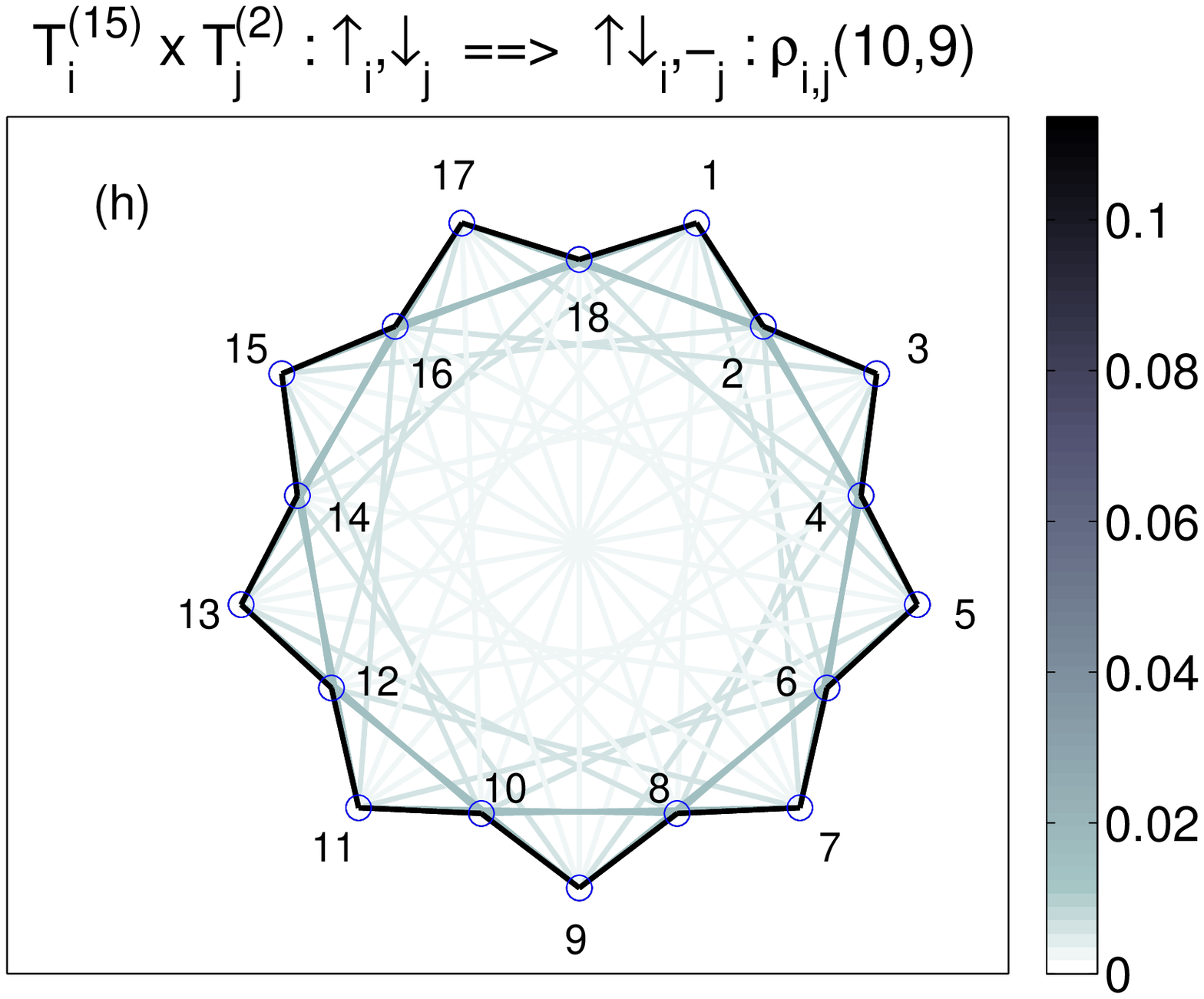}
}
\centerline{
\includegraphics[scale=0.25]{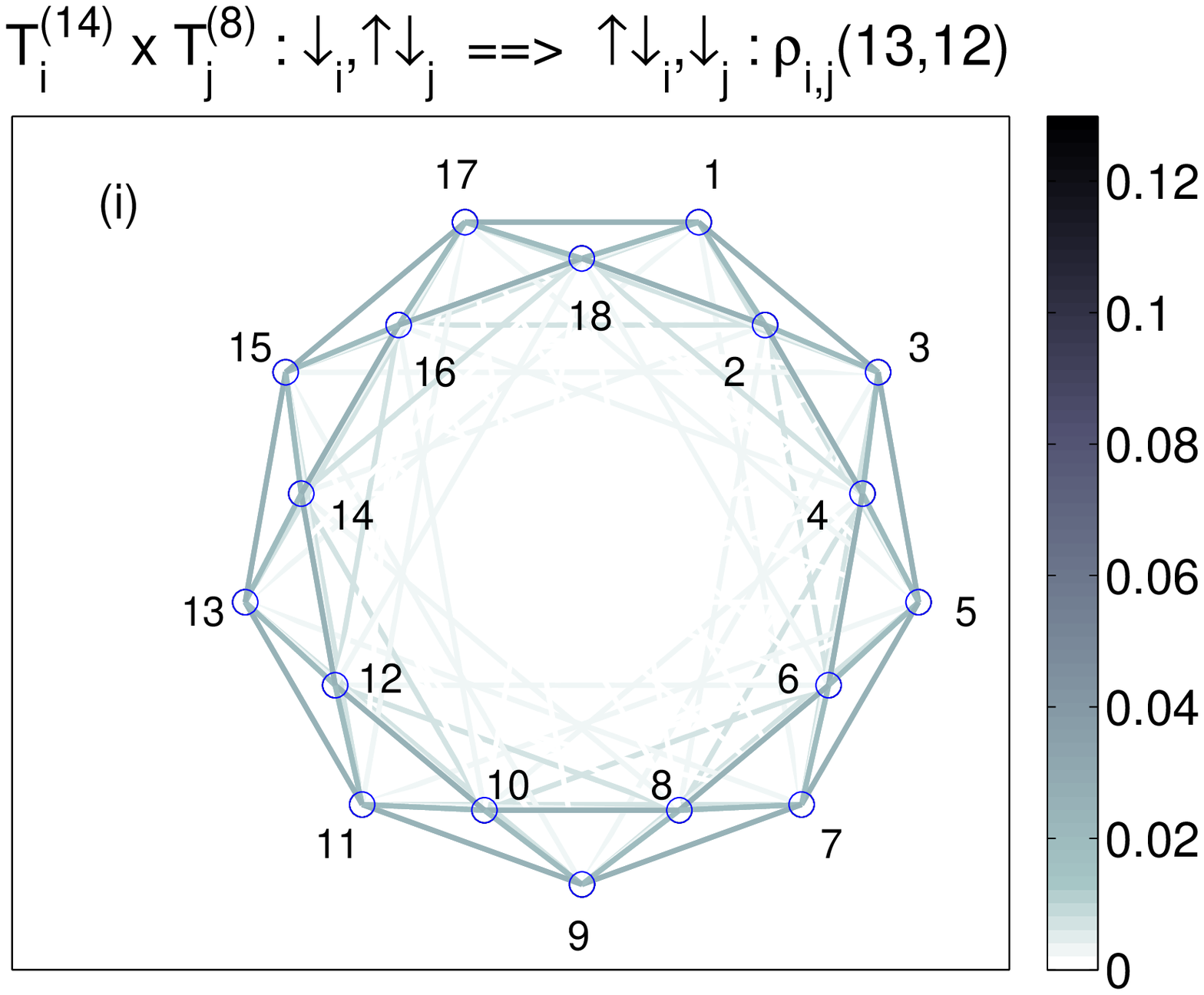}
\includegraphics[scale=0.25]{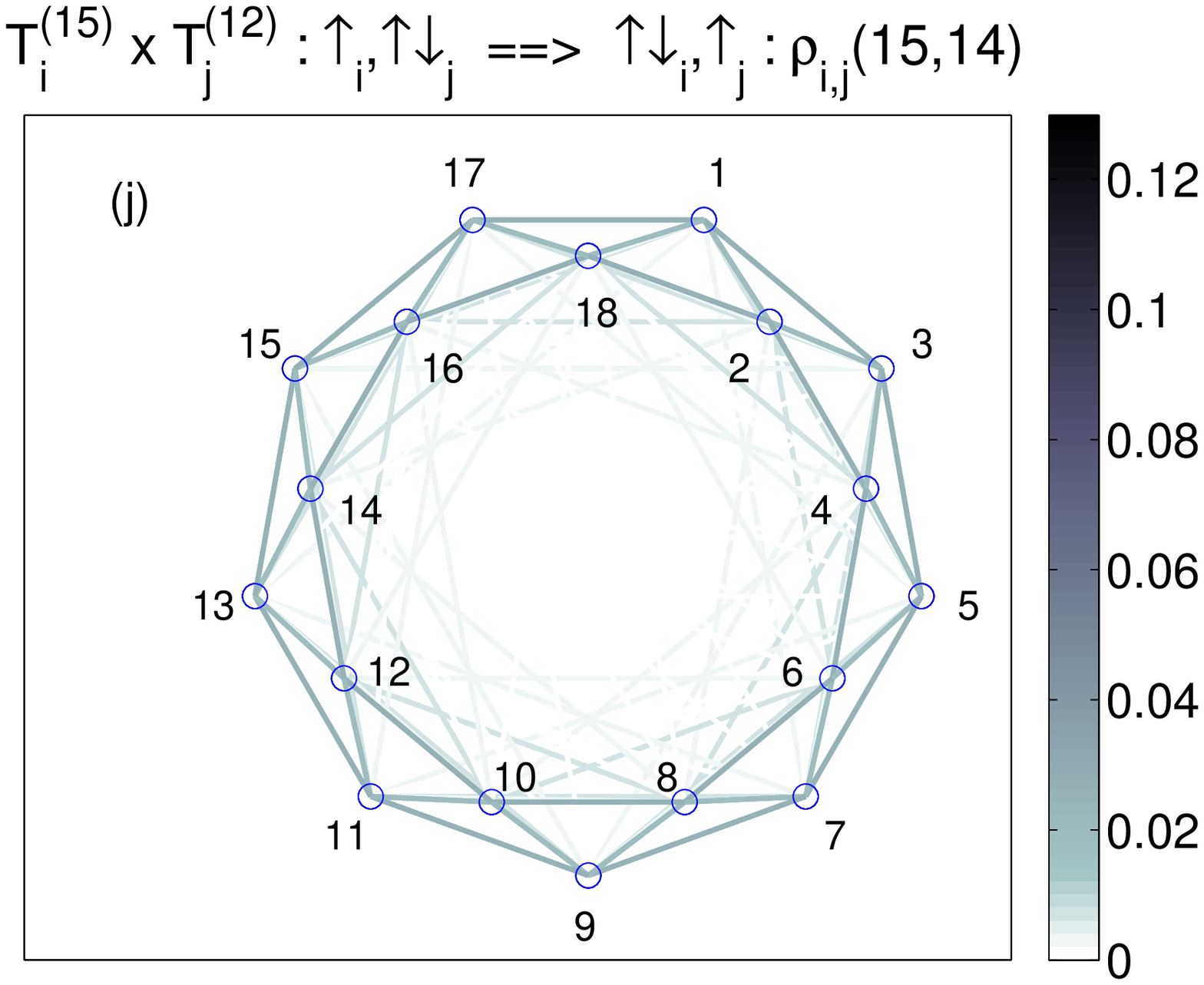}
}
\end{minipage}
\hskip 1.0cm
\begin{minipage}{8cm}
\centerline{
\includegraphics[scale=0.25]{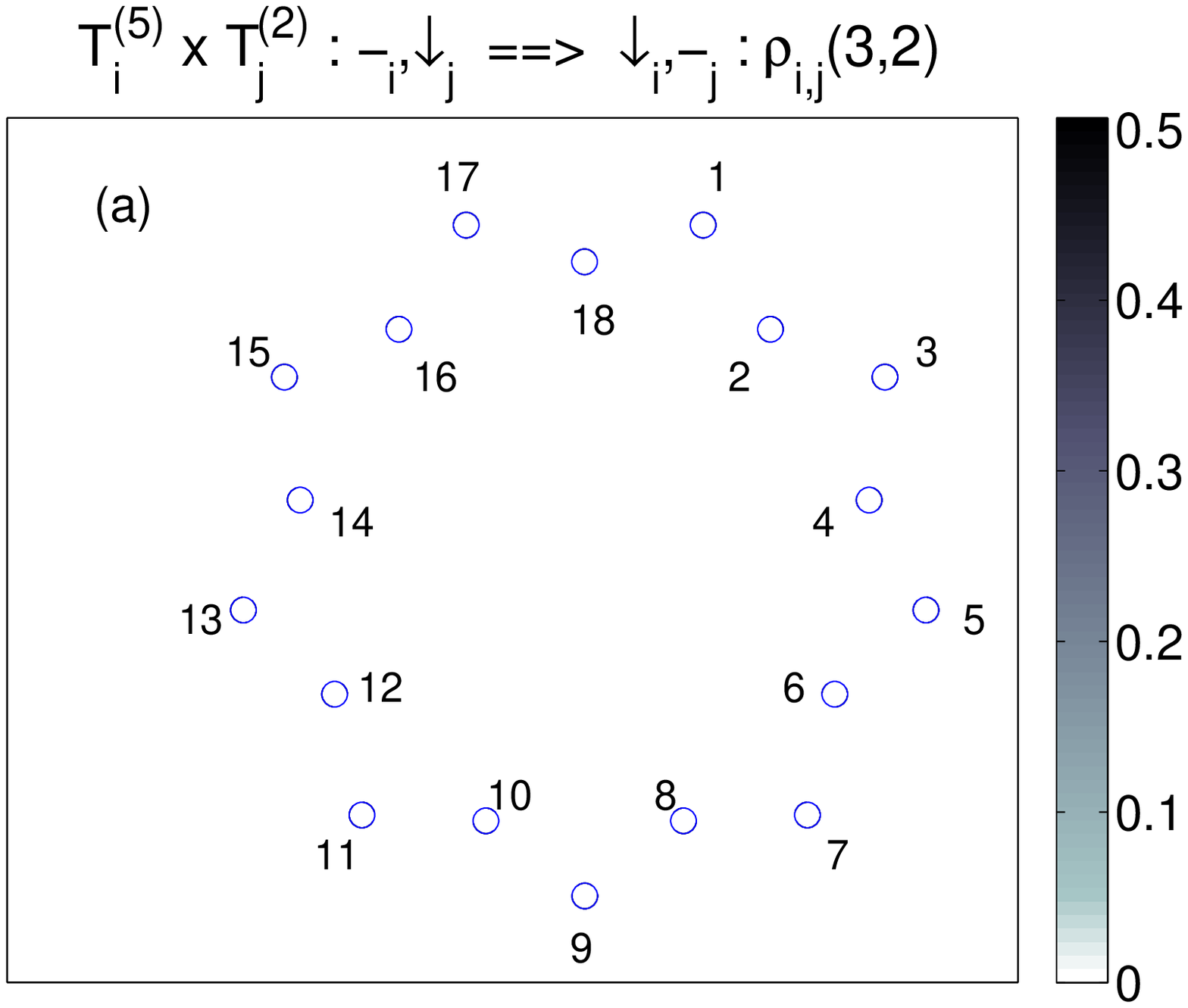}
\includegraphics[scale=0.25]{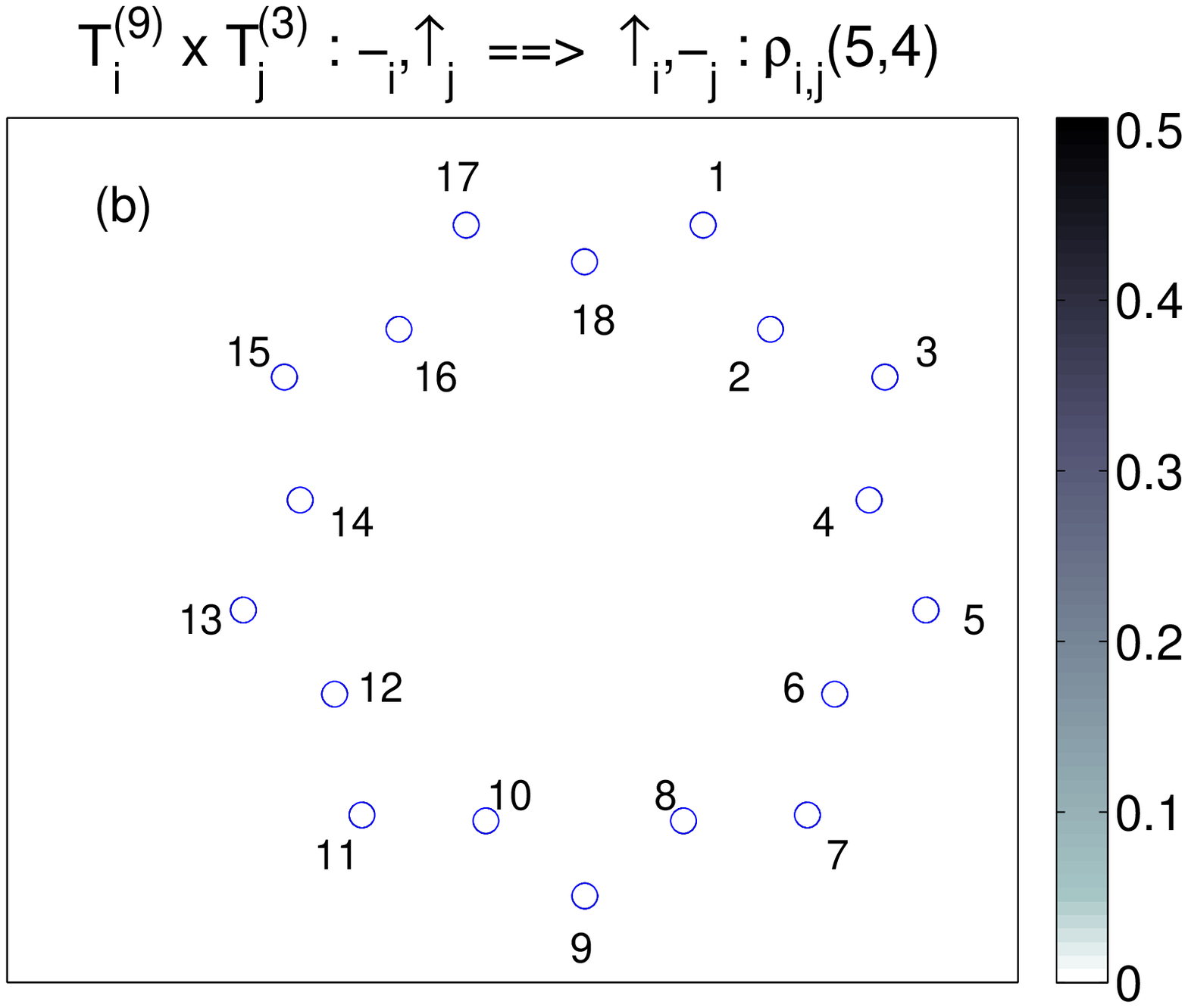}
}
\centerline{
\includegraphics[scale=0.25]{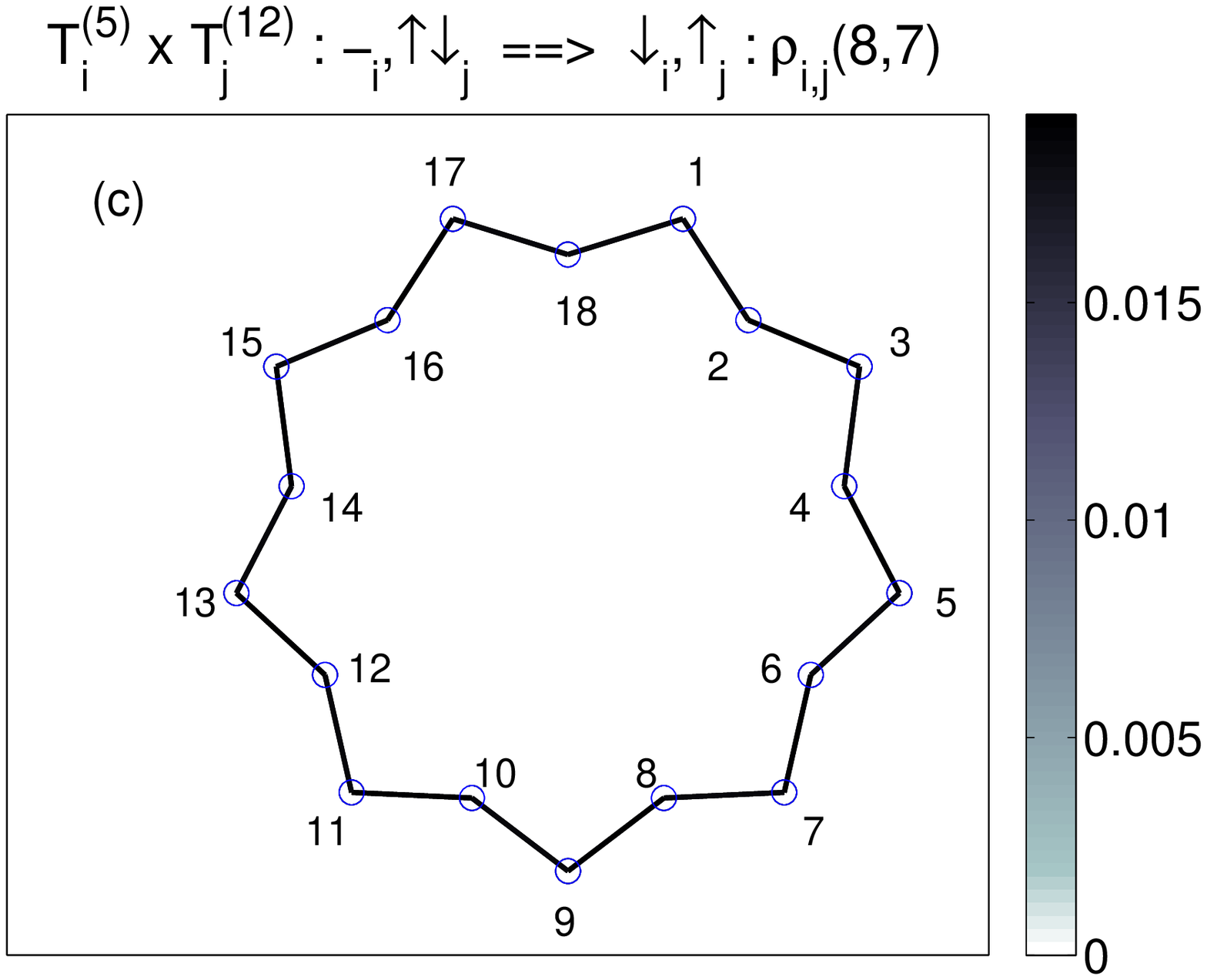}
\includegraphics[scale=0.25]{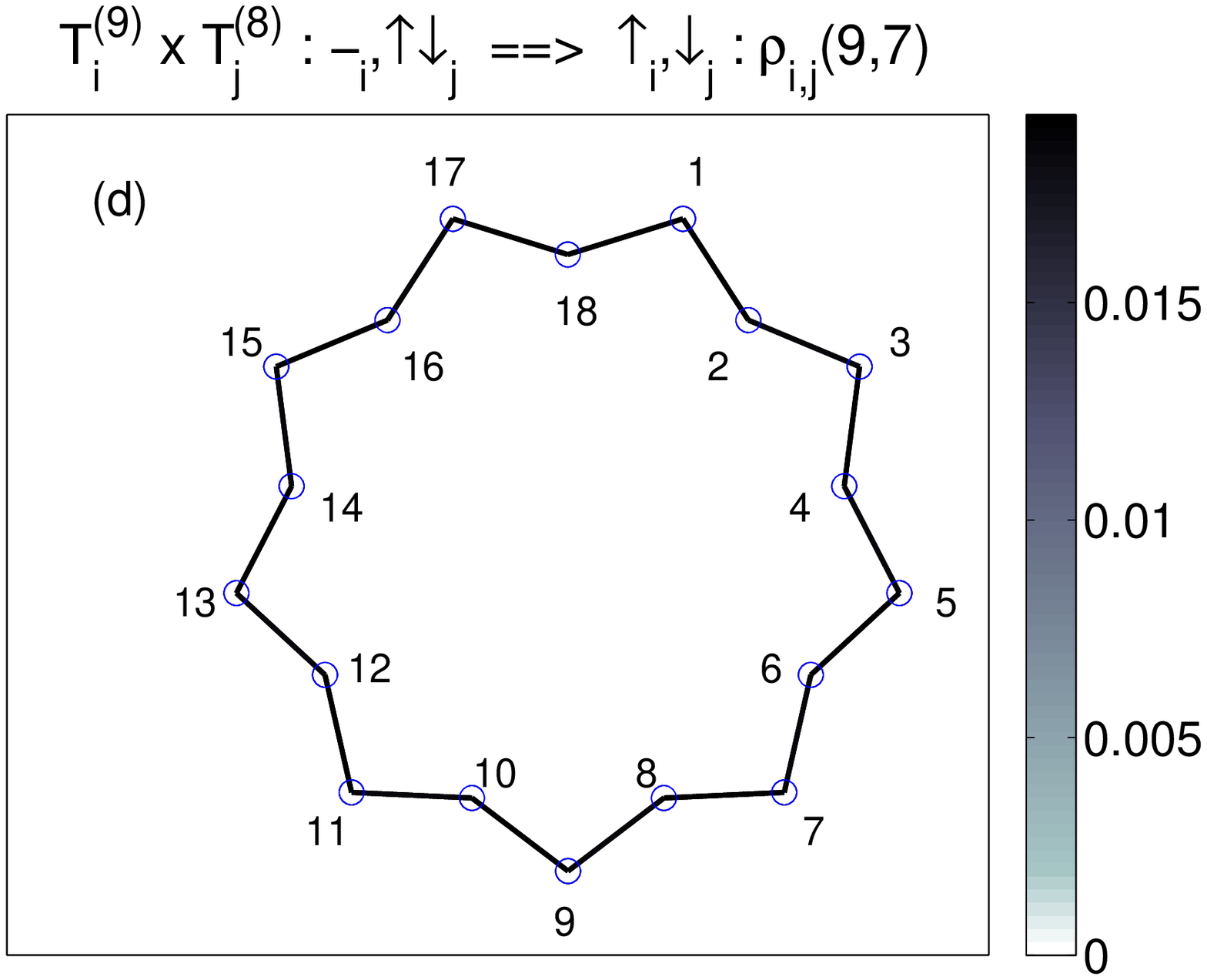}
}
\centerline{
\includegraphics[scale=0.25]{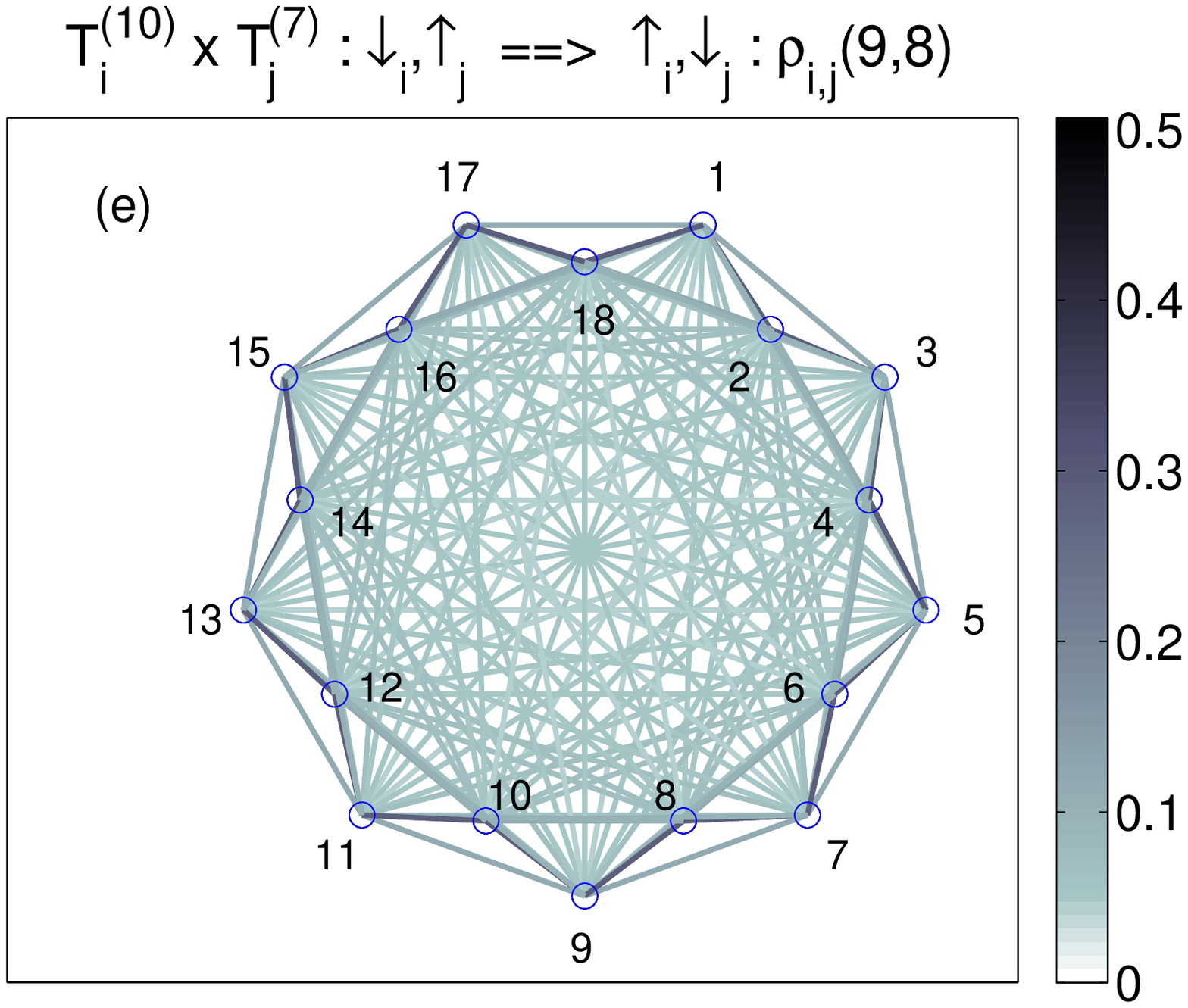}
\includegraphics[scale=0.25]{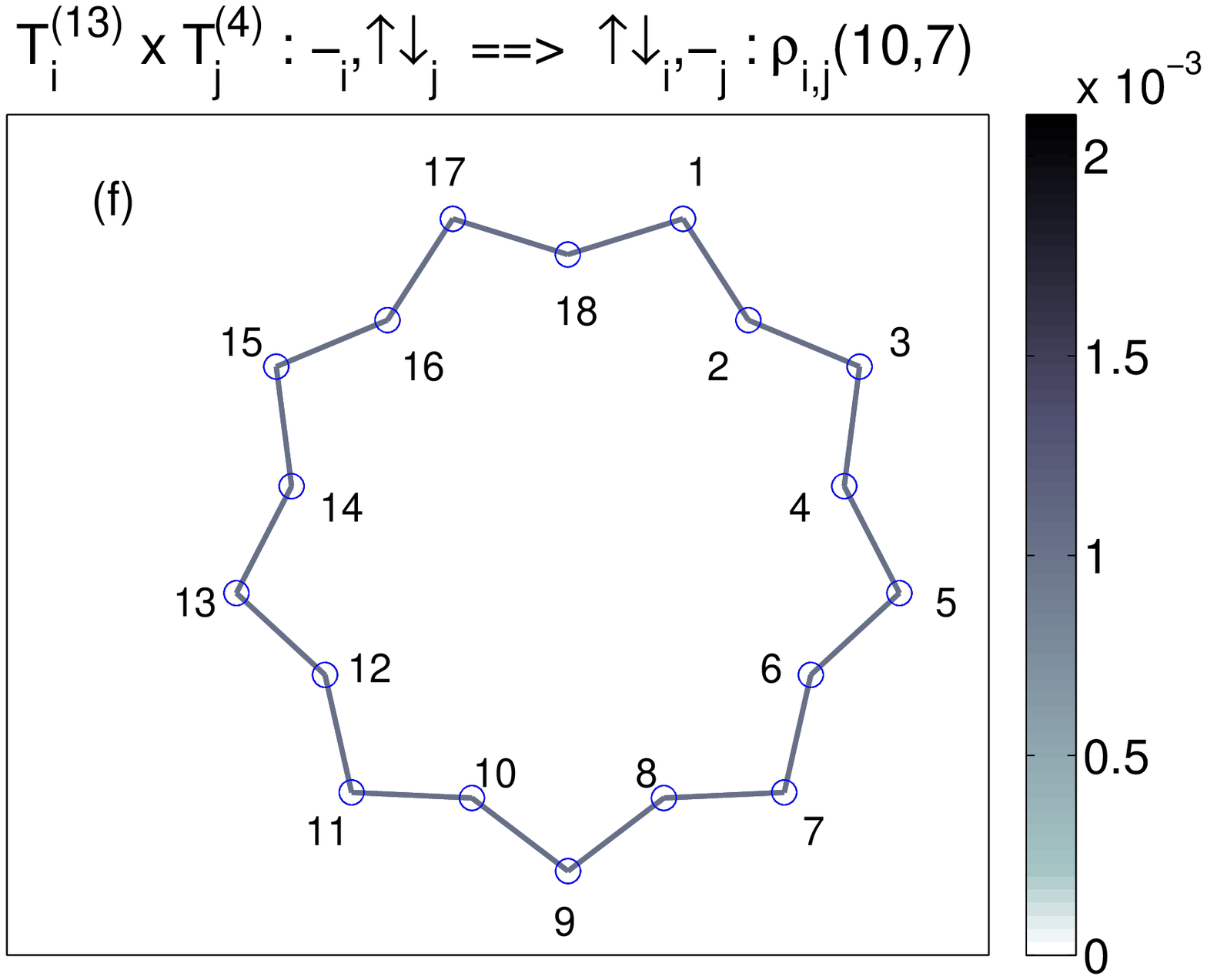}
}
\centerline{
\includegraphics[scale=0.25]{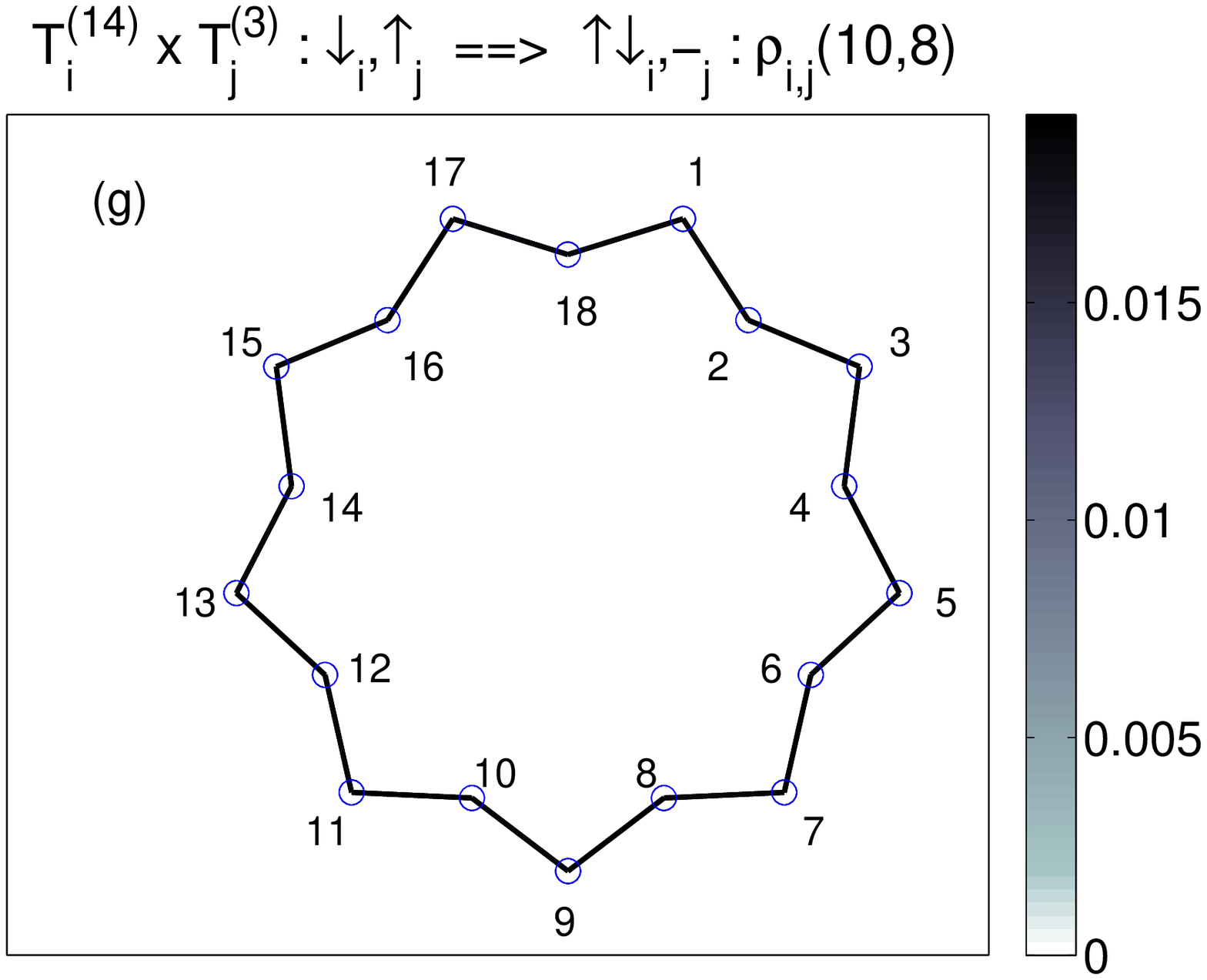}
\includegraphics[scale=0.25]{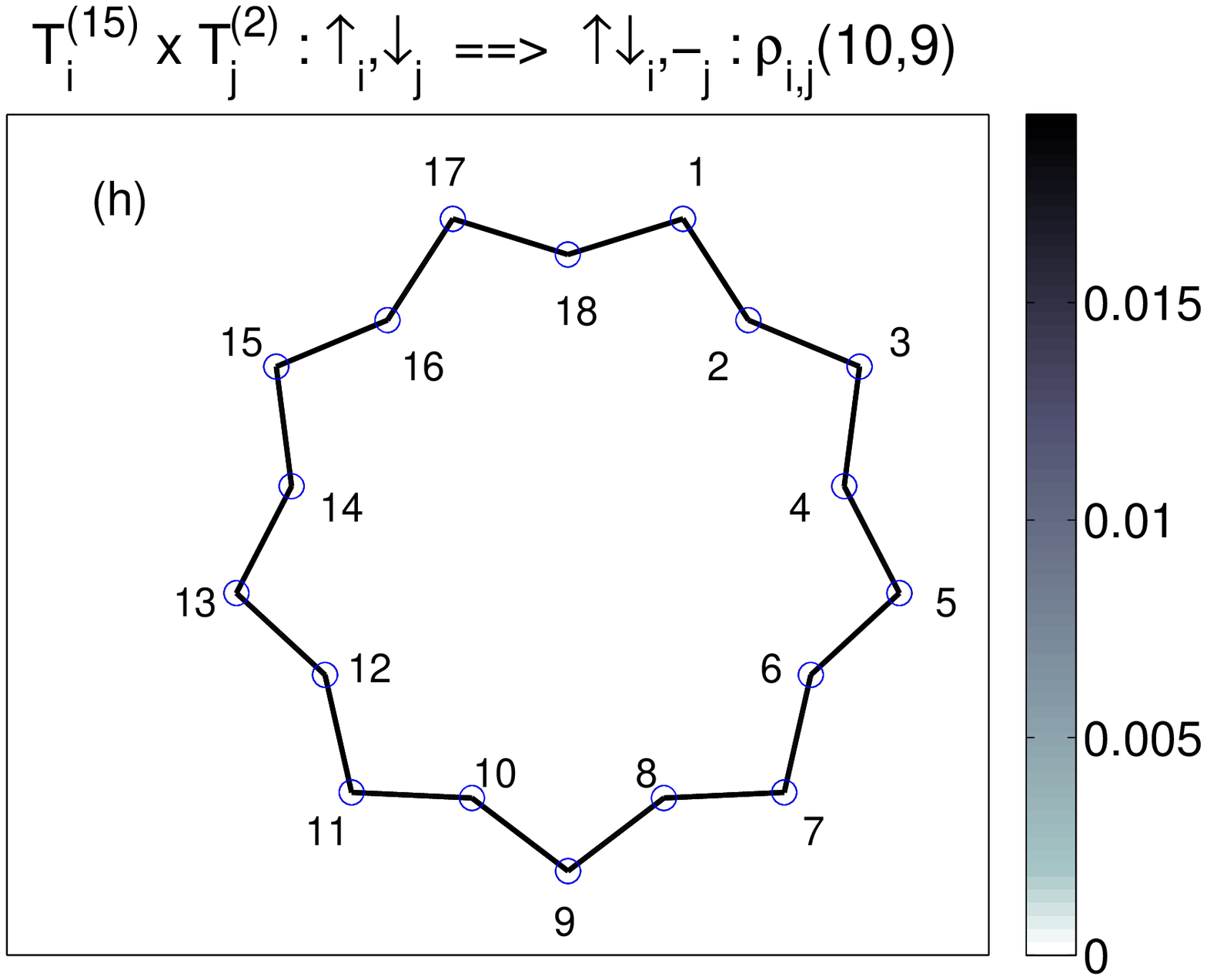}
}
\centerline{
\includegraphics[scale=0.25]{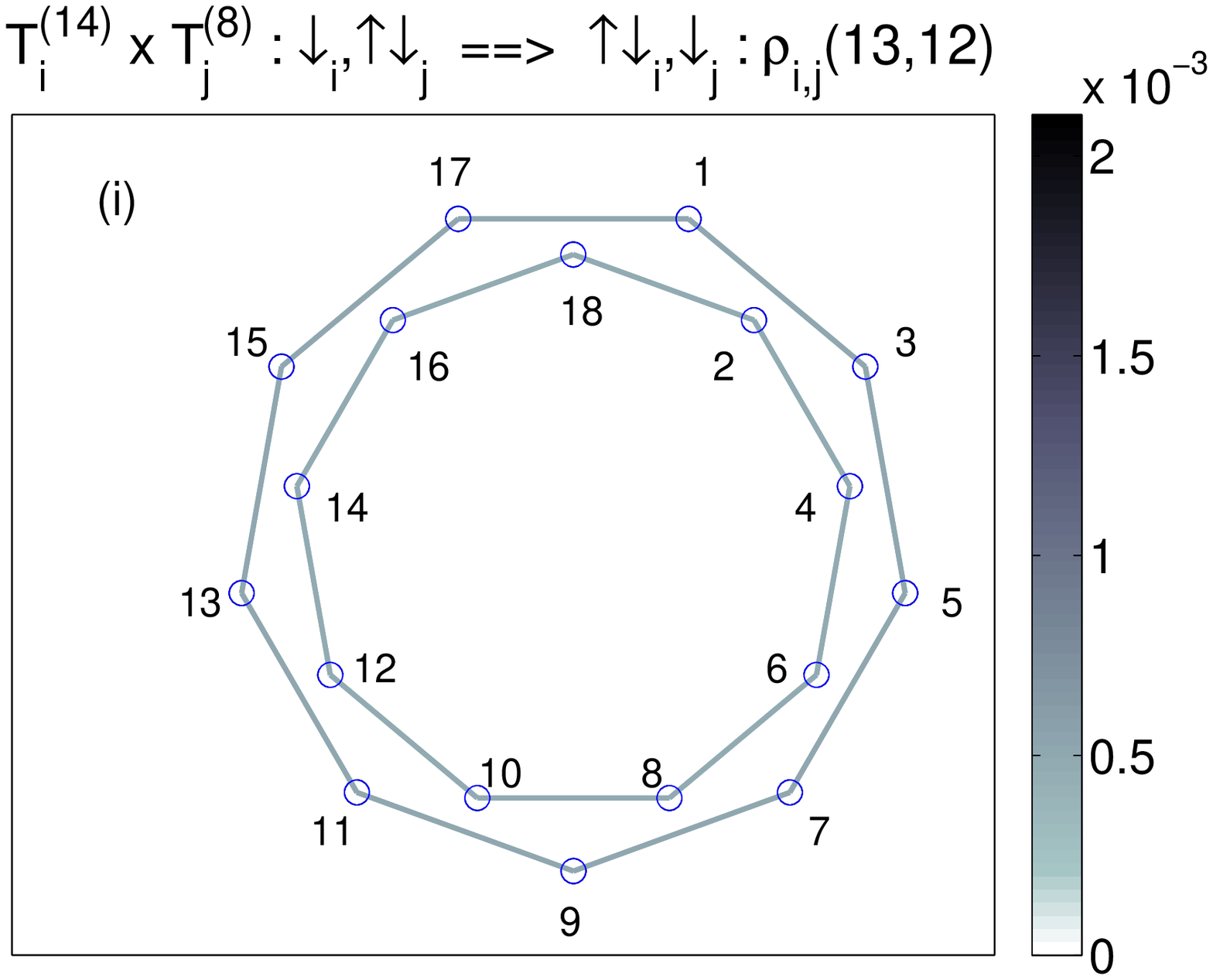}
\includegraphics[scale=0.25]{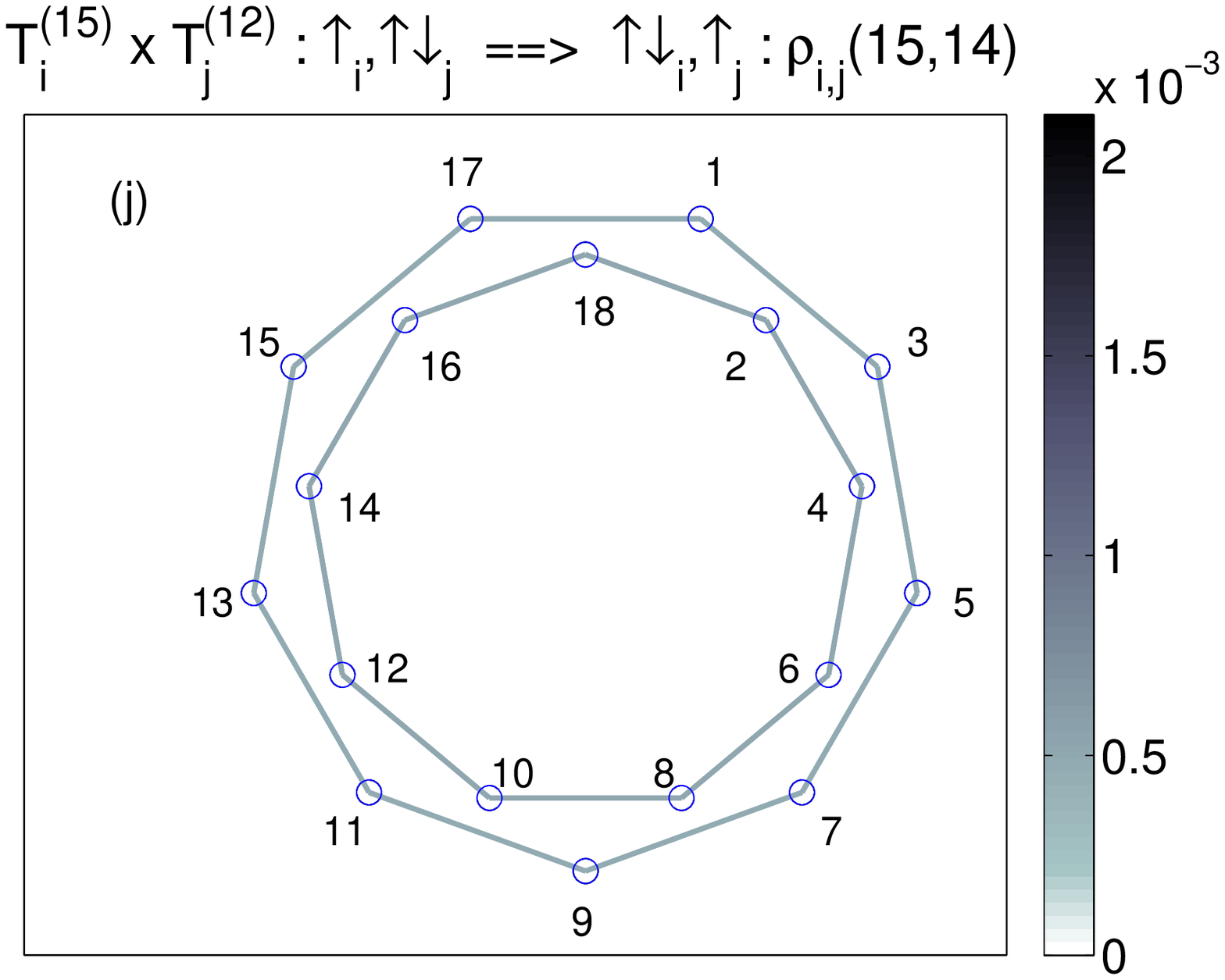}
}
\end{minipage}
\end{minipage}
\caption{Pictorial representation of the absolute value of the generalized correlation functions used to construct the lower-triangular elements of the two-orbital reduced density matrix for Li$_{18}$ using only 2$s$ atomic functions at $d_{\rm Li-Li}=3.05${\AA } and at $d_{\rm Li-Li}=6.00${\AA }. Strength of transition amplitues between initial ($|\alpha_i\rangle|\beta_j\rangle$) and final states ($|\alpha^\prime_i\rangle|\beta^\prime_j\rangle$) on orbital $i$ and $j$ are indicated with different line colors.}
\label{fig:li_rho_3.05}
\end{figure*}
\begin{figure*}
\begin{minipage}{20cm}
\hskip -2.0cm
\begin{minipage}{8cm}
\centerline{
\includegraphics[scale=0.25]{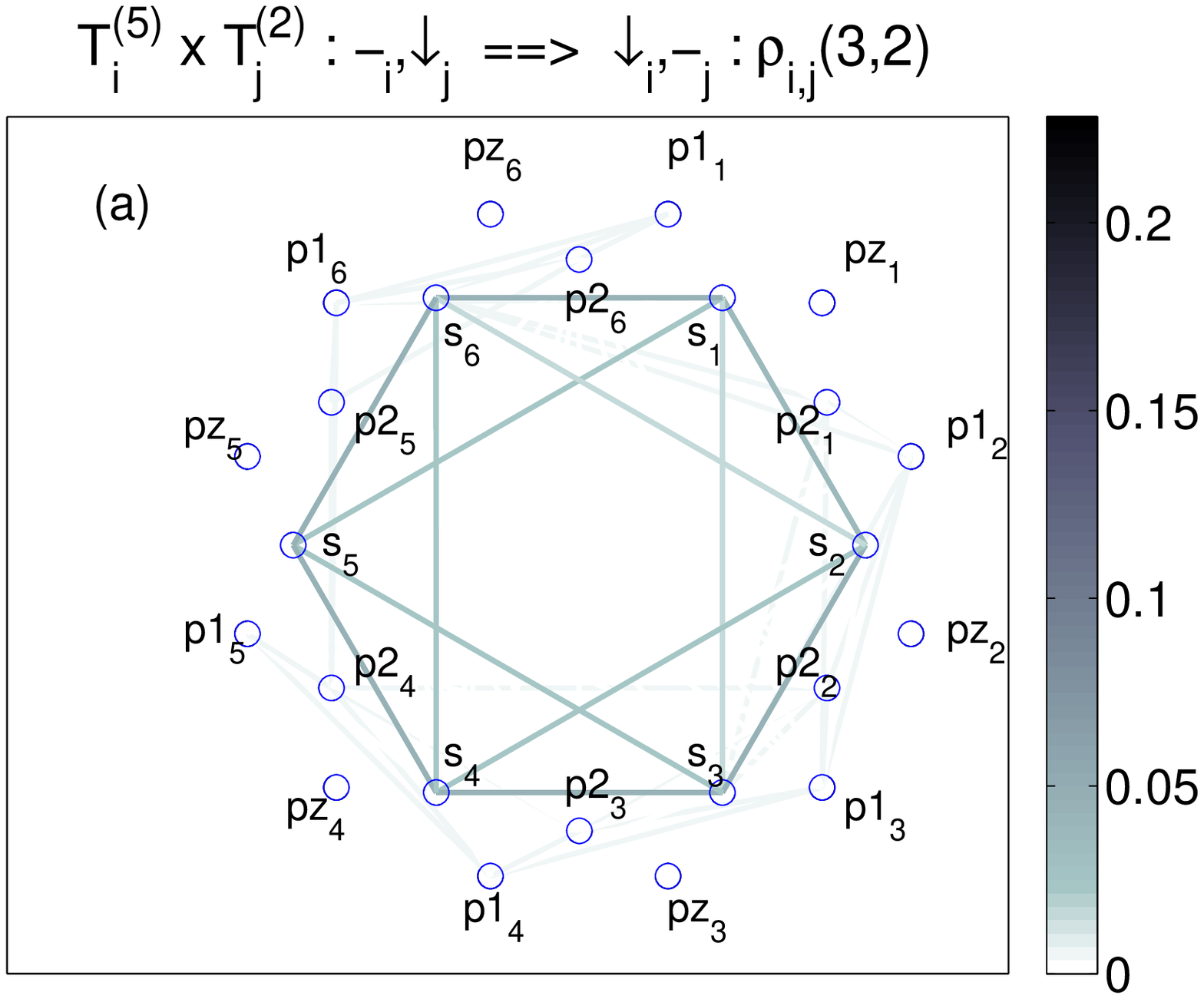}
\includegraphics[scale=0.25]{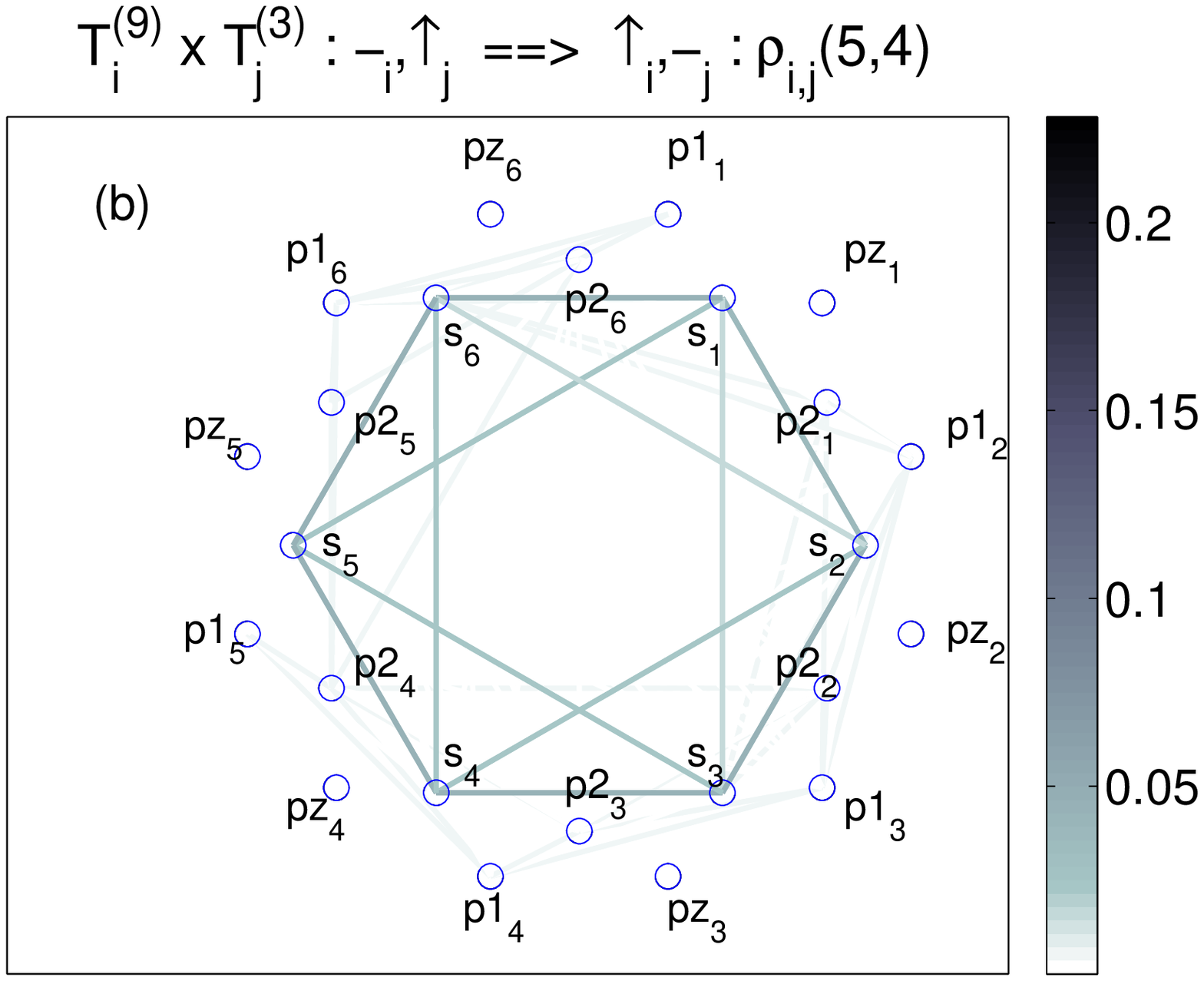}
}
\centerline{
\includegraphics[scale=0.25]{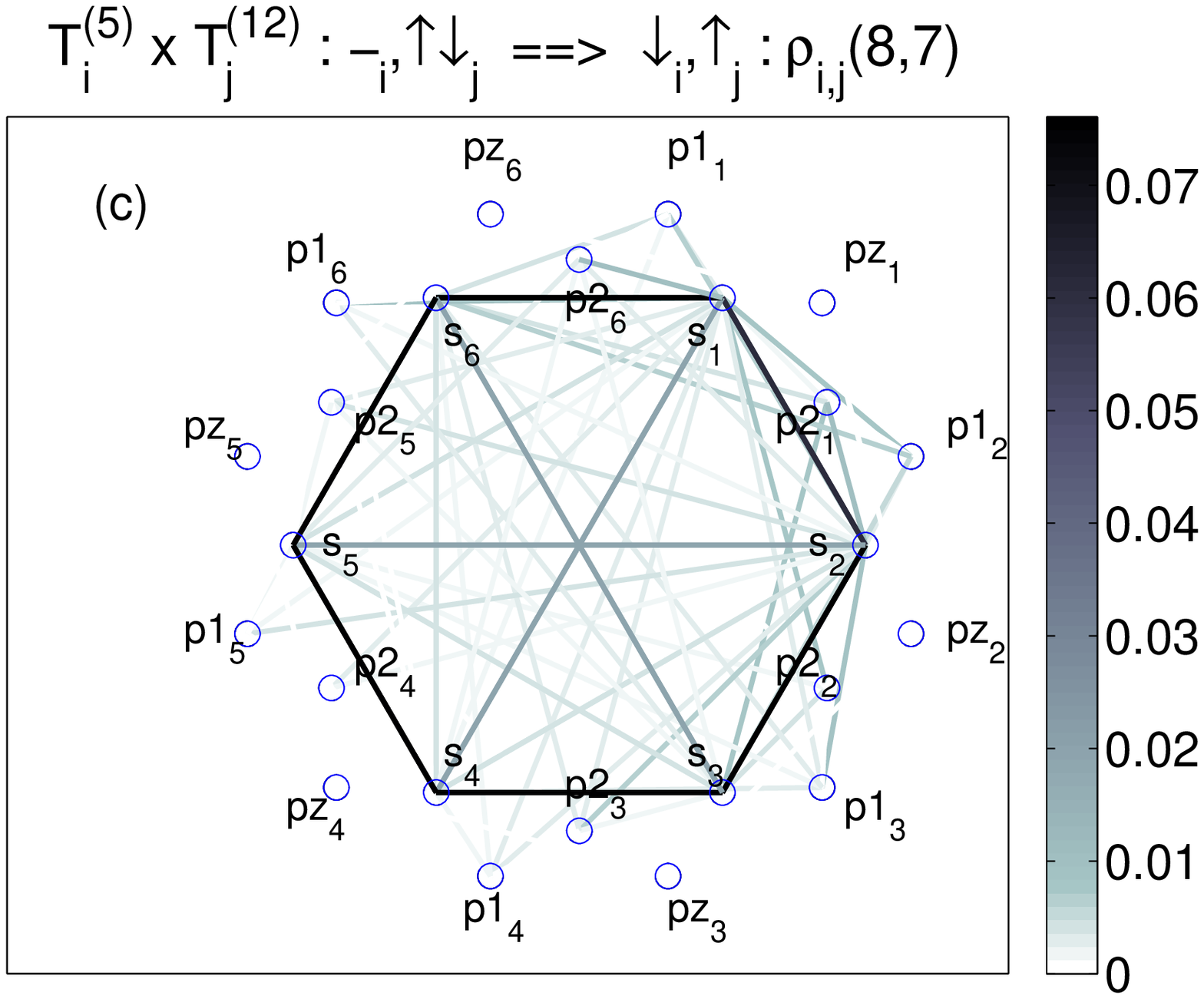}
\includegraphics[scale=0.25]{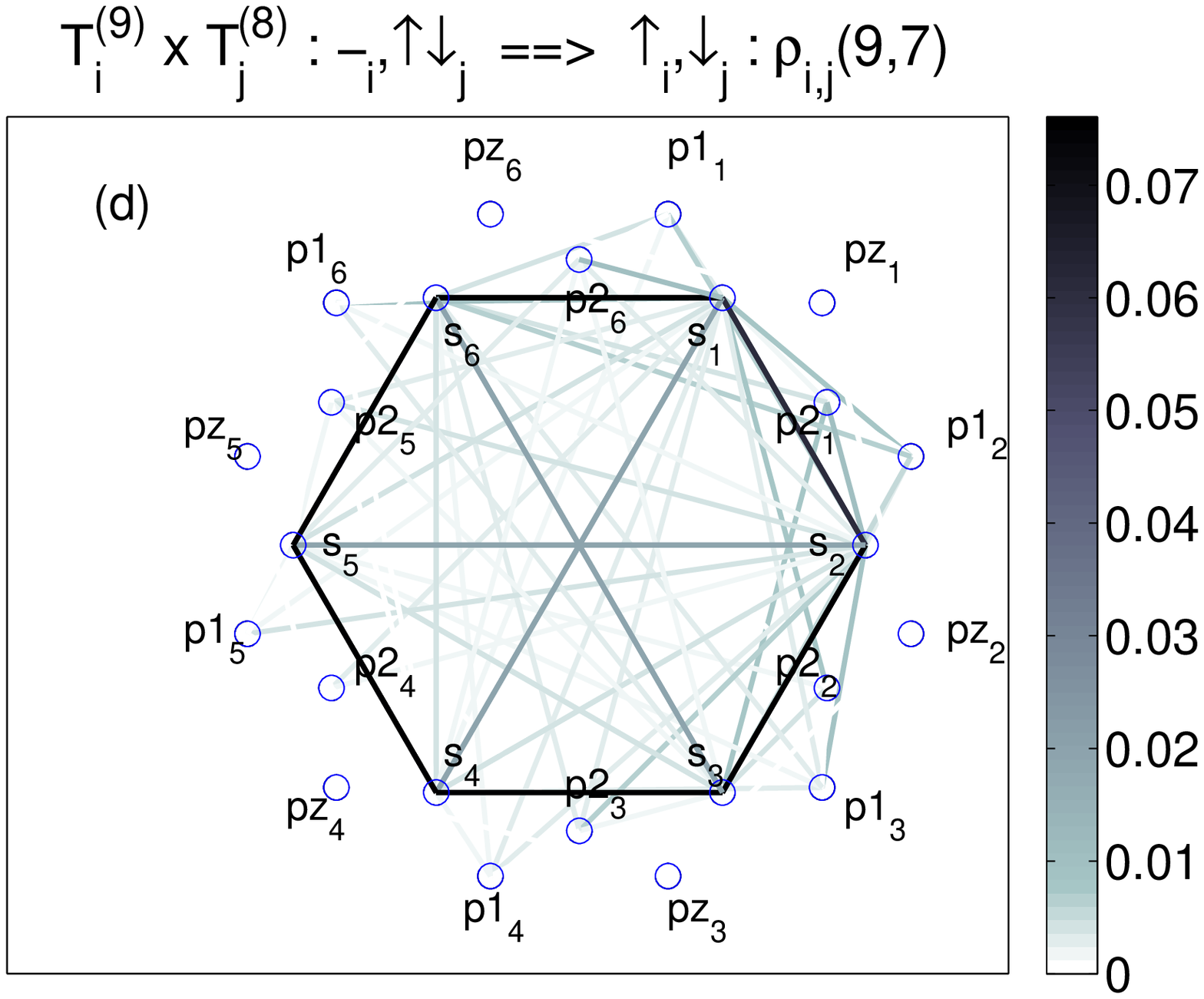}
}
\centerline{
\includegraphics[scale=0.25]{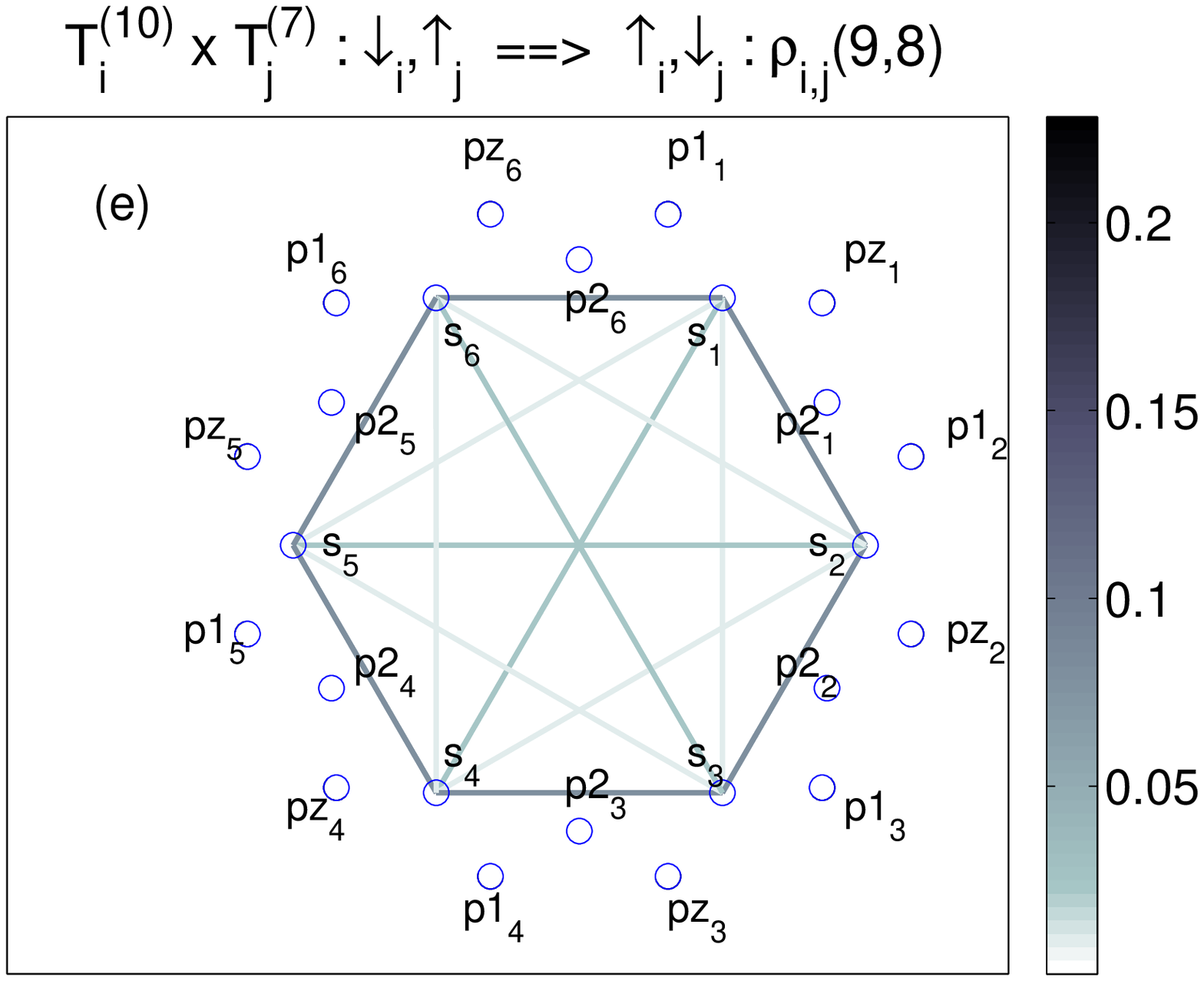}
\includegraphics[scale=0.25]{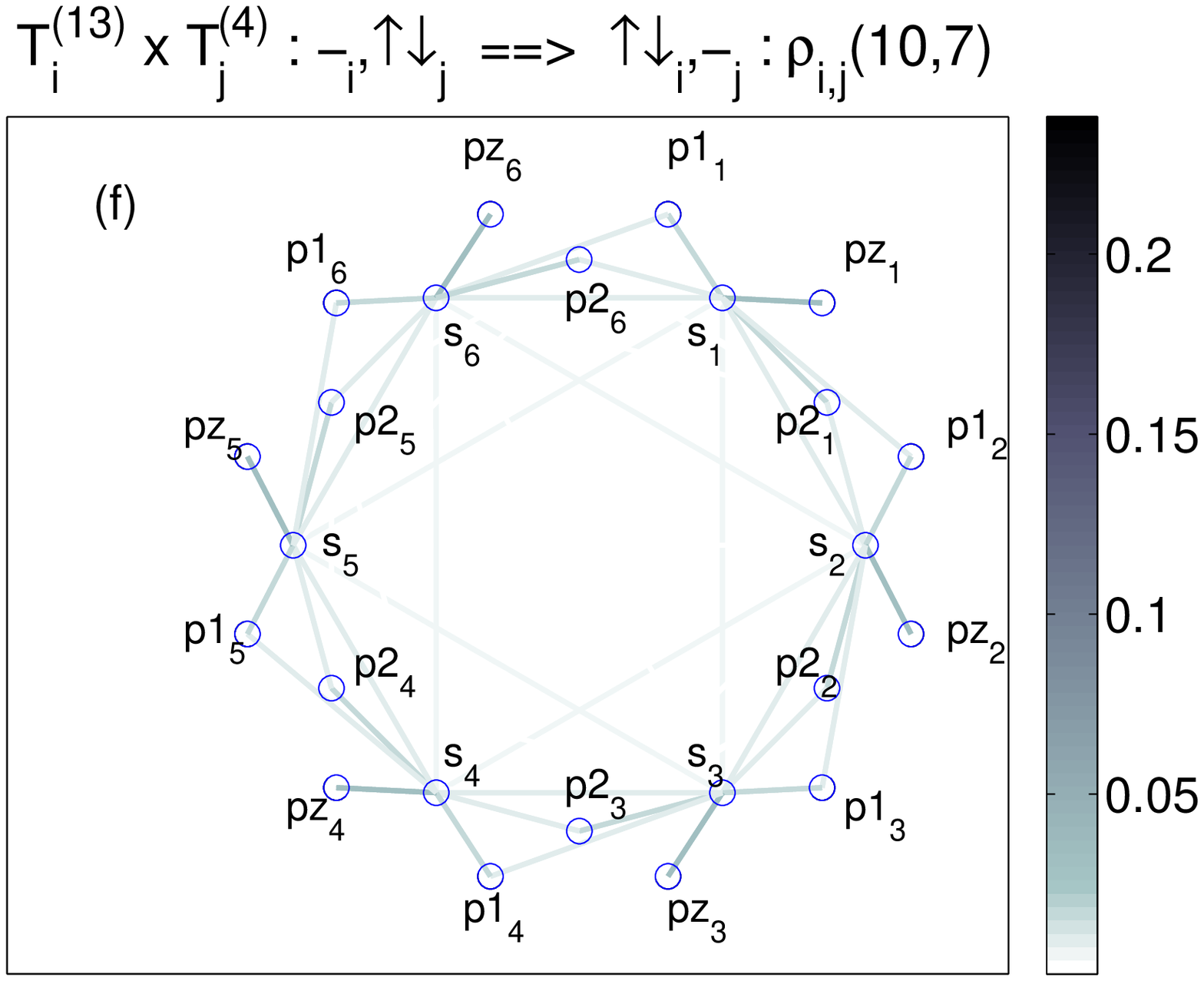}
}
\centerline{
\includegraphics[scale=0.25]{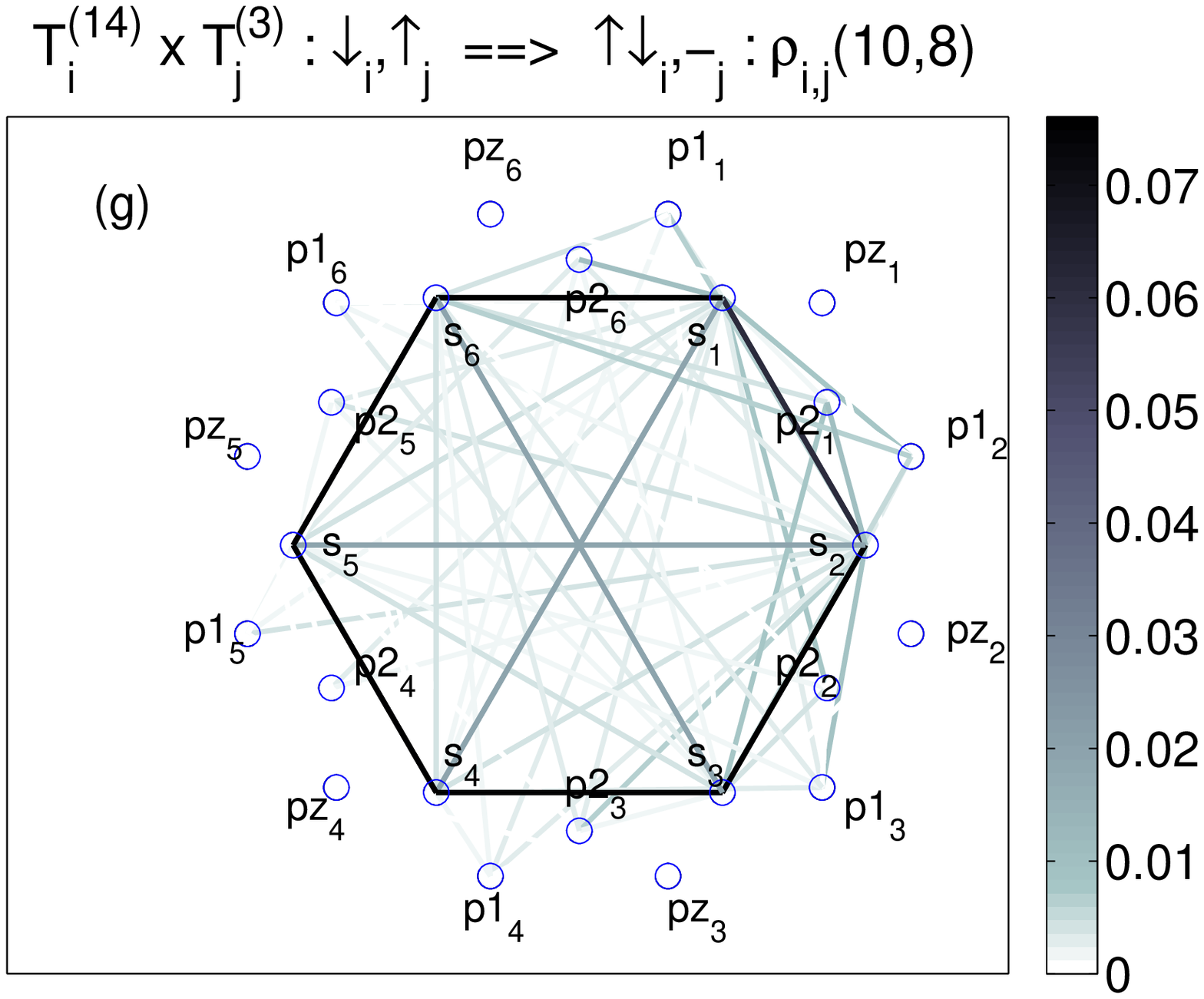}
\includegraphics[scale=0.25]{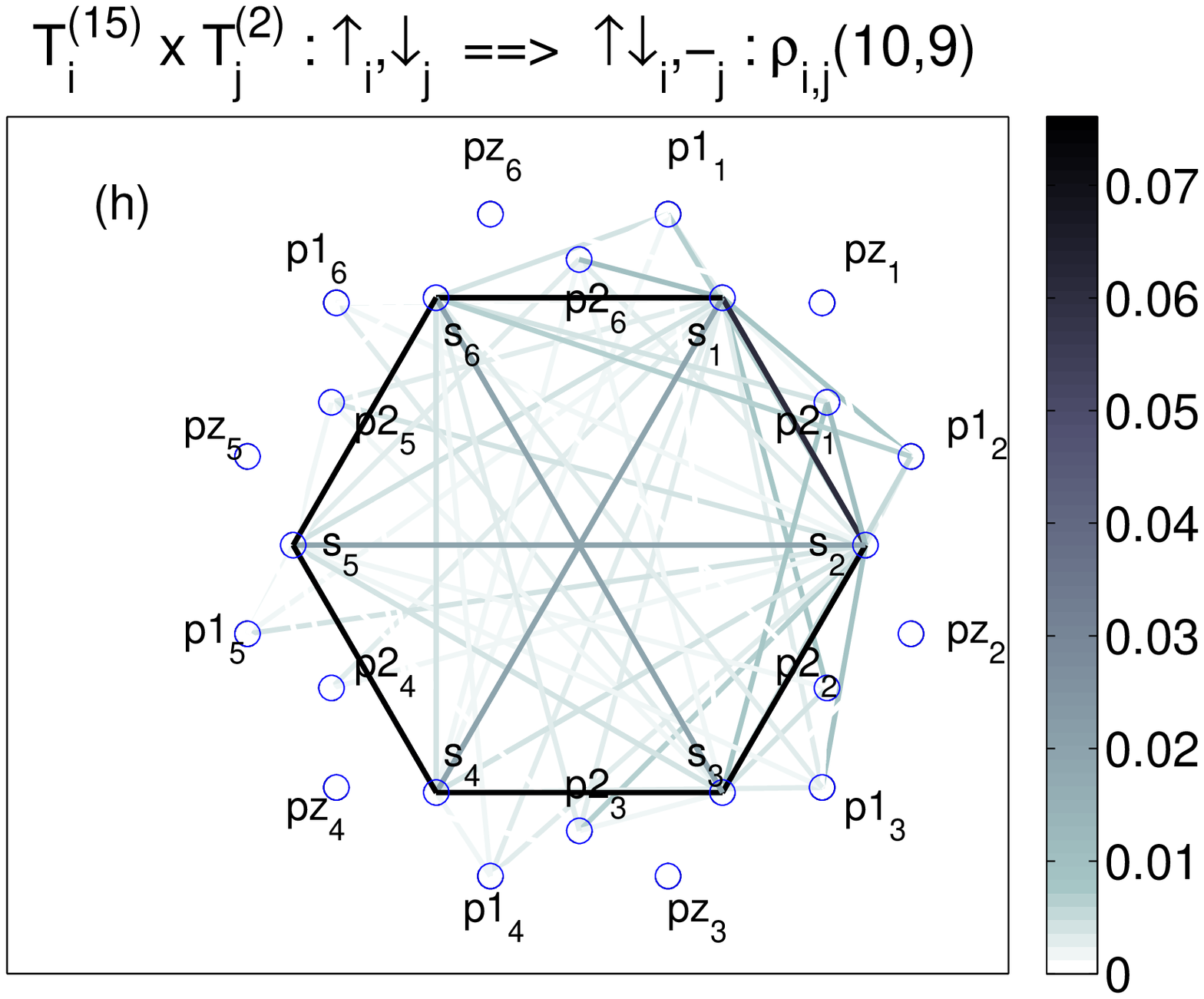}
}
\centerline{
\includegraphics[scale=0.25]{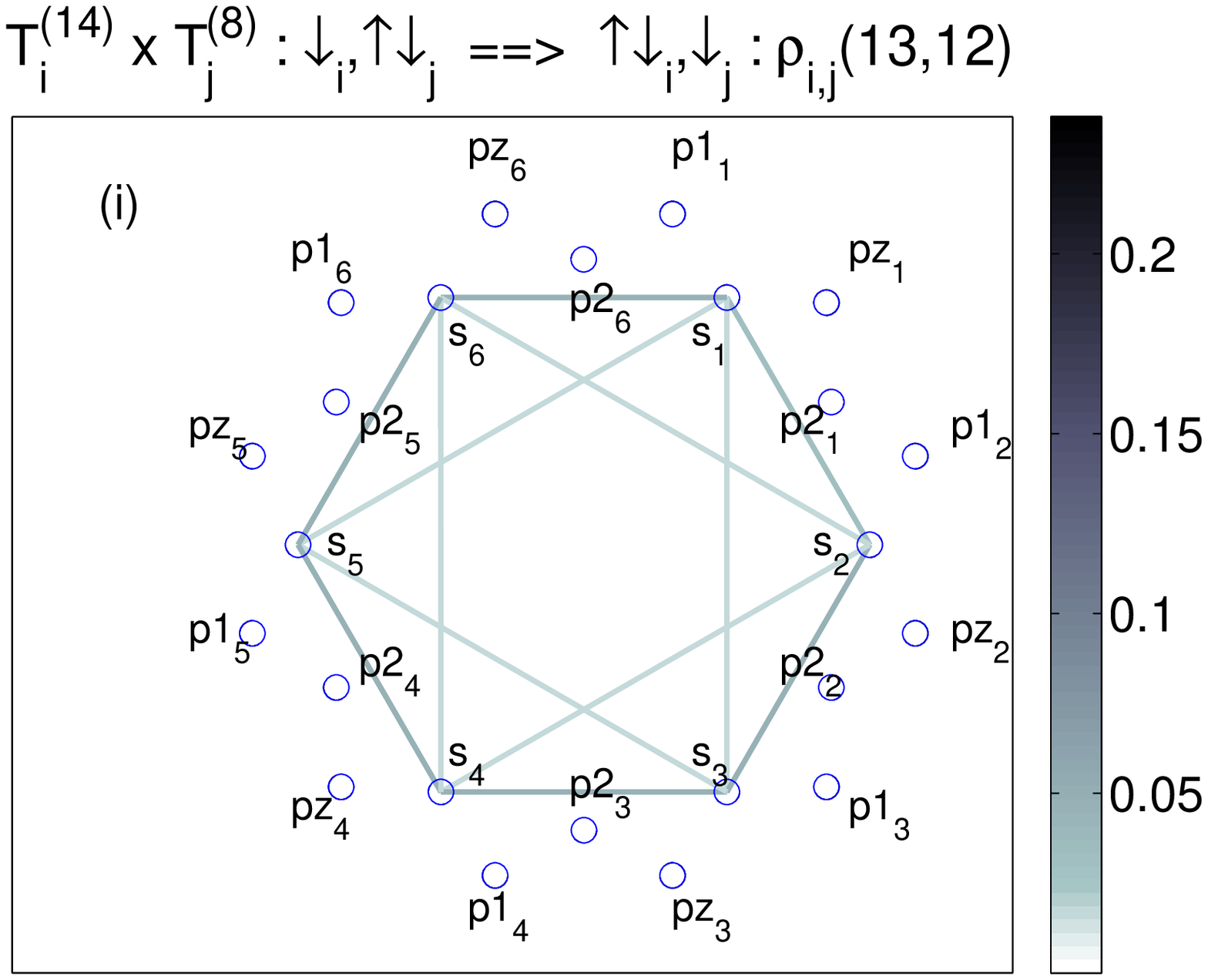}
\includegraphics[scale=0.25]{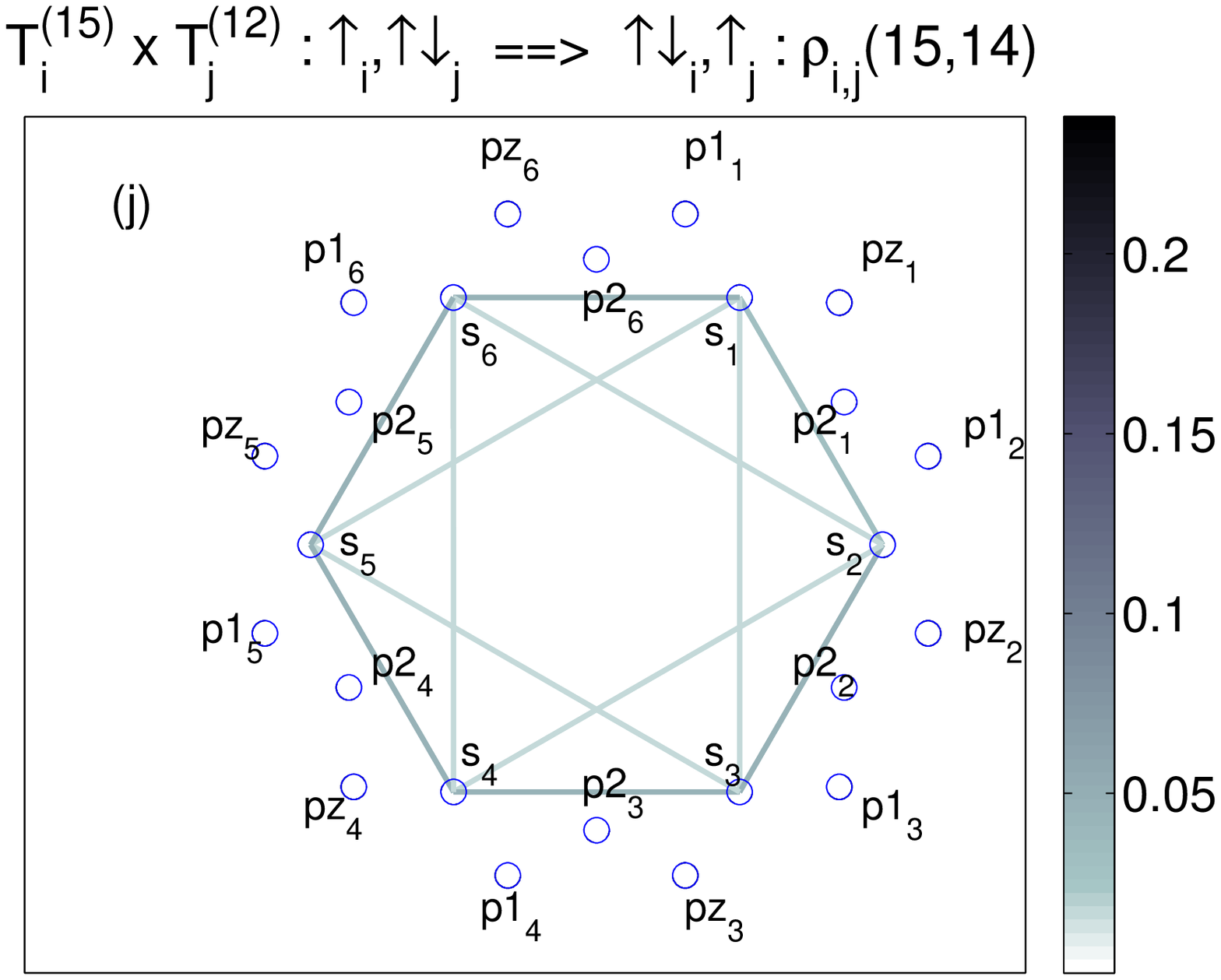}
}
\end{minipage}
\hskip 1.0cm
\begin{minipage}{8cm}
\centerline{
\includegraphics[scale=0.25]{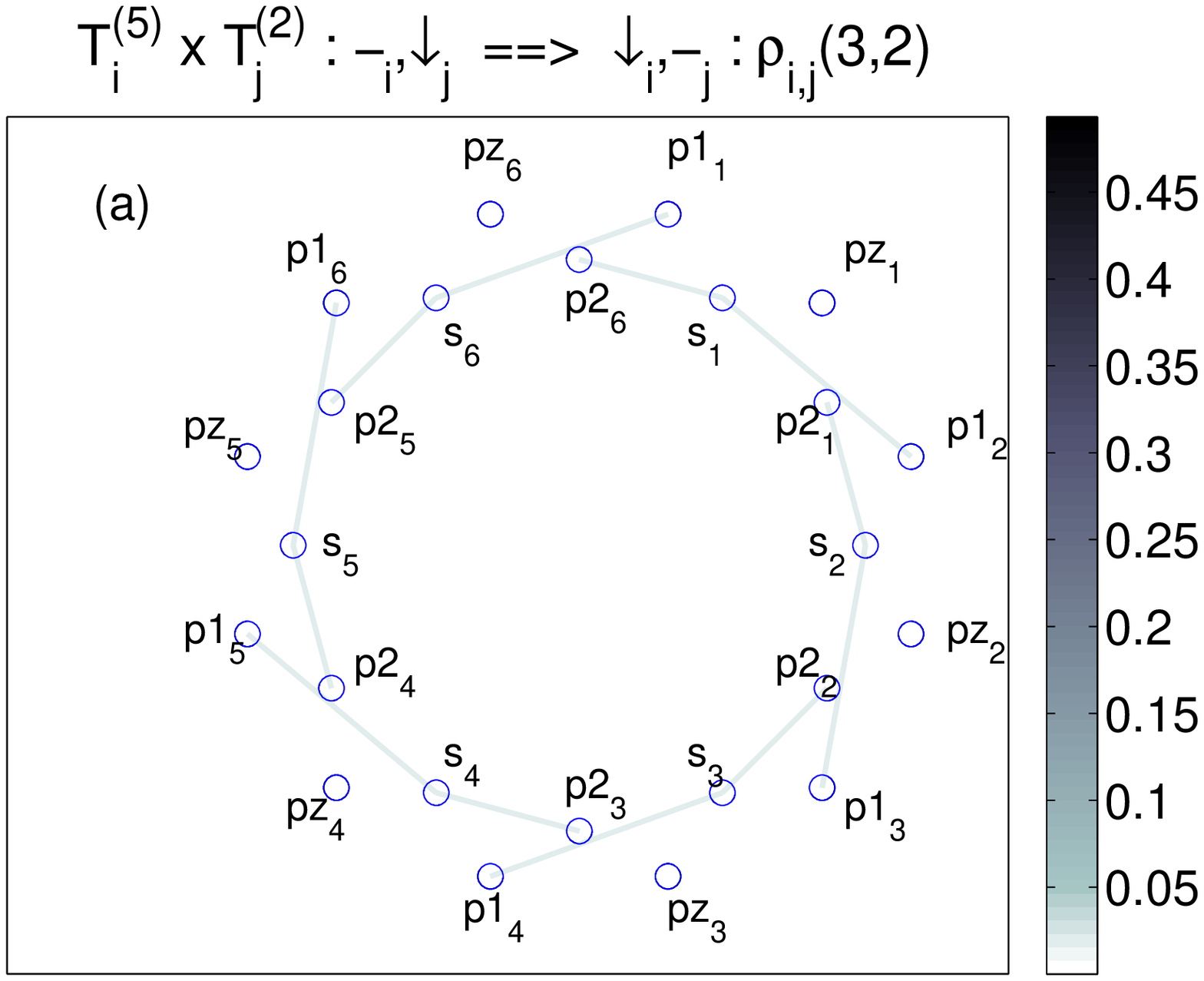}
\includegraphics[scale=0.25]{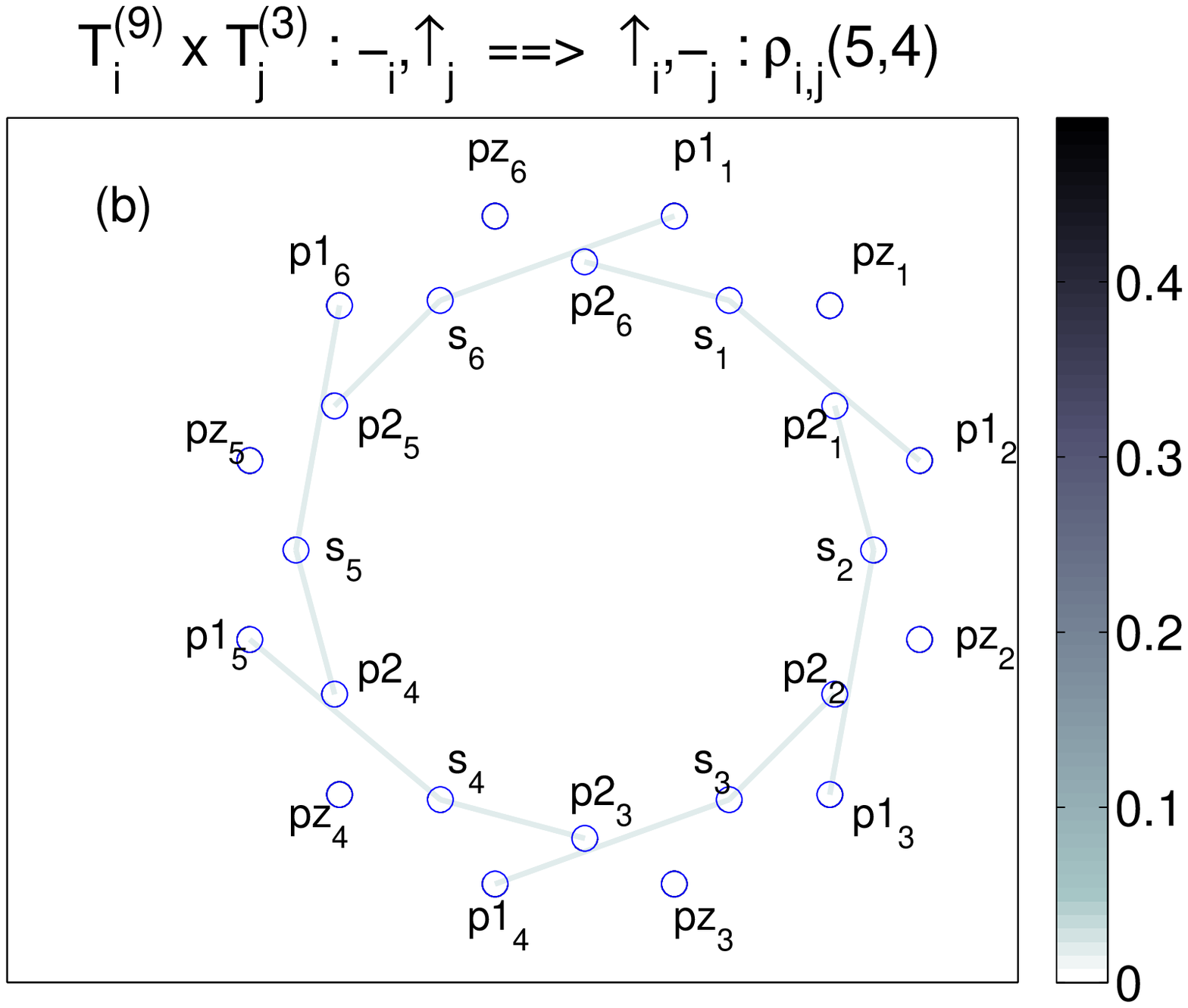}
}
\centerline{
\includegraphics[scale=0.25]{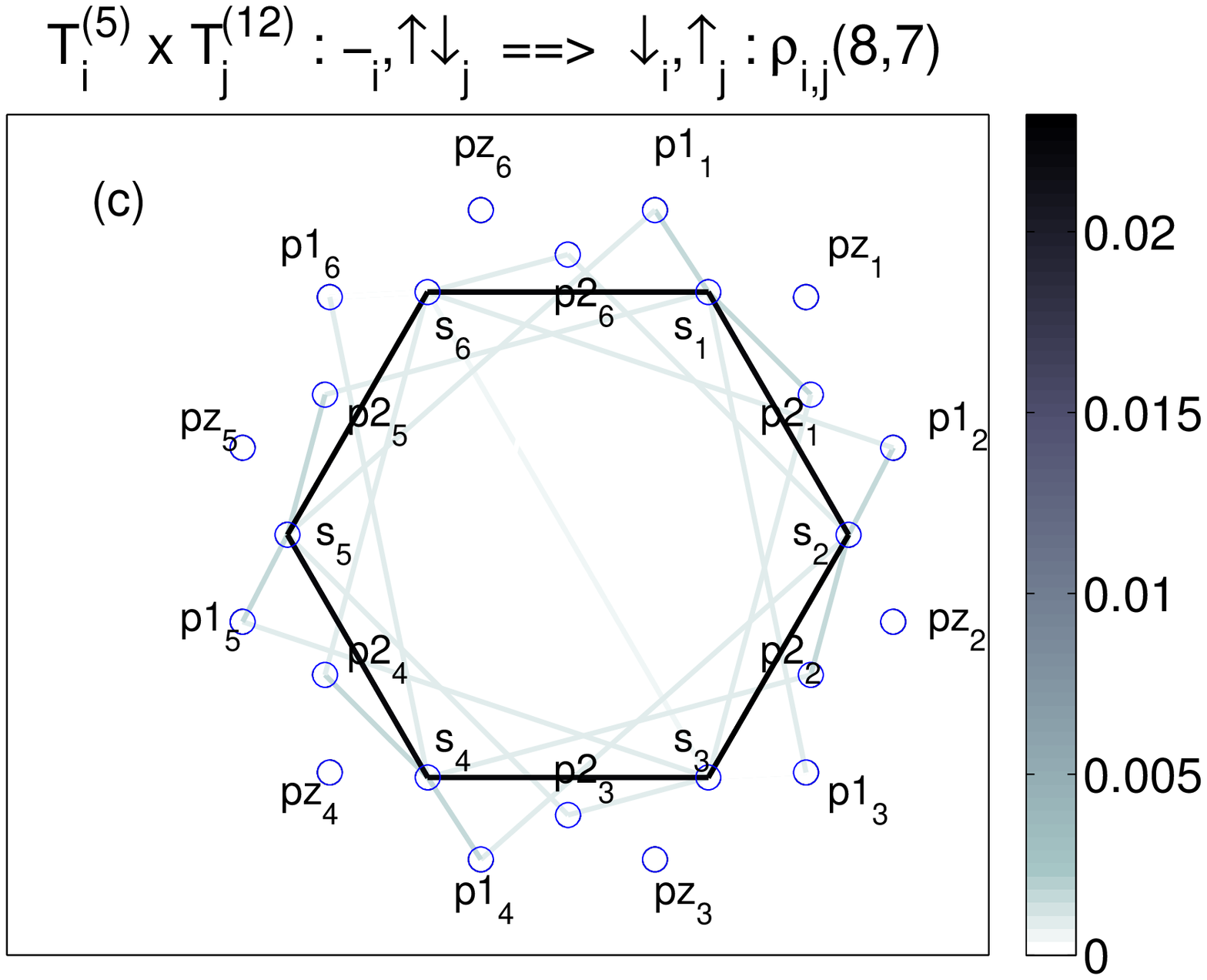}
\includegraphics[scale=0.25]{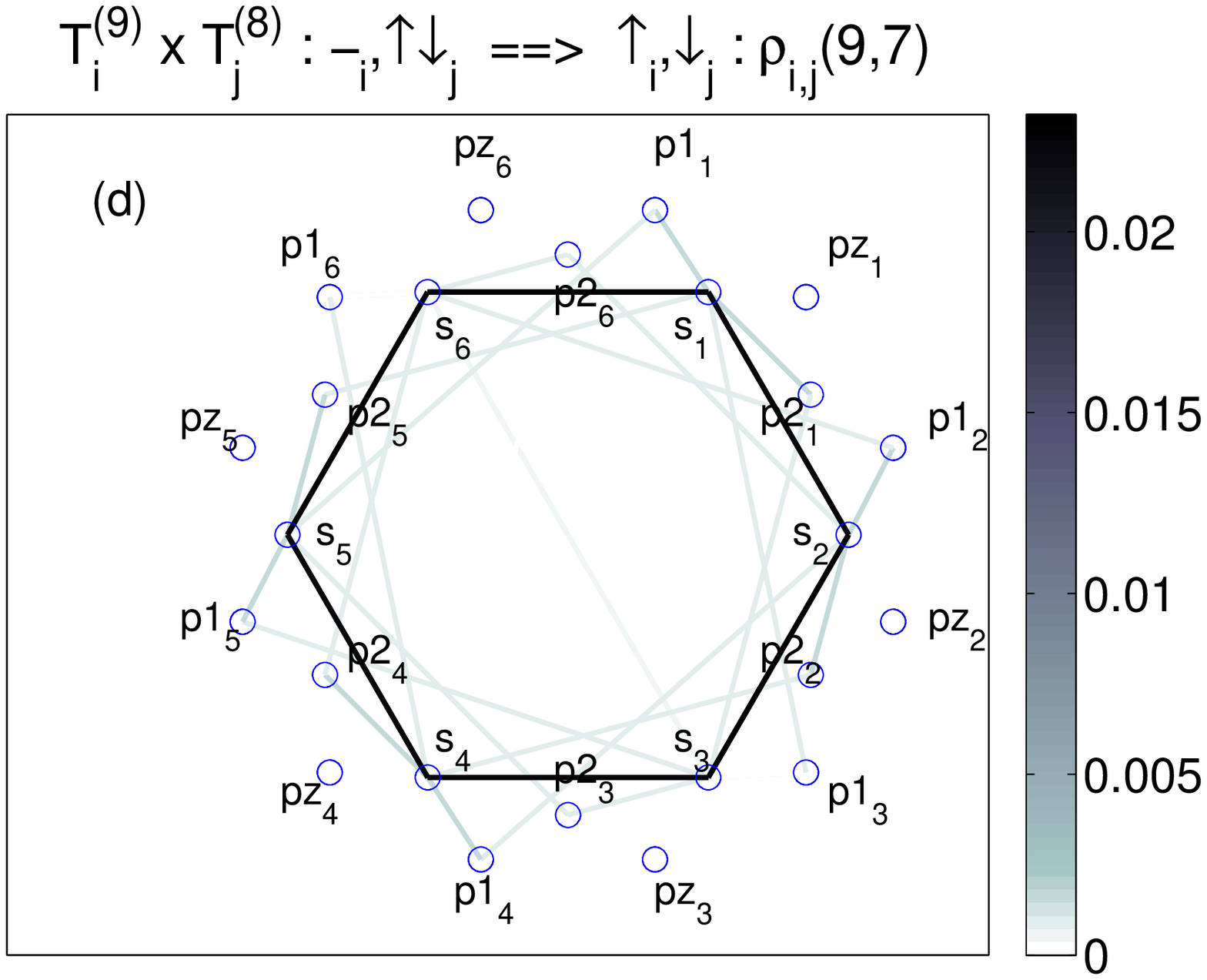}
}
\centerline{
\includegraphics[scale=0.25]{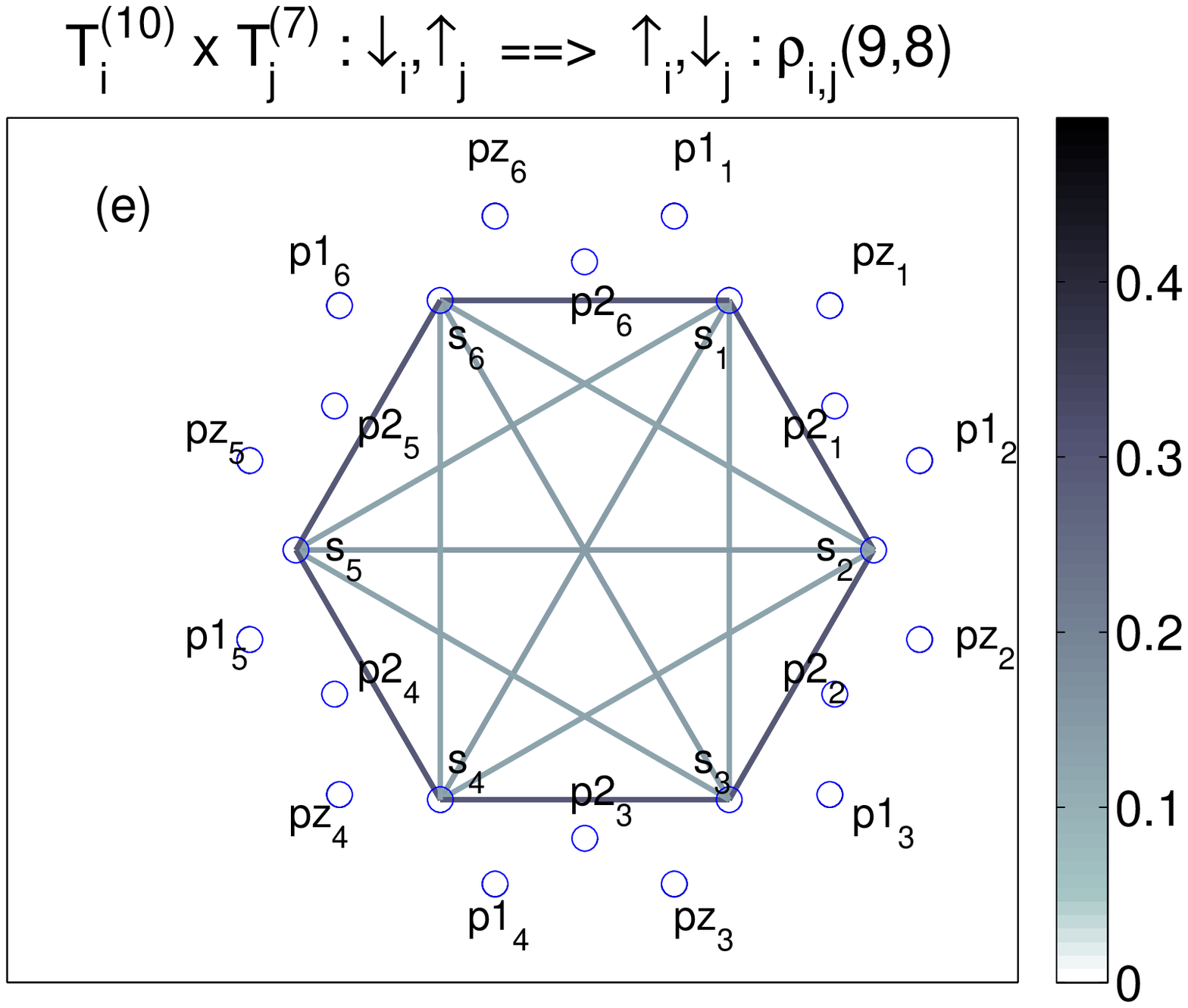}
\includegraphics[scale=0.25]{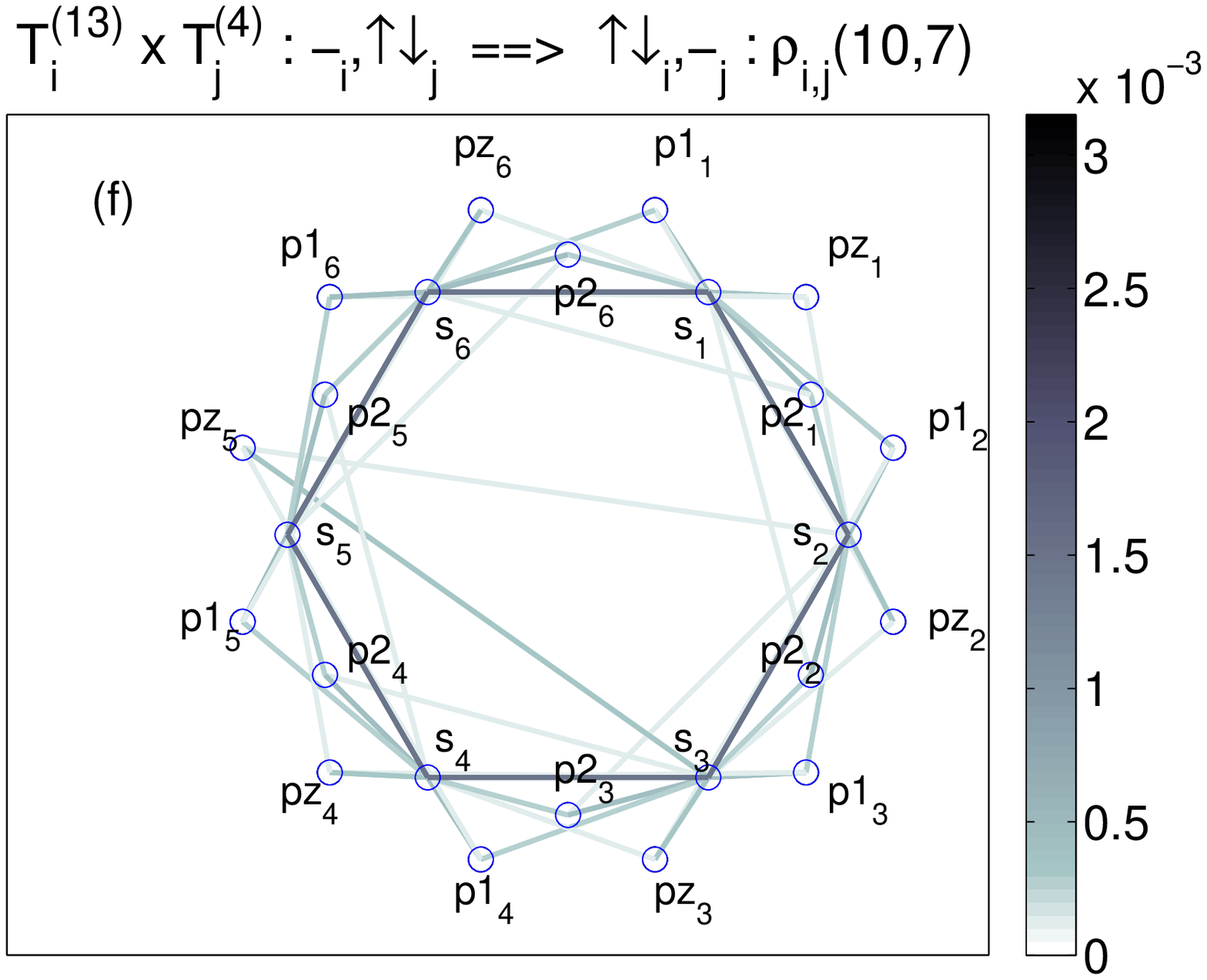}
}
\centerline{
\includegraphics[scale=0.25]{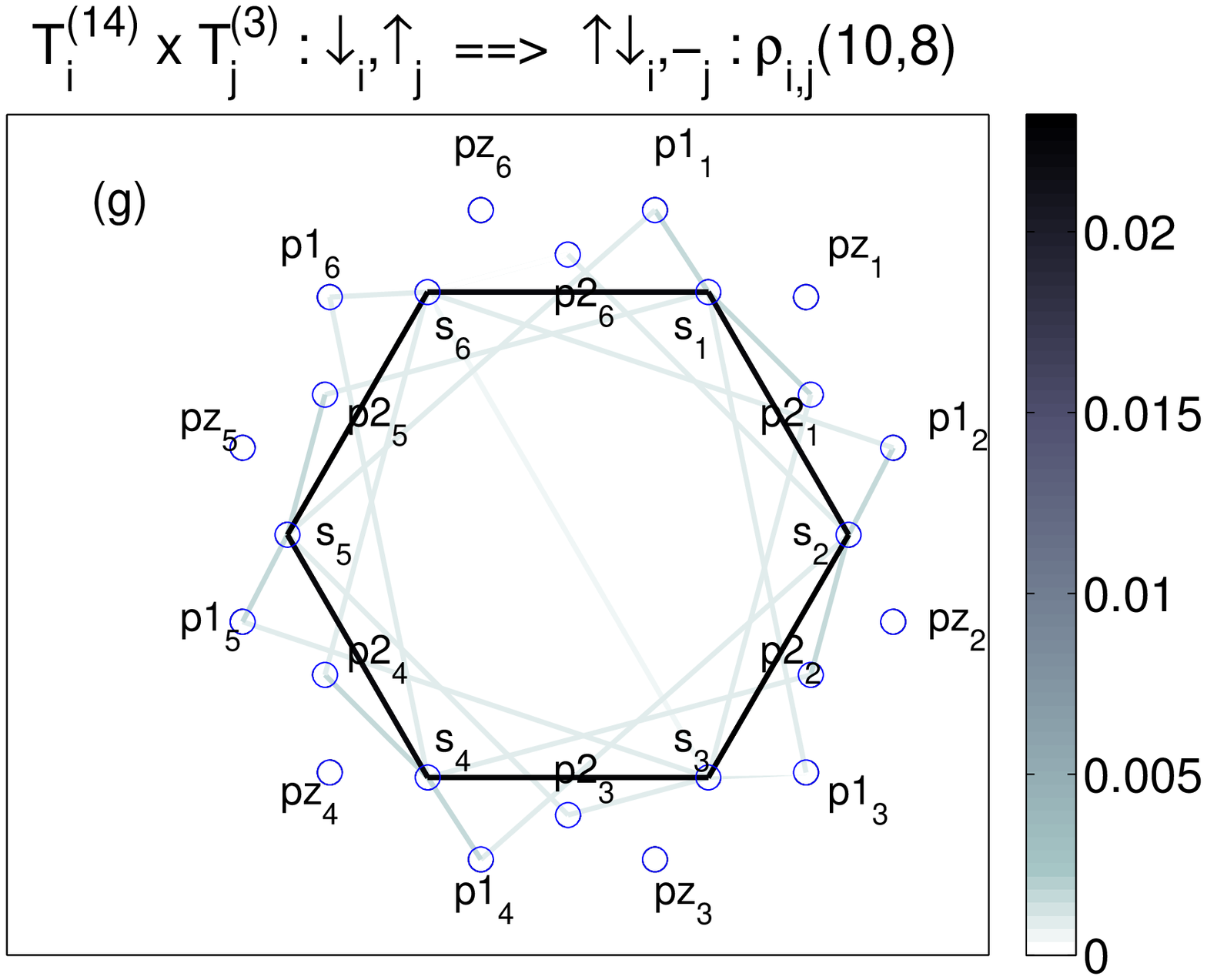}
\includegraphics[scale=0.25]{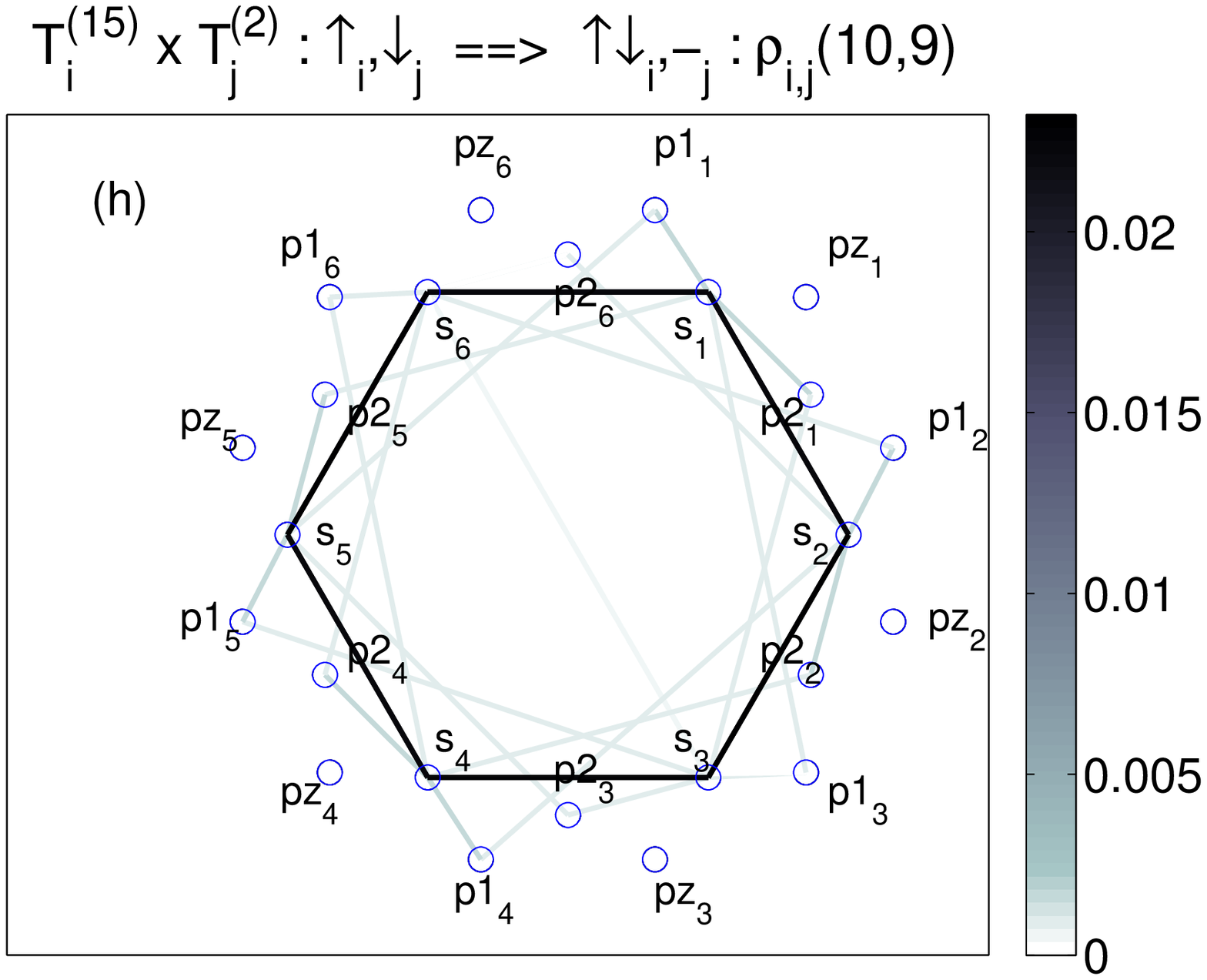}
}
\centerline{
\includegraphics[scale=0.25]{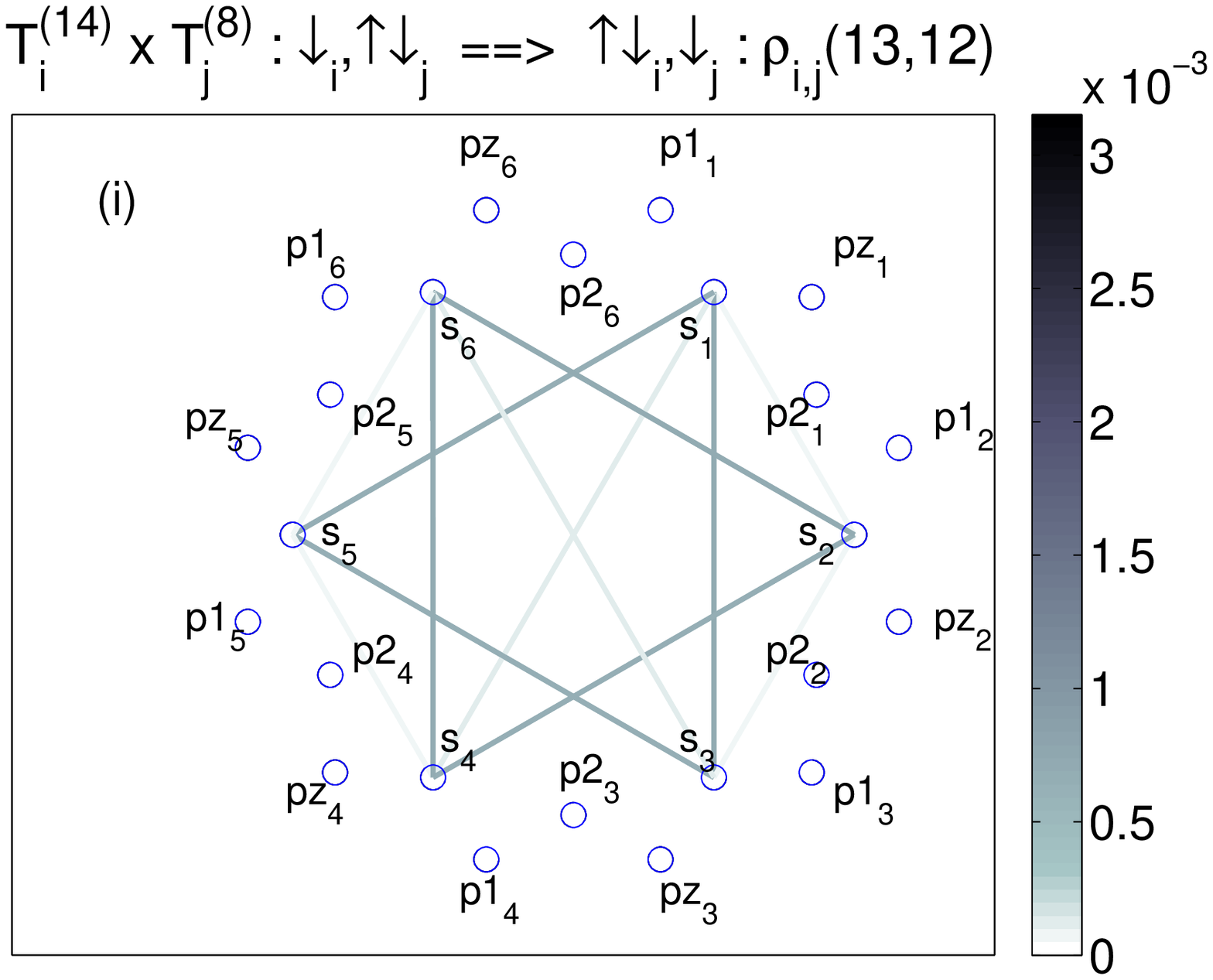}
\includegraphics[scale=0.25]{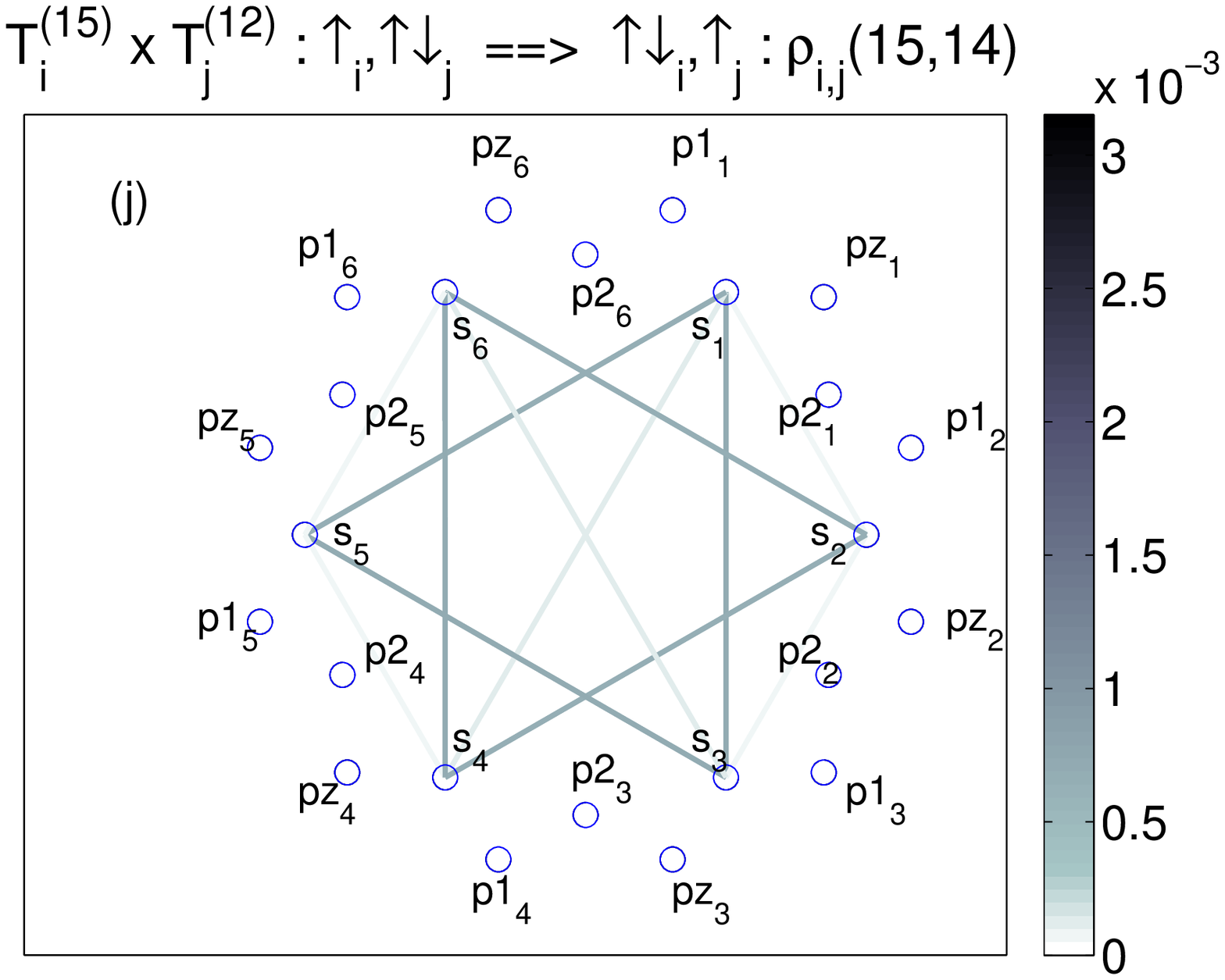}
}
\end{minipage}
\end{minipage}
\caption{Pictorial representation of the absolute value of the generalized correlation functions used to construct the lower-triangular elements of the two-orbital reduced density matrix for Li$_{6}$ using 2$s$ and 2$p$ atomic functions at $d_{\rm Li-Li}=3.05${\AA } and at $d_{\rm Li-Li}=6.00${\AA }. Strength of transition amplitues between initial ($|\alpha_i\rangle|\beta_j\rangle$) and final states ($|\alpha^\prime_i\rangle|\beta^\prime_j\rangle$) on orbital $i$ and $j$ are indicated with different line colors.}
\label{fig:li_rho_sto-3g}
\end{figure*}
\begin{figure*}
\begin{minipage}{20cm}
\hskip -2.0cm
\begin{minipage}{8cm}
\centerline{
\includegraphics[scale=0.25]{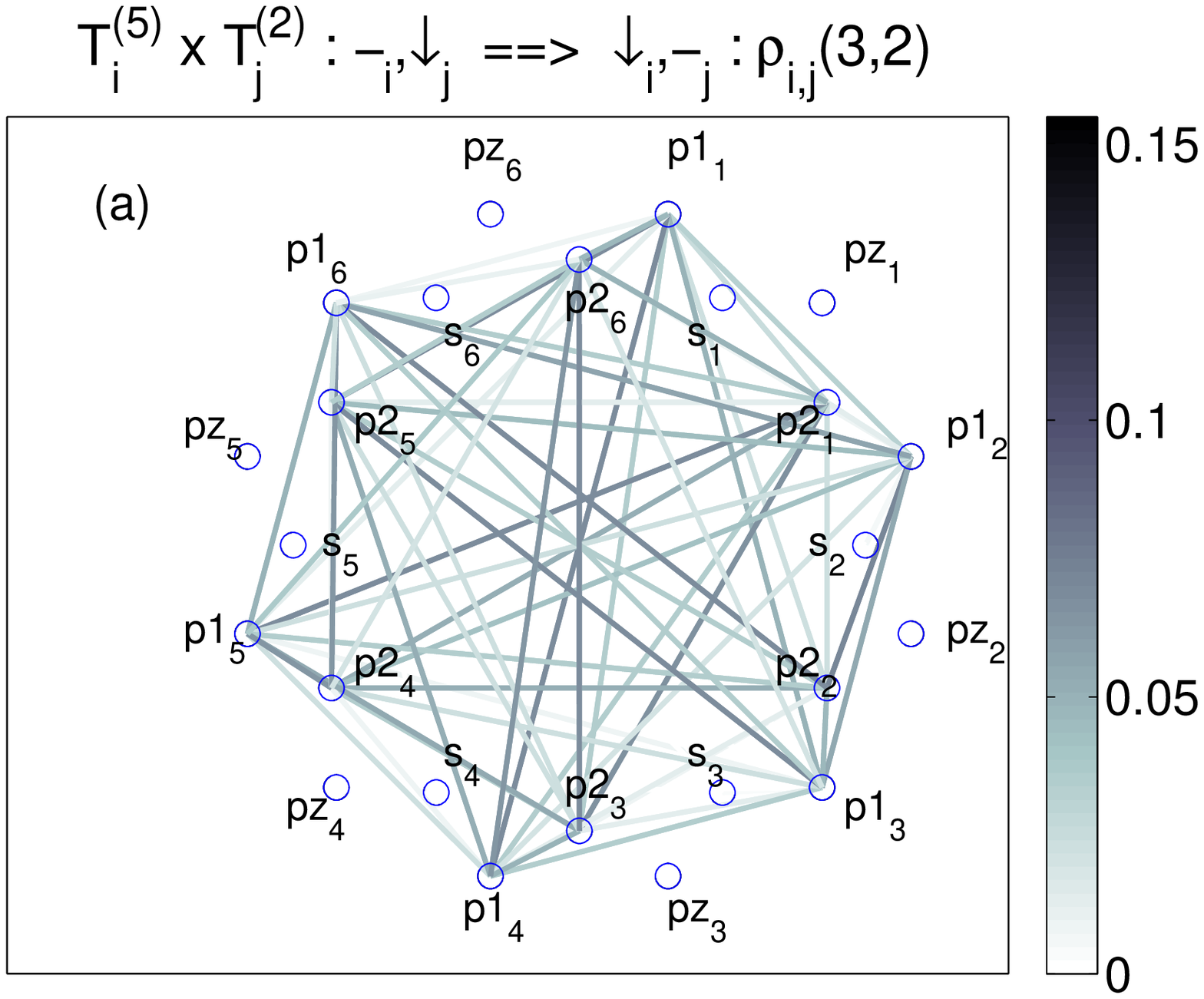}
\includegraphics[scale=0.25]{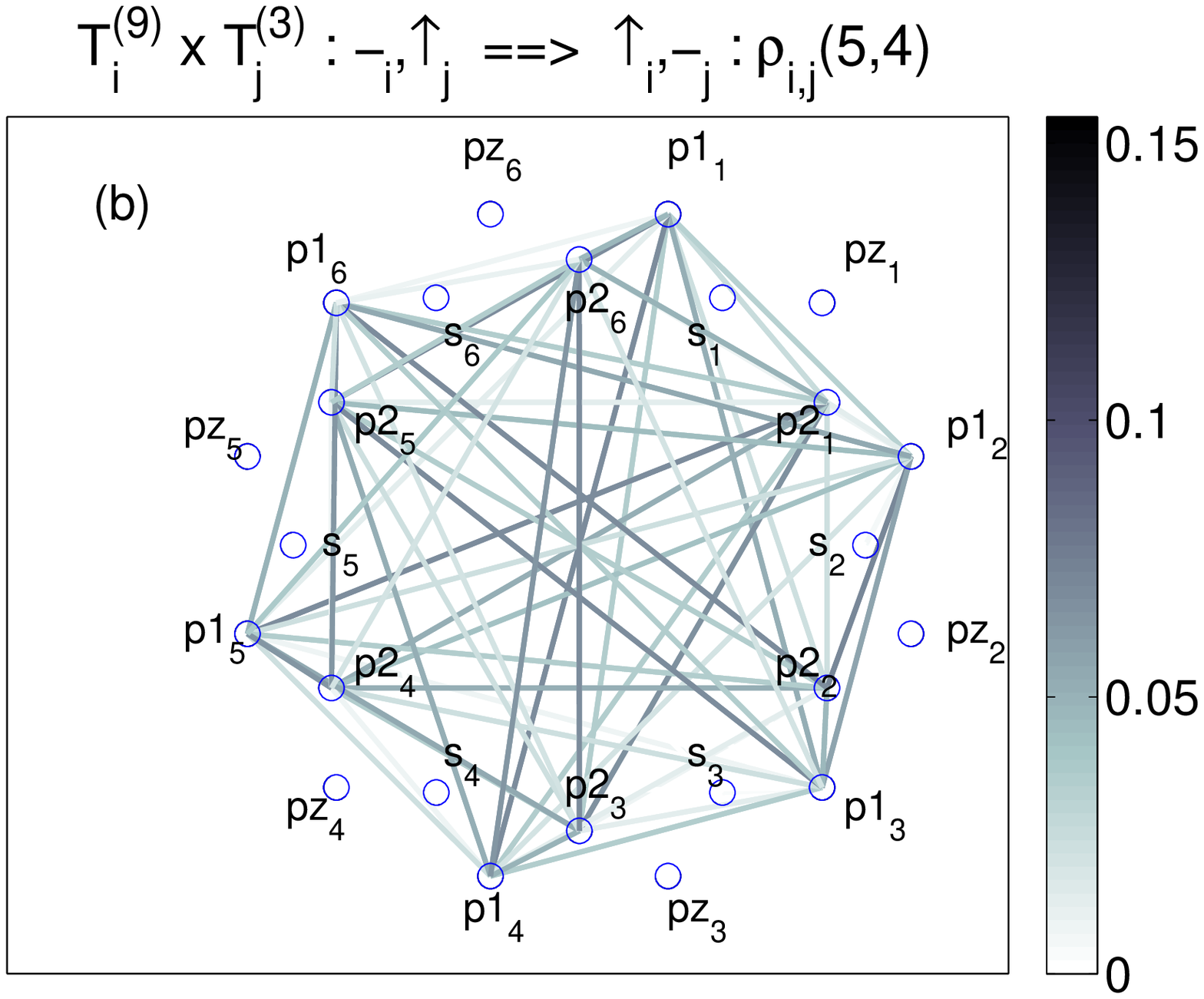}
}
\centerline{
\includegraphics[scale=0.25]{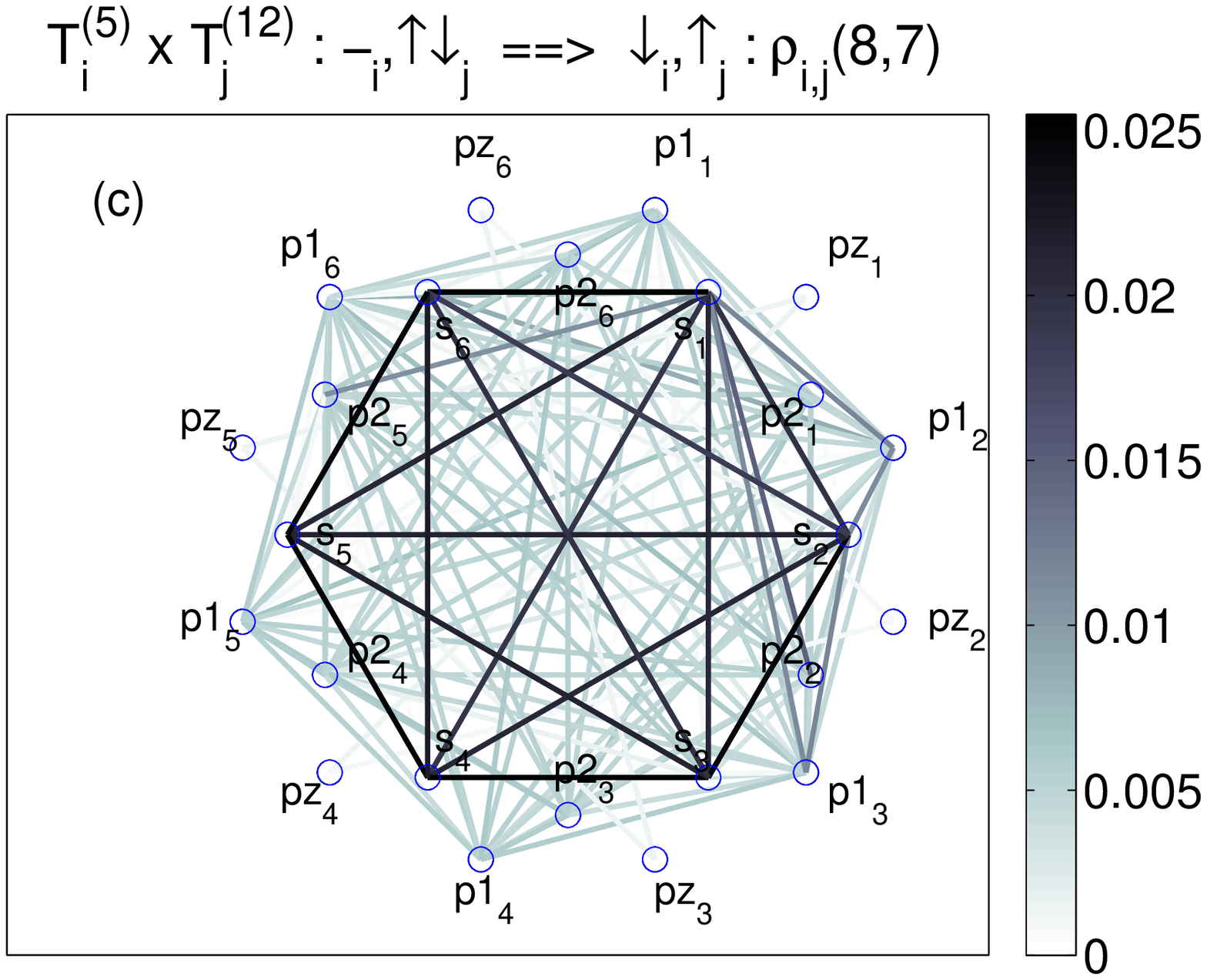}
\includegraphics[scale=0.25]{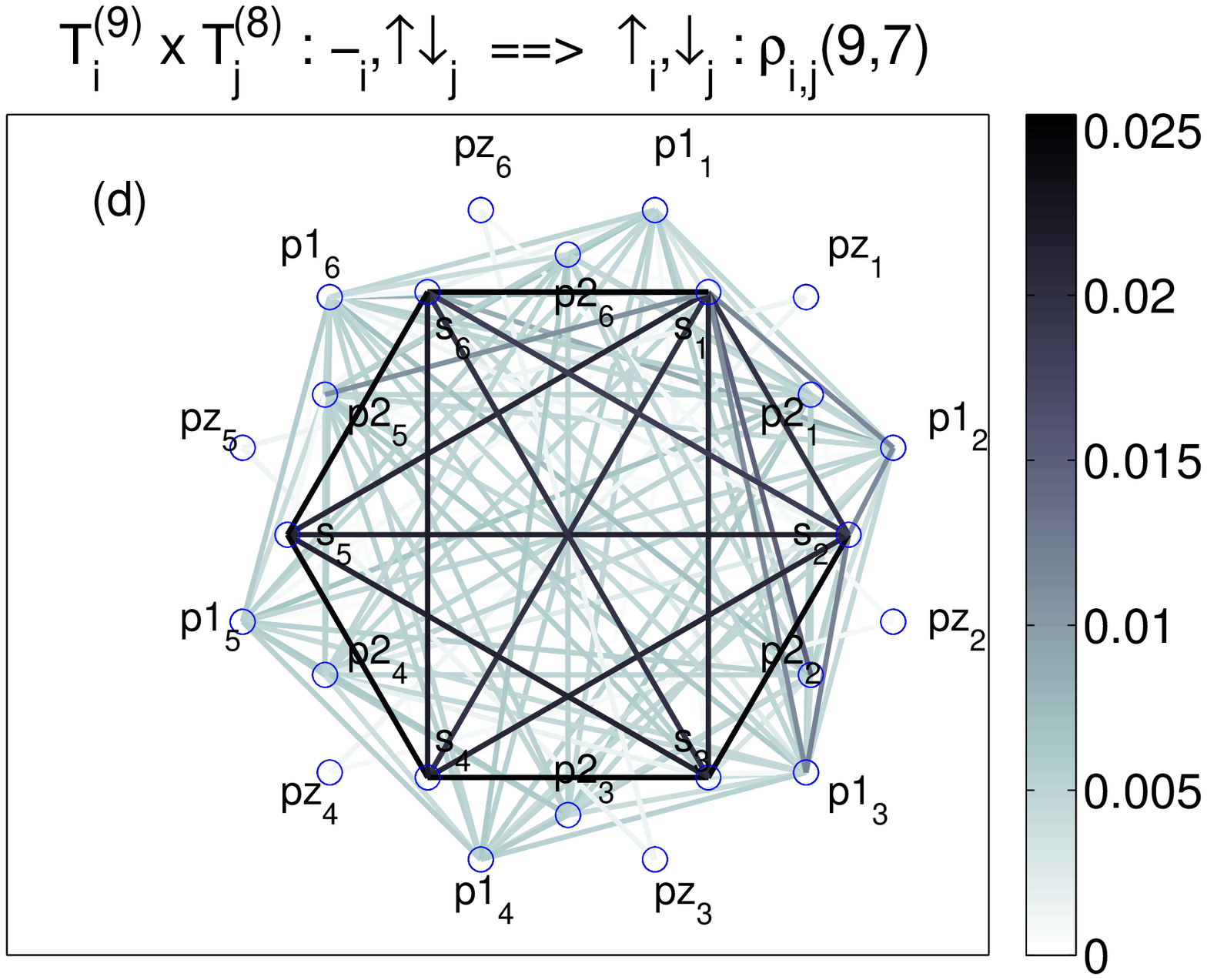}
}
\centerline{
\includegraphics[scale=0.25]{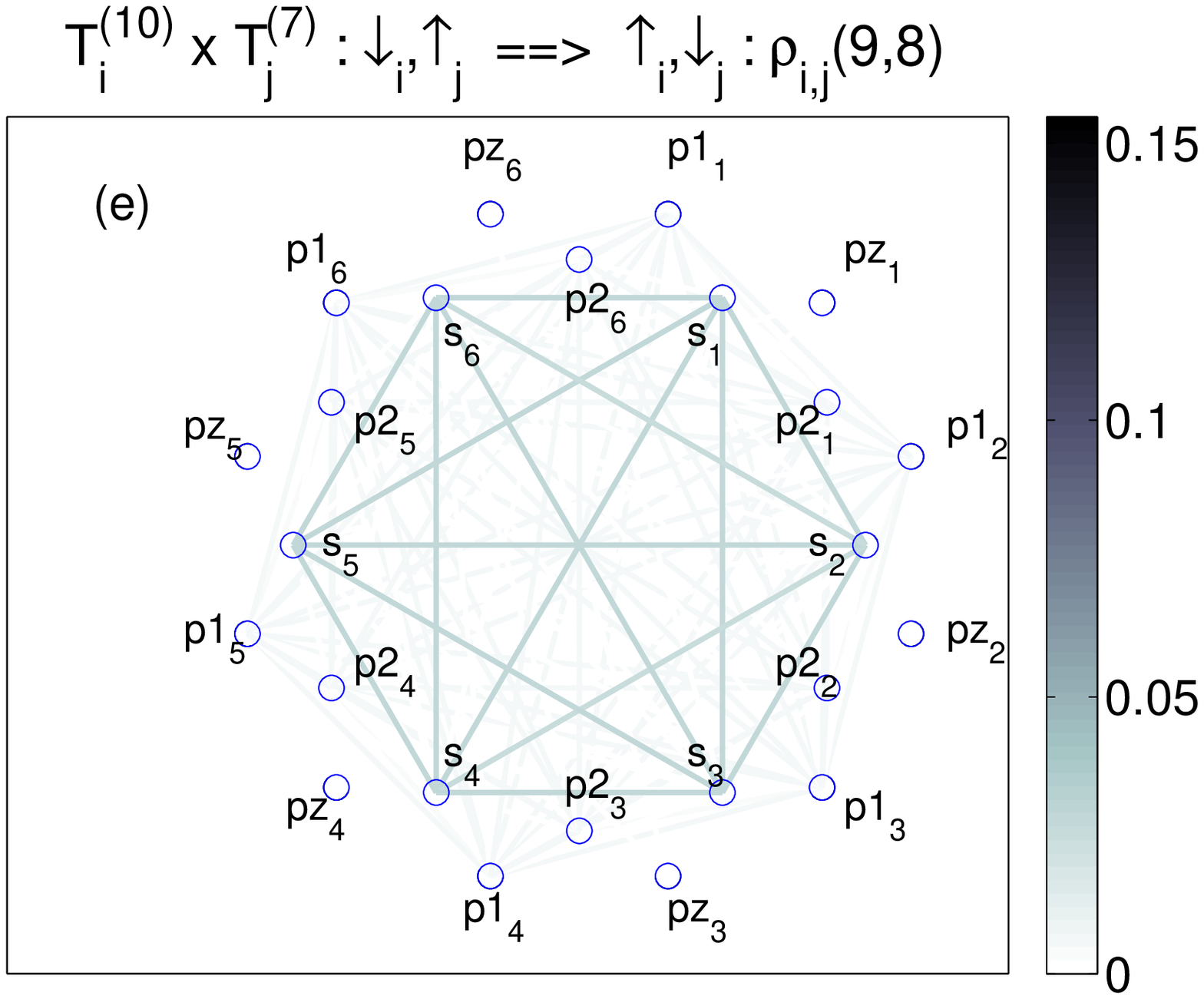}
\includegraphics[scale=0.25]{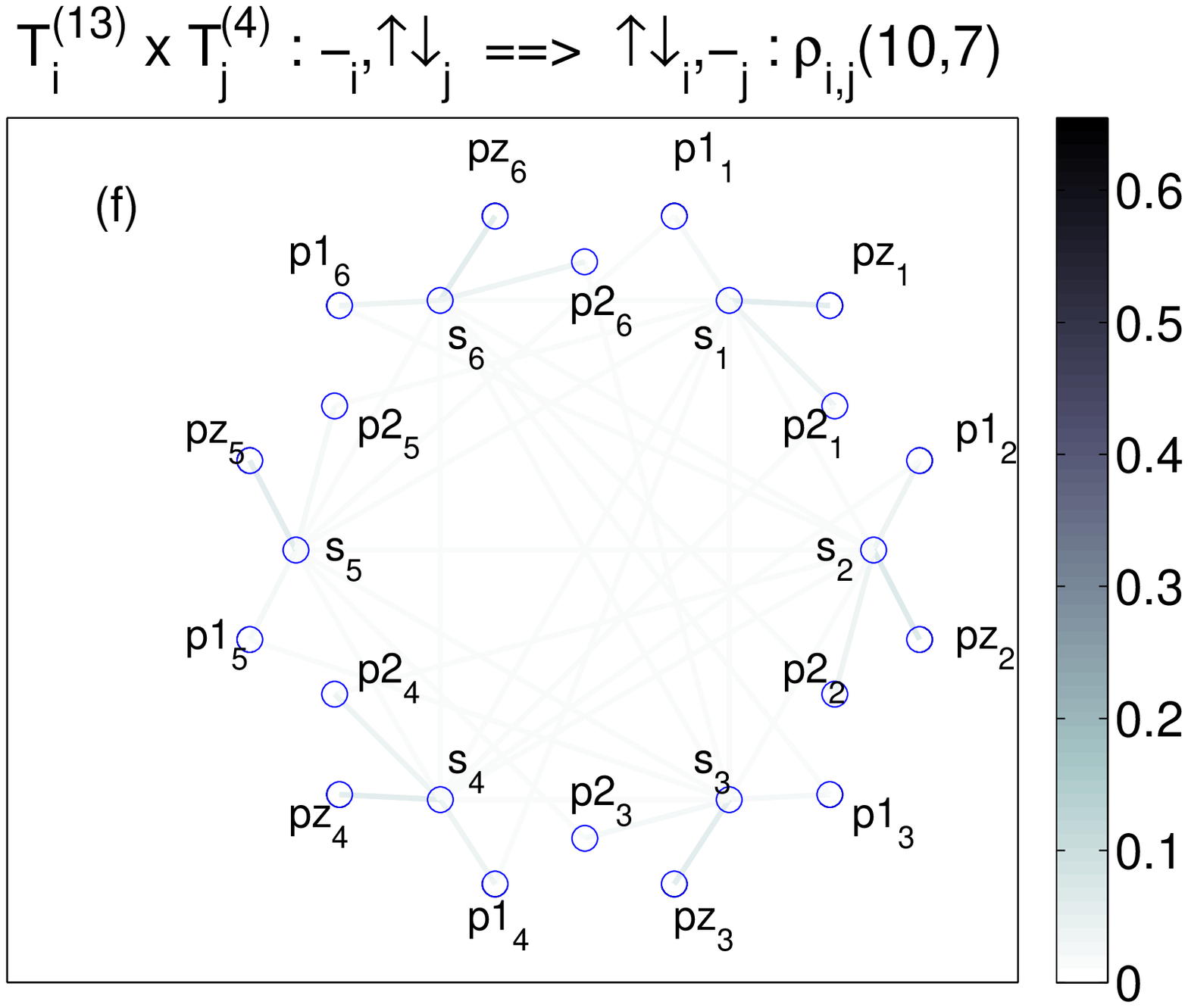}
}
\centerline{
\includegraphics[scale=0.25]{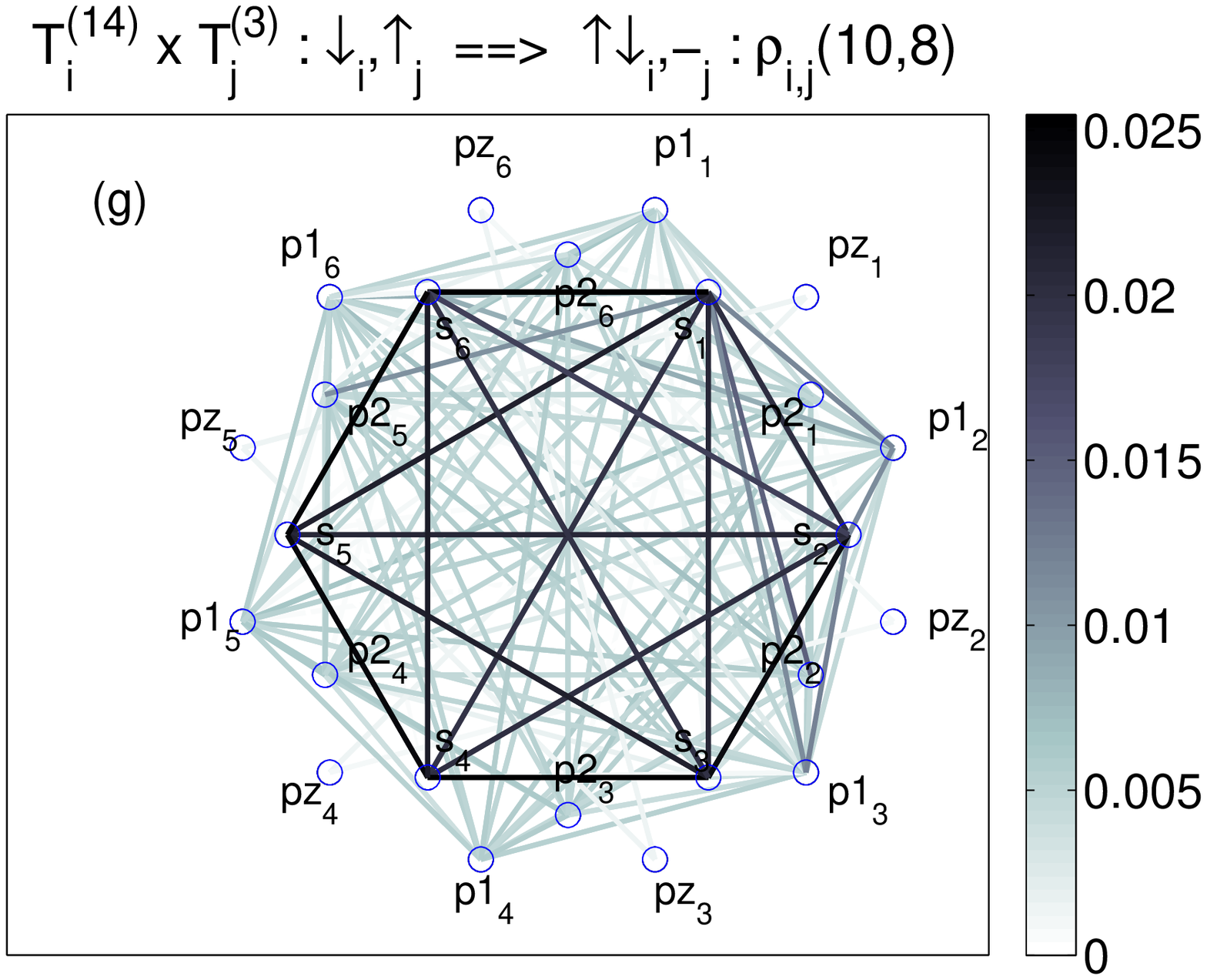}
\includegraphics[scale=0.25]{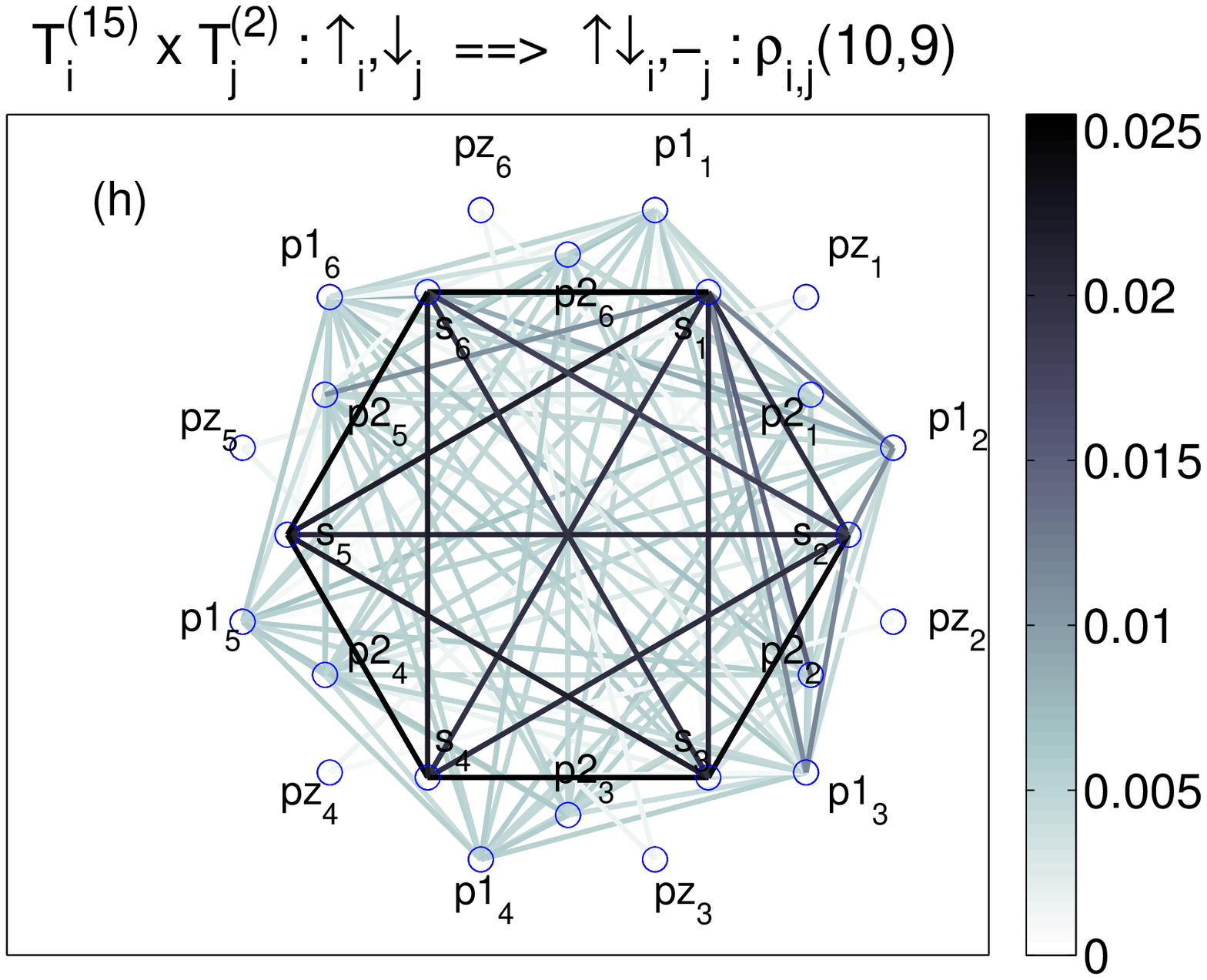}
}
\centerline{
\includegraphics[scale=0.25]{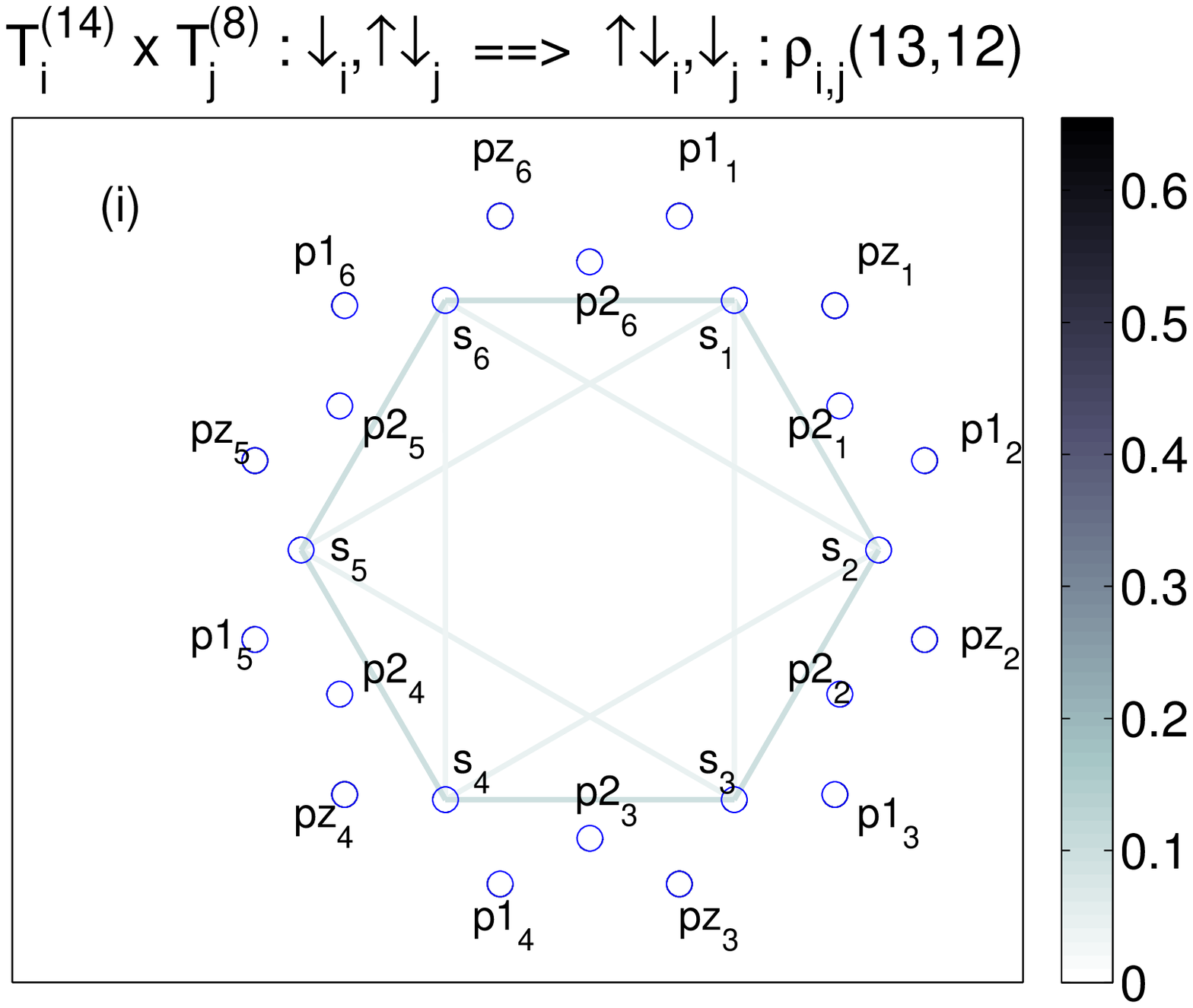}
\includegraphics[scale=0.25]{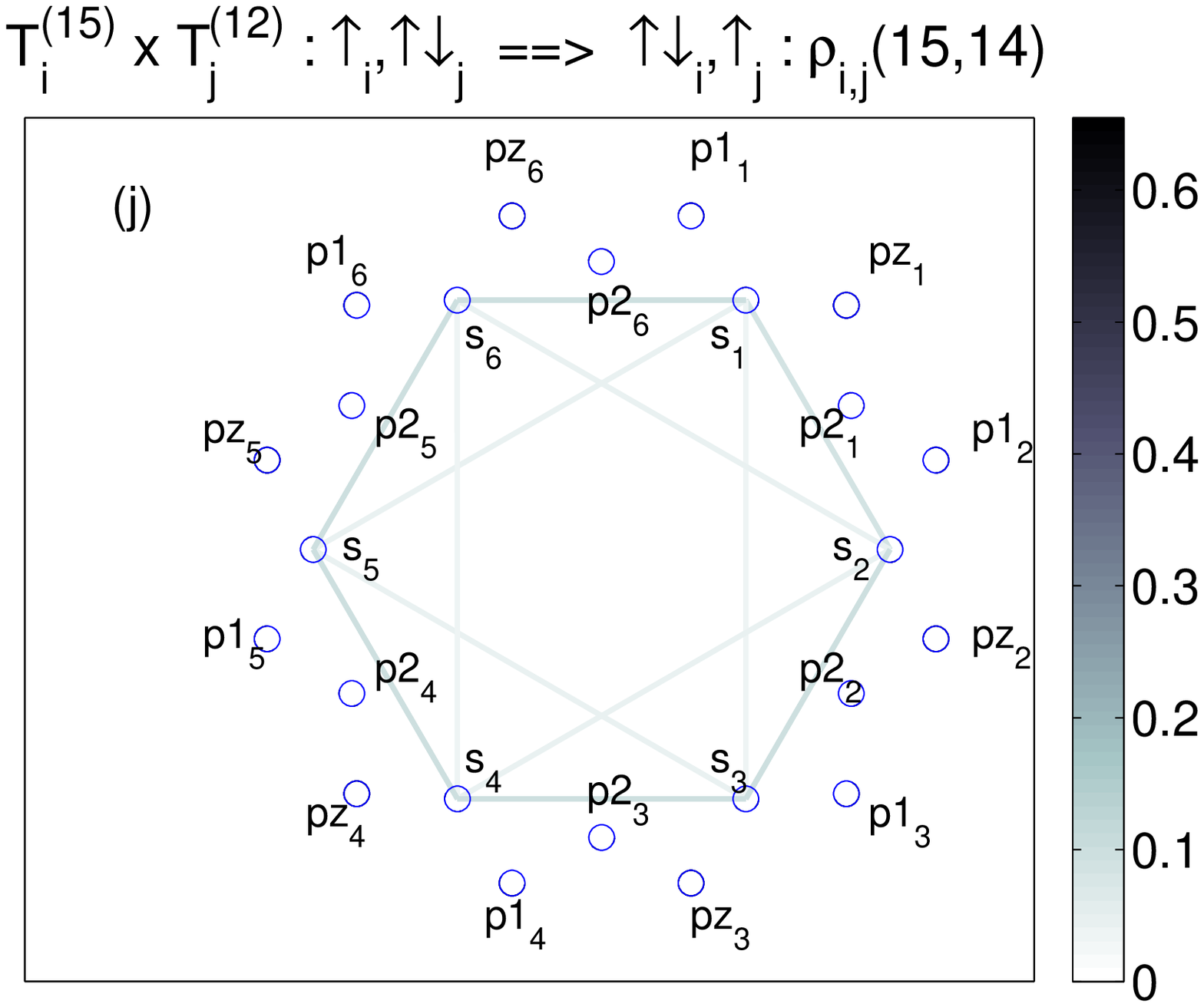}
}
\end{minipage}
\hskip 1.0cm
\begin{minipage}{8cm}
\centerline{
\includegraphics[scale=0.25]{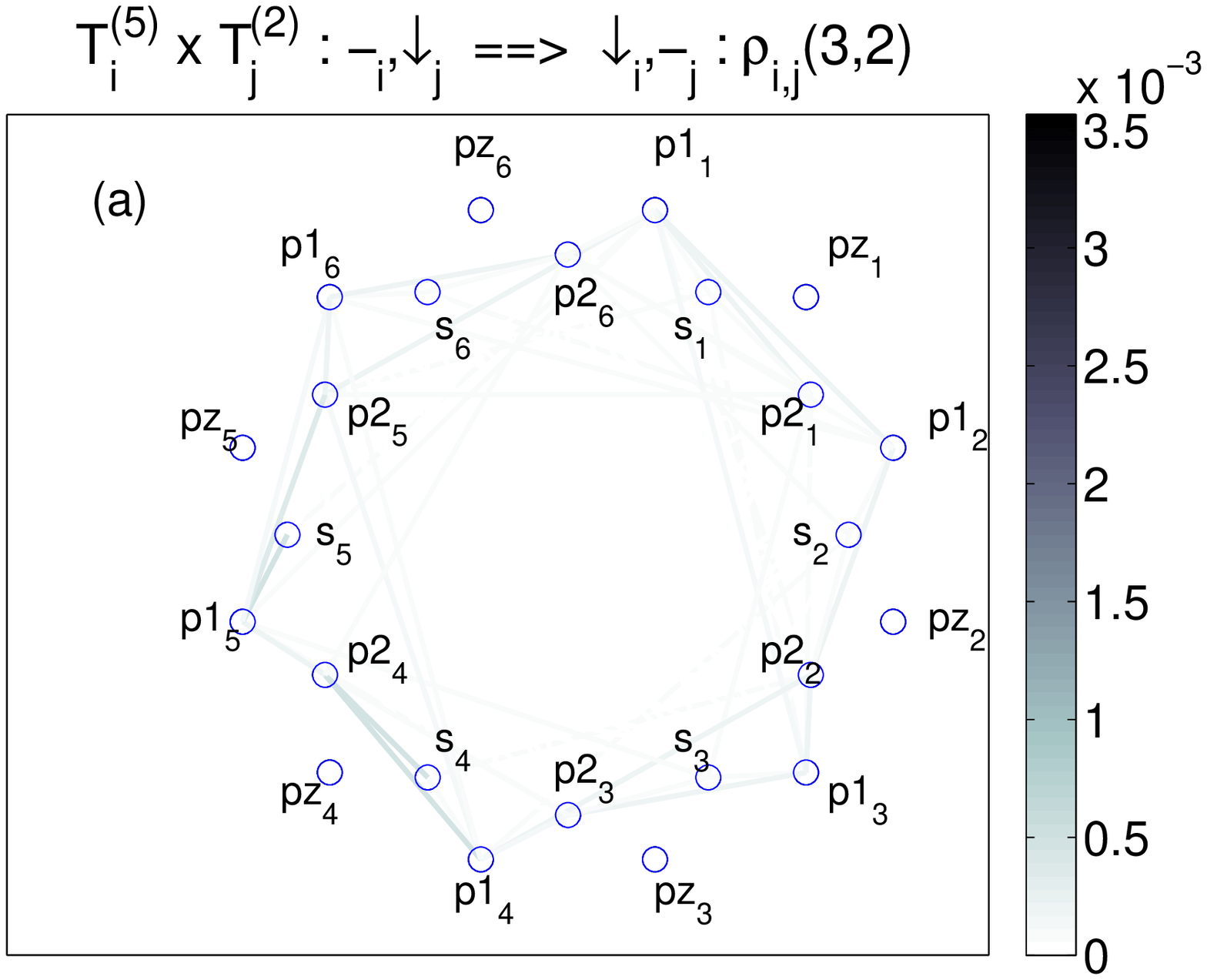}
\includegraphics[scale=0.25]{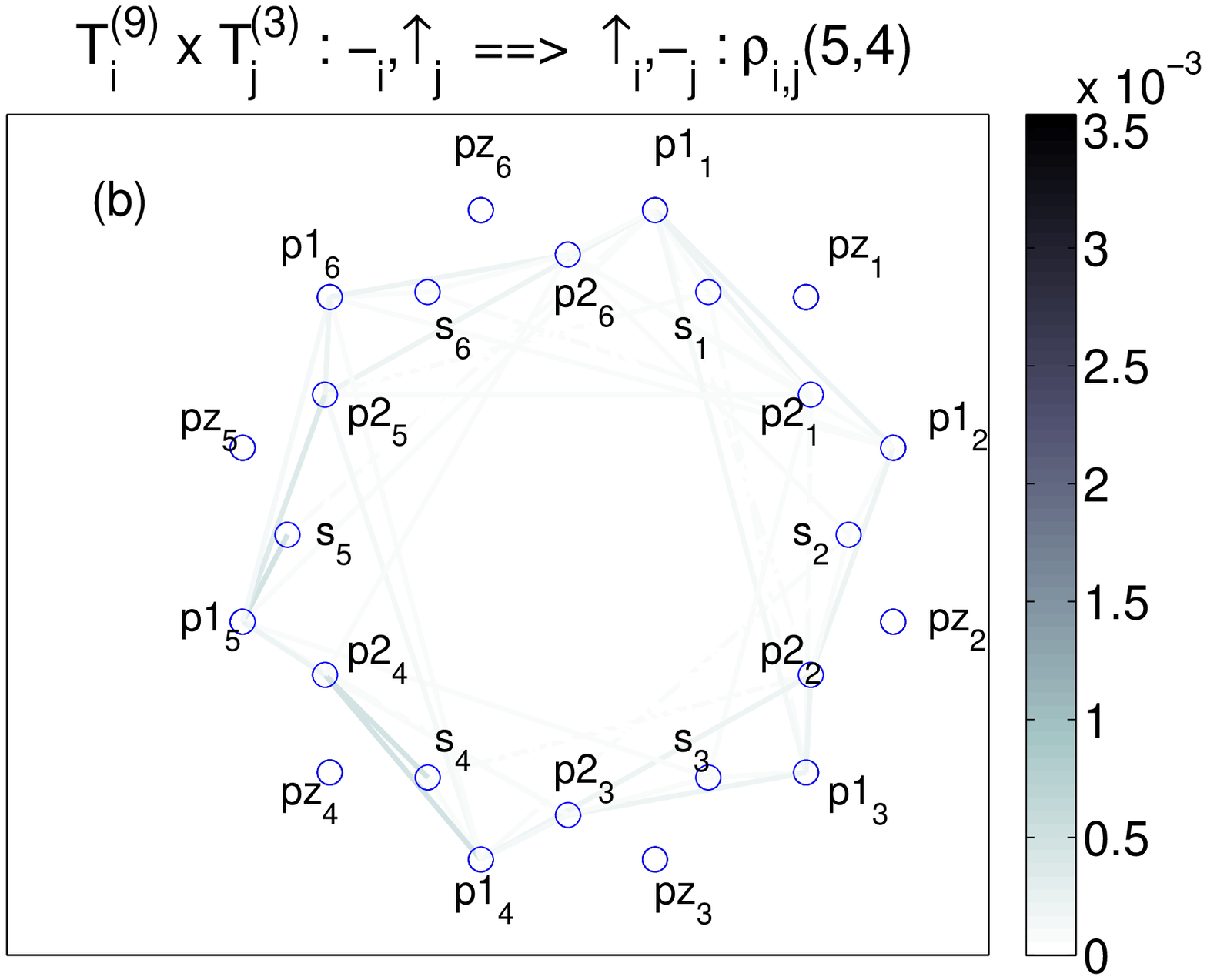}
}
\centerline{
\includegraphics[scale=0.25]{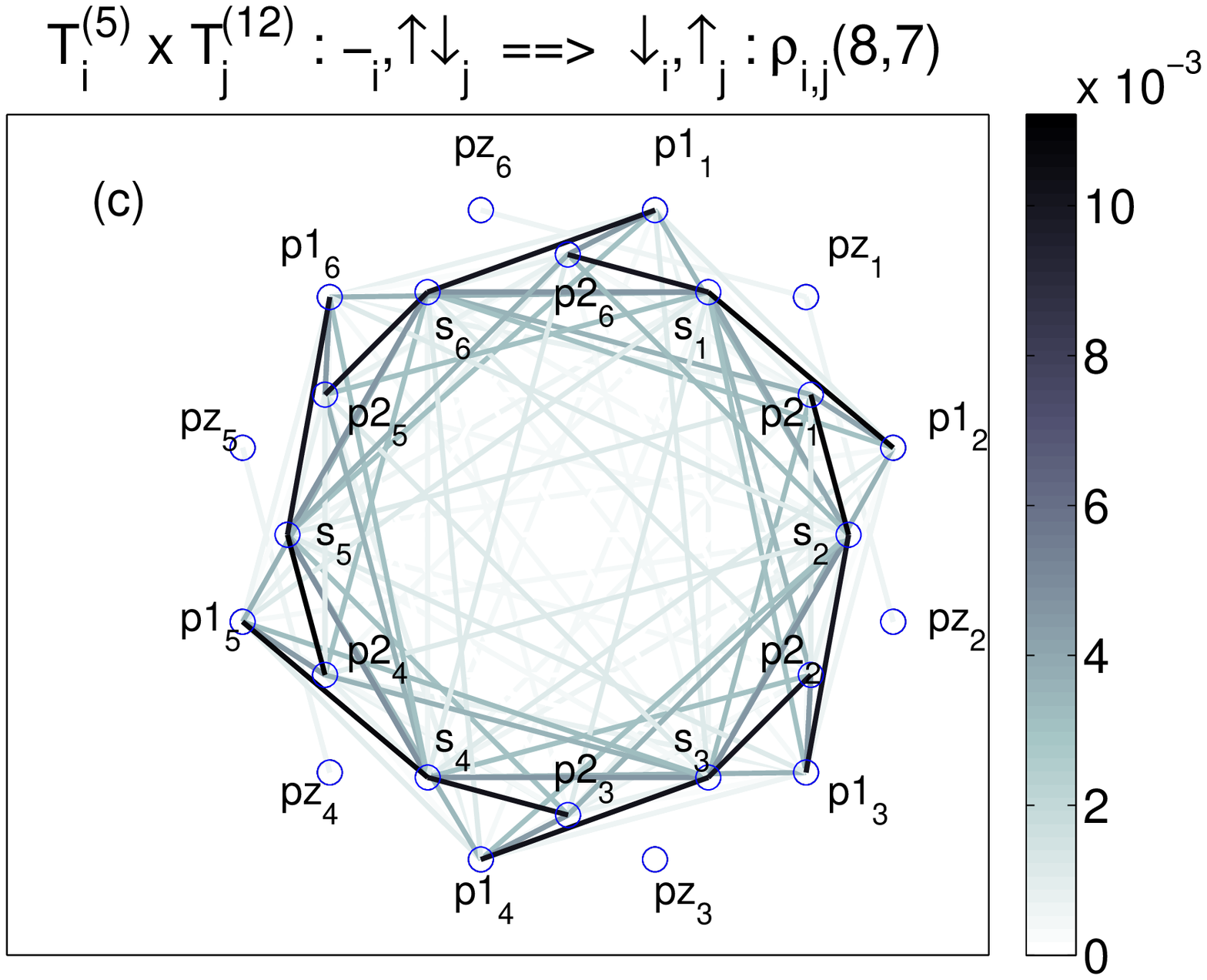}
\includegraphics[scale=0.25]{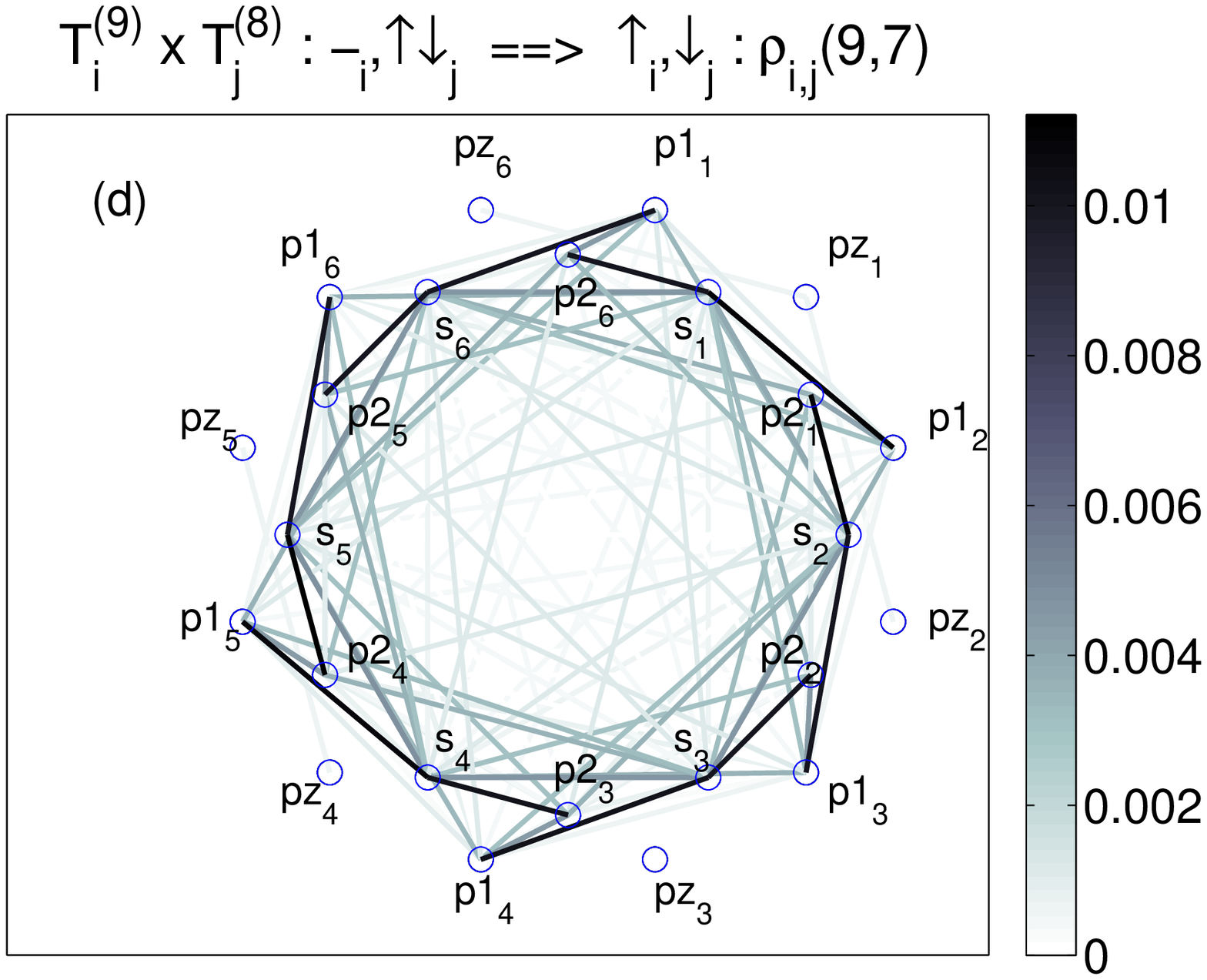}
}
\centerline{
\includegraphics[scale=0.25]{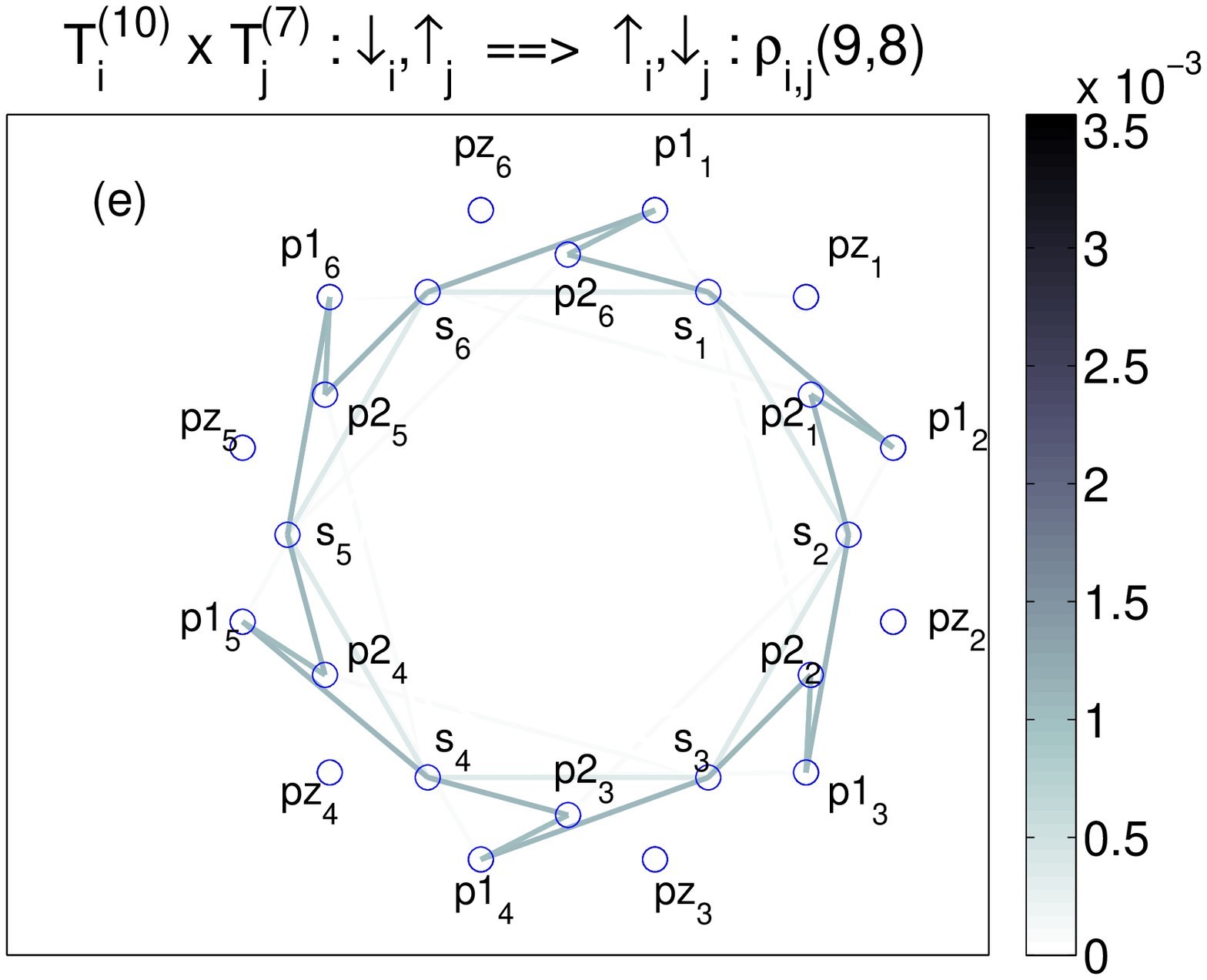}
\includegraphics[scale=0.25]{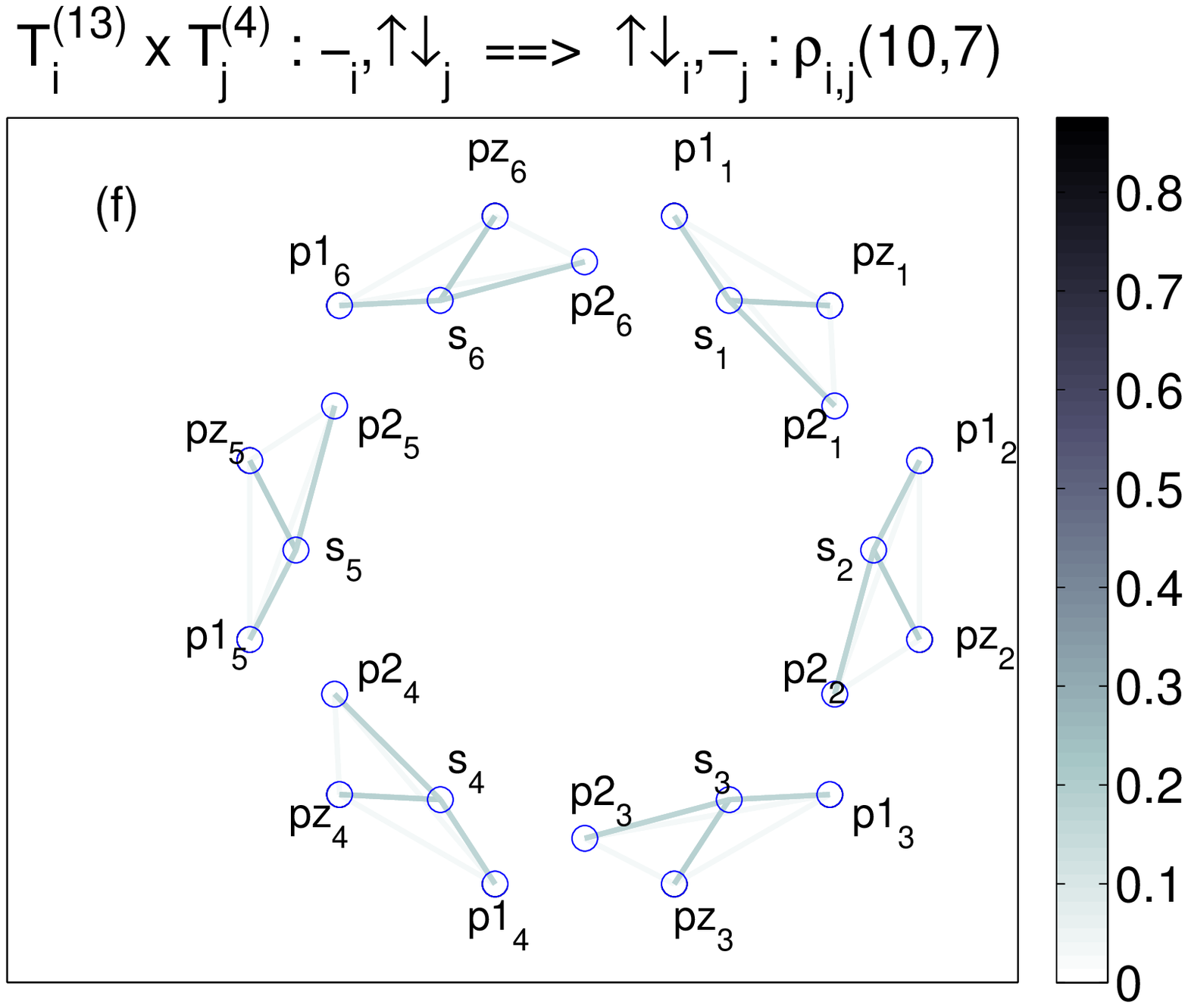}
}
\centerline{
\includegraphics[scale=0.25]{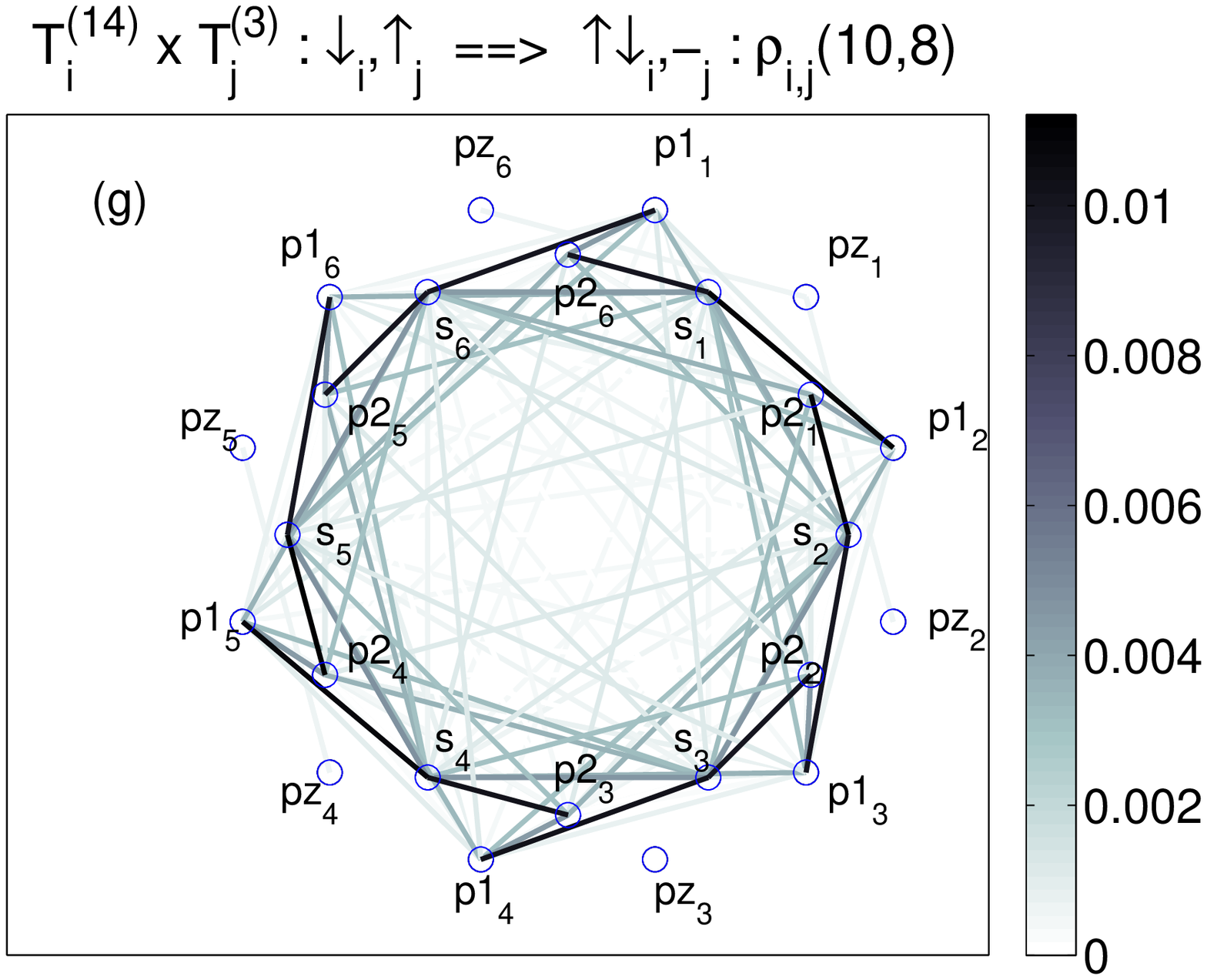}
\includegraphics[scale=0.25]{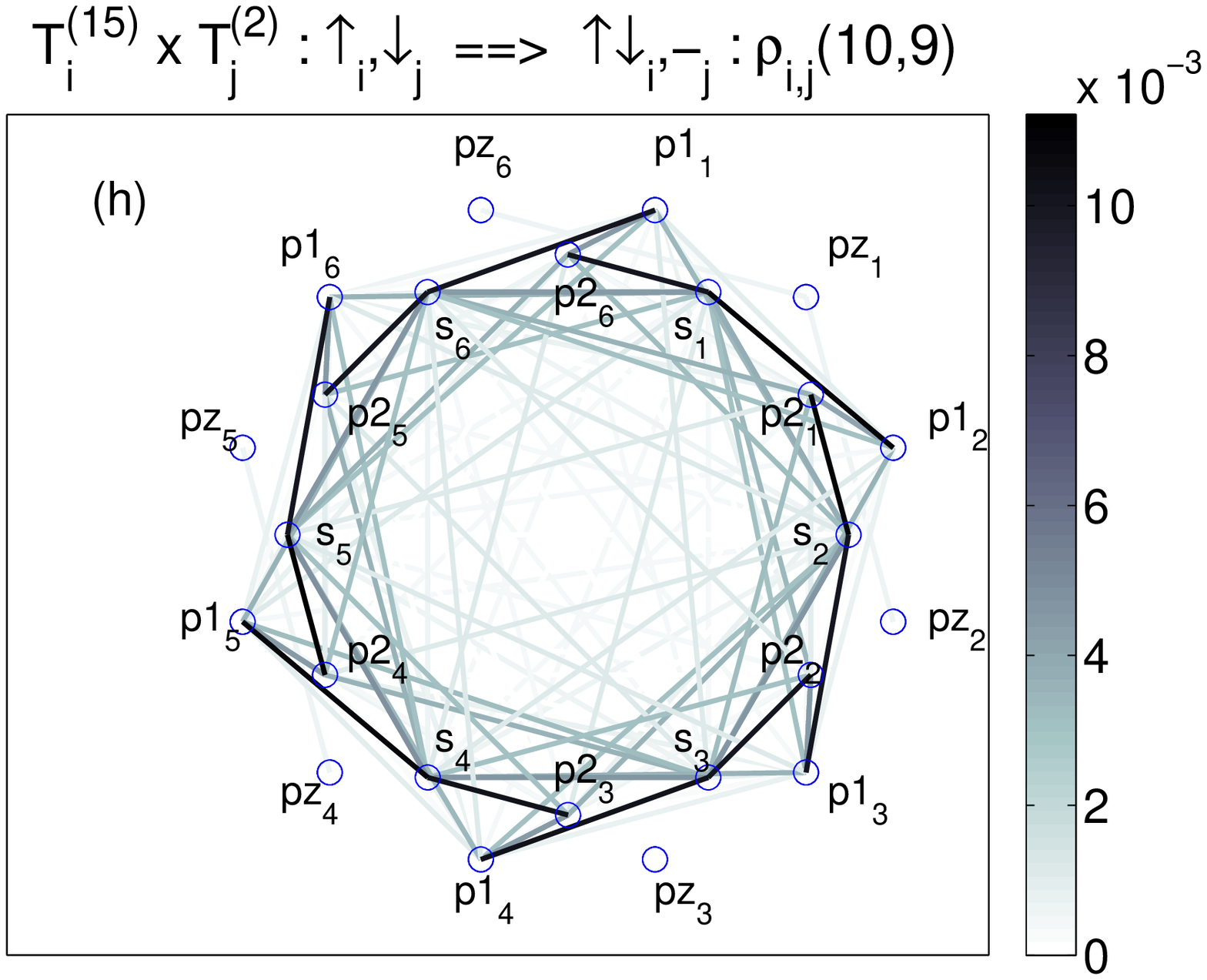}
}
\centerline{
\includegraphics[scale=0.25]{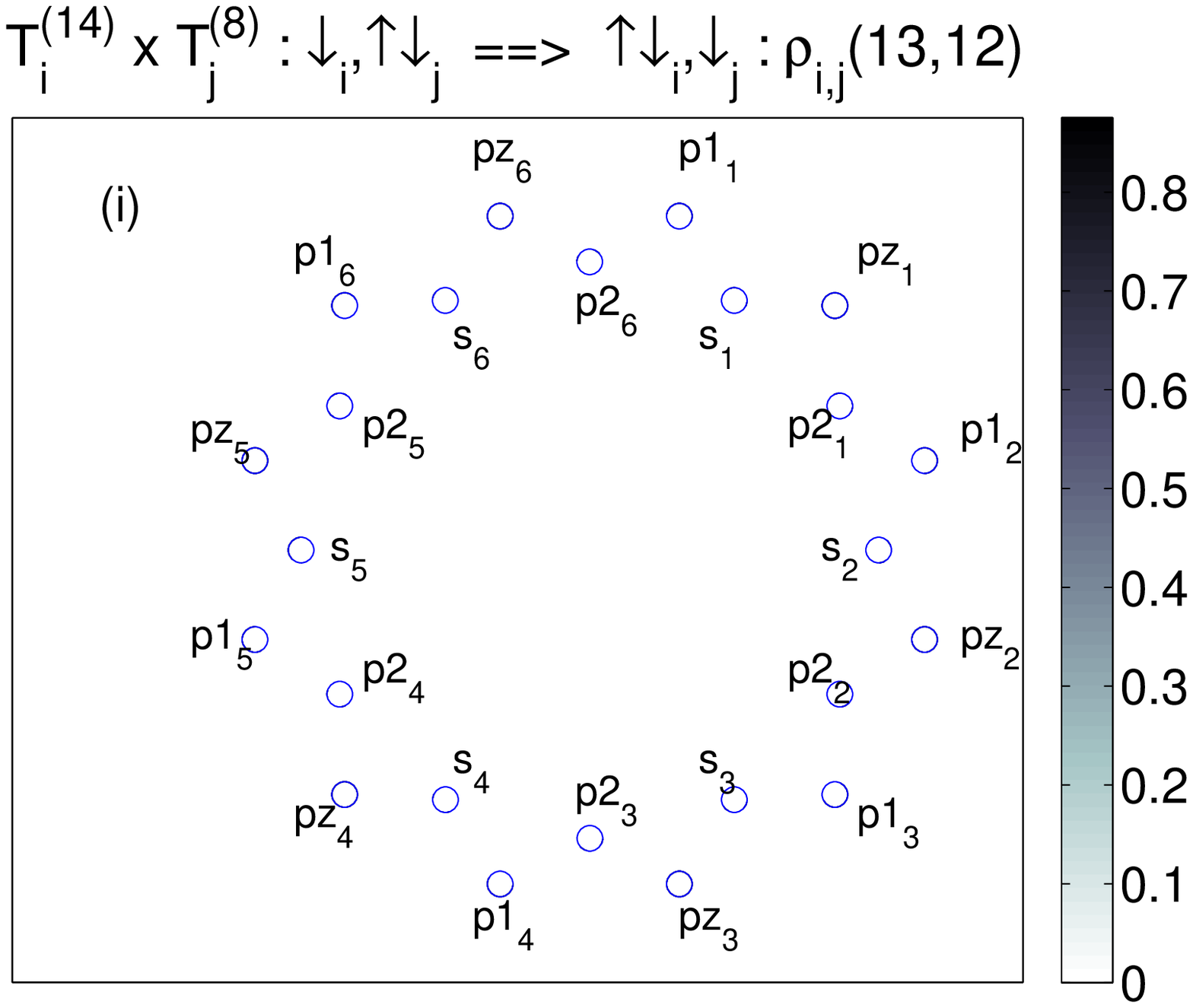}
\includegraphics[scale=0.25]{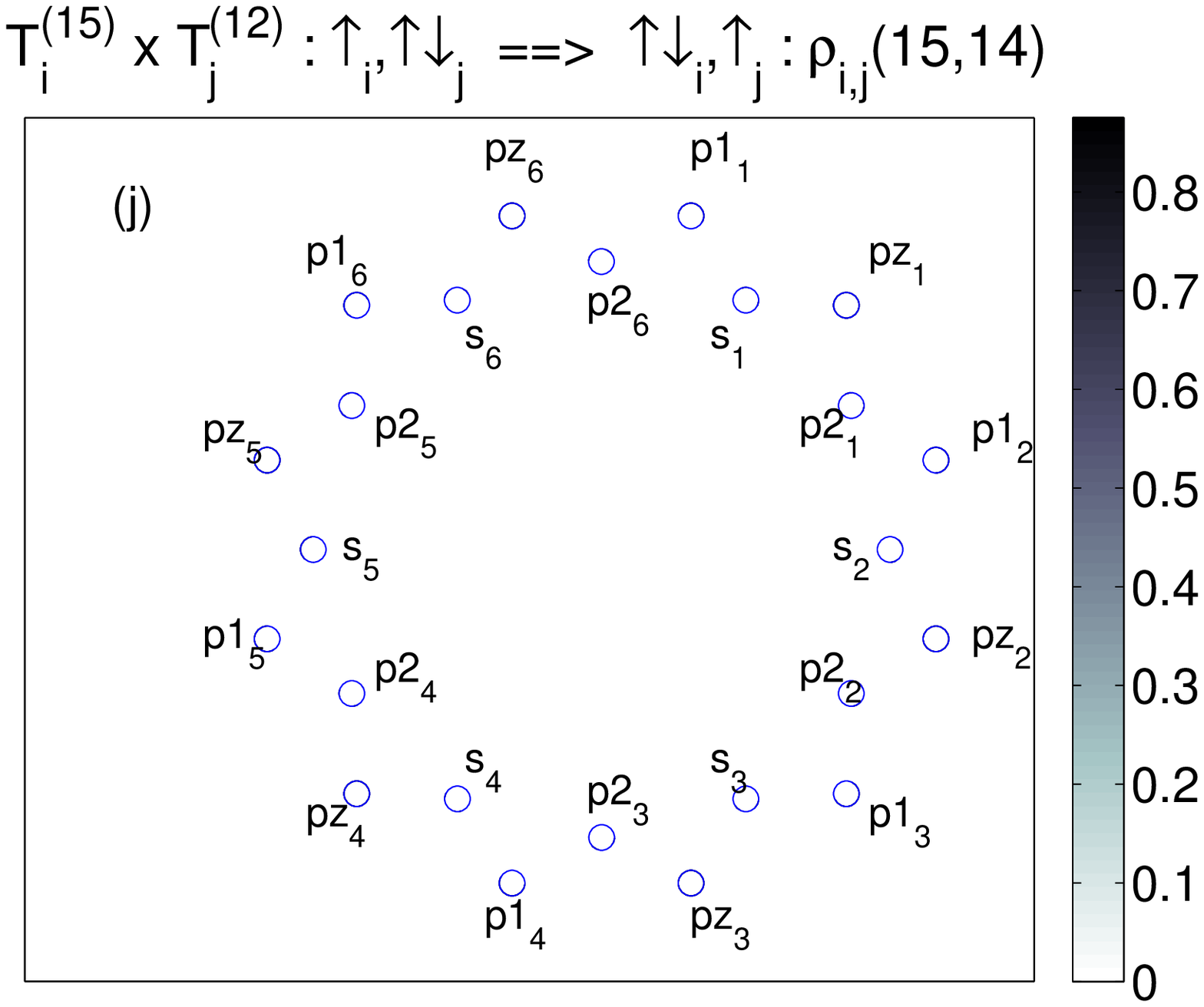}
}
\end{minipage}
\end{minipage}
\caption{Pictorial representation of the absolute value of the generalized correlation functions used to construct the lower-triangular elements of the two-orbital reduced density matrix for Be$_{6}$ using 2$s$ and 2$p$ atomic functions at $d_{\rm Be-Be}=2.15${\AA } and at $d_{\rm Be-Be}=3.30${\AA }. Strength of transition amplitues between initial ($|\alpha_i\rangle|\beta_j\rangle$) and final states ($|\alpha^\prime_i\rangle|\beta^\prime_j\rangle$) on orbital $i$ and $j$ are indicated with different line colors.}\label{fig:be_rho}
\end{figure*}

\end{document}